%% file: HIG-22-001_temp.tex
\pdfoutput=1
\documentclass[11pt,twoside,a4paper,cmspaper,final,collab]{cms-tdr}
\def\svnVersion{de3ef52-D}\def\svnDate{2022/06/30}\def\cmsCernNoTag{CERN-EP-2022-039}\def\cmsCernDate{\today}\def\cmsMessage{Published in Nature as \href{http://dx.doi.org/10.1038/s41586-022-04892-x}{\doi{10.1038/s41586-022-04892-x}.}}
\begin{document}\cmsNoteHeader{HIG-22-001}

\newcommand{\ttH}{\ensuremath{\PQt\PQt\PH}\xspace}
\newcommand{\tH}{\ensuremath{\PQt\PH}\xspace}
\newcommand{\bbH}{\ensuremath{\PQb\PQb\PH}\xspace}
\newcommand{\ggH}{\ensuremath{\Pg\Pg\PH}\xspace}
\newcommand{\VH}{\ensuremath{\PV\PH}\xspace}
\newcommand{\WH}{\ensuremath{\PW\PH}\xspace}
\newcommand{\ZH}{\ensuremath{\PZ\PH}\xspace}
\newcommand{\VVHH}{\ensuremath{\PV\PV\PH\PH}\xspace}
\newcommand{\bb}{\ensuremath{\PQb\PQb}\xspace}
\newcommand{\ditau}{\ensuremath{\PGt\PGt}\xspace}
\newcommand{\ZZ}{\ensuremath{\PZ\PZ}\xspace}
\newcommand{\WW}{\ensuremath{\PW\PW}\xspace}
\newcommand{\VZ}{\ensuremath{\PV\PZ}\xspace}
\newcommand{\cmsq}{\ensuremath{\,\text{cm}^{2}}\xspace}
\newcommand{\meter}{\unit{m}}
\newcommand{\kappav}{\ensuremath{\kappa_\mathrm{V}}}
\newcommand{\kappaf}{\ensuremath{\kappa_\mathrm{f}}}
\newcommand{\kappal}{\ensuremath{\kappa_\lambda}}
\newcommand{\kappavv}{\ensuremath{\kappa_\mathrm{2V}}}
\newcommand{\lint}{\mathcal{L}} 
\newcommand{\desc}[1]{\textbf{#1:}}
\providecommand{\cmsTable}[1]{\resizebox{\textwidth}{!}{#1}}
\ifthenelse{\boolean{cms@external}}{\providecommand{\cmsTableInt}[1]{#1}}{\providecommand{\cmsTableInt}[1]{\cmsTable{#1}}}
\newcolumntype{d}[1]{D{.}{.}{#1}}
\newlength\cmsTabSkip\setlength{\cmsTabSkip}{3ex}

\newcommand{\figtitle}[1]{ {\bf #1\\[0.5ex]}}

\cmsNoteHeader{HIG-22-001}
\title{A portrait of the Higgs boson by the CMS experiment ten years after the discovery}
\titlerunning{CMS: The Higgs at 10}

\author*[cern]{CMS Collaboration}

\date{\today}

\abstract{In July 2012, the ATLAS and CMS Collaborations at the CERN Large Hadron Collider announced the observation of a Higgs boson at a mass of around 125\GeV. Ten years later, and with the data corresponding to the production of 30 times larger number of Higgs bosons, we have learnt much more about the properties of the Higgs boson. The CMS experiment has observed the Higgs boson in numerous fermionic and bosonic decay channels, established its spin-parity quantum numbers, determined its mass and measured its production cross sections in various modes. Here the CMS Collaboration reports the most up-to-date combination of results on the properties of the Higgs boson, including the most stringent limit on the cross section for the production of a pair of Higgs bosons, on the basis of data from proton-proton collisions at a centre-of-mass energy of 13\TeV. Within the uncertainties, all these observations are compatible with the predictions of the standard model of elementary particle physics. Much evidence points to the fact that the standard model is a low-energy approximation of a more comprehensive theory. Several of the standard model issues originate in the sector of Higgs boson physics. An order of magnitude larger number of Higgs bosons, expected to be examined over the next fifteen years, will help deepen our understanding of this crucial sector.
}

\hypersetup{
pdfauthor={CMS Collaboration},
pdftitle={A portrait of the Higgs boson by the CMS experiment},
pdfsubject={CMS},
pdfkeywords={CMS, Higgs boson, Higgs boson pair production}}

\maketitle 

The established theory of elementary particle physics, commonly referred to as the standard model (SM), provides a complete description of the electromagnetic (EM), weak, 
and strong interactions of matter particles, which are spin-1/2 fermions, through three different sets of mediators, which are spin-1 
bosons. (In quantum mechanics, spin is an intrinsic form of angular momentum carried by elementary particles.)
These vector bosons are the massless photons (gluons) for the EM (strong) interaction, and the heavy \PW and \PZ bosons for the weak interaction.
The SM has been very successful in providing accurate predictions for essentially all particle physics experiments carried out 
so far. 
In 2012, the final missing particle of the SM, the Higgs boson, was observed by the ATLAS \cite{ATLAS:2012yve} 
and CMS \cite{CMS:2012qbp,CMS:2013btf} Collaborations at CERN.

The Higgs boson is a prediction of a mechanism that
took place in the early universe, less than a picosecond after the Big Bang, which led to the EM and the weak
interactions becoming distinct in their actions.
In the SM, this mechanism, labelled as the Brout--Englert--Higgs (BEH) mechanism, introduces a complex scalar (spin-0)
field that permeates the entire universe. Its quantum manifestation is known as the SM Higgs boson. 
Scalar fields are described only by a number at every point in space that is invariant under 
Lorentz transformations. An analogy can be drawn of a map of an area where temperature is shown at various 
positions mimicking a scalar field. The same map, where instead the wind speed and direction are shown, would correspond to a vector field.

\section{The long road to the Higgs boson}
\label{story}

The BEH mechanism was first proposed in 1964 in the works of Brout and Englert~\cite{Englert:1964et}, Higgs \cite{Higgs:1964ia,Higgs:1964pj}, and Guralnik, Hagen, and Kibble~\cite{Guralnik:1964eu}. Further details of the mechanism were presented in 1966 by Higgs~\cite{Higgs:1966ev} and in 1967 by Kibble~\cite{Kibble:1967sv}.
In 1967, Weinberg~\cite{Weinberg:1967tq} and Salam~\cite{Salam:1968rm}, extending the 1961 work of
Glashow~\cite{Glashow:1961tr}, proposed the use of the BEH mechanism for a theory of the unification of the EM and weak interactions, 
labelled as the electroweak (EW) interaction.
The key element in this work was the conjecture that nature possesses 
an EW symmetry, mathematically described by the Lagrangian of the theory, which is spontaneously broken, 
granting mass to the \PW and \PZ bosons. 
An additional feature of this model is that it provides a mechanism for granting masses to fermions as well, 
through the so-called Yukawa interactions~\cite{Weinberg:1967tq,Nambu:1961tp}.  
Thus, the elementary particles interacting with the BEH field acquire mass.
The impact is far reaching: for example, electrons become massive, allowing atoms to form, and endowing our universe with the observed complexity. 

Salam and Weinberg had further conjectured that the model they put forward might be renormalizable (that is, give finite answers). 
In 1971, 't Hooft and Veltman~\cite{tHooft:1971akt,tHooft:1972tcz}  showed how indeed this theory could be renormalized. 
This development put the Glashow--Salam--Weinberg model on a firm basis deserving serious experimental scrutiny.

After the \PW and \PZ  bosons were discovered by the UA1 and UA2 experiments at CERN in 1983~\cite{UA1:1983crd,UA2:1983tsx,UA1:1983mne,UA2:1983mlz},
the search for the Higgs boson became a central thrust in particle physics and an important motivation for 
the CERN Large Hadron Collider (LHC)~\cite{Evans:2008zzb}, and the ATLAS and CMS experiments.

Finding the Higgs boson has been demanding. This is a consequence of its large mass, which puts it beyond the 
reach of previous electron-positron colliders, such as LEP~\cite{LEPWorkingGroupforHiggsbosonsearches:2003ing} at CERN, and 
low cross section modes coupled with unfavourable decay channels in the range
of mass in which it was eventually found, which made it challenging to observe at previous hadron colliders, 
such as the Tevatron~\cite{CDF:2012laj} at Fermilab. 
In the SM, the Higgs boson is an elementary scalar particle, a type that had never been observed before.
Fundamental scalar particles are subject to quantum corrections that can be as large as the scale of the physics beyond the SM (BSM).
As this scale can be many orders of magnitude larger than the EW scale, which is $\sim 100$ \GeV, the measured mass of the Higgs boson is puzzlingly 
small. How to resolve this puzzle is part of the motivation for future work and accelerators.

The BEH mechanism does not predict the mass of the Higgs boson, but once the mass is fixed, all its other properties 
are precisely defined. 
The Higgs boson, once produced, decays directly to the heaviest allowed elementary particles.
However, decays to massless particles can also occur through quantum loops. 
At the LHC, the production of Higgs bosons is dominated by gluon-gluon fusion ($\ggH$) proceeding via a 
virtual  top quark loop. 
The mass of a real particle is defined as $m^2=E^2-p^2$, where $E$ is the energy and $p$ is the momentum vector of the particle. For a virtual particle this equation is not valid and thus a virtual particle does not have a defined value of the mass.  A virtual particle is denoted by an asterisk, \eg $\PW^*$ denoting a virtual \PW boson. Henceforth the distinction between 
real and virtual particles will be dropped, unless mentioned otherwise. 
At a mass of around 125\GeV the Higgs boson decays dominantly into 
a $\PQb$ quark and its antiquark. 
Henceforth, the distinction between a particle and its antiparticle will be dropped. 

From the accurate observation and measurement of the products of the Higgs boson decays and of those associated with its production,
experiments are able to 
infer its properties, including the strength of its self-interaction ($\lambda$)~\cite{th-paper} and, potentially, decays into beyond-the-SM (BSM) 
particles.

This paper presents the combination of results from single Higgs boson production and decay, and its pair production, 
using data sets corresponding to an integrated luminosity ($\lint$) up to 138\fbinv~\cite{CMS-LUM-17-003}, collected by CMS in 2016--2018.
An integrated luminosity of 1\fbinv corresponds to $\approx$100 trillion proton-proton collisions at a centre-of-mass energy of 13\TeV. 

In addition, a few projections are made for an assumed data sample corresponding to $\lint =3000\fbinv$, recorded at $\sqrt{s}=14~\TeV$, expected to be accumulated by the end of the next decade during the high-luminosity operation of the LHC accelerator (HL-LHC).

\section{The CMS experiment and data sets}
The CMS apparatus~\cite{CMS:2008xjf}, illustrated in Extended Data Fig.~\ref{Extended-Fig1}, is a multipurpose, 
nearly hermetic detector, 
designed to trigger on~\cite{CMS:2020cmk,CMS:2016ngn} and identify electrons (\Pe), muons (\PGm), photons (\PGg), 
and (charged and neutral)
 hadrons~\cite{CMS:2020uim,CMS:2018rym,CMS:2014pgm}. 
A trigger is a filter that selects interesting events, where \textit{event} refers to the 
result of the selected interaction in a beam crossing, as observed in the detector.
A global event reconstruction algorithm~\cite{CMS:2017yfk} combines the information provided 
by the all-silicon inner tracker, crystal electromagnetic calorimeter, and brass and scintillator hadron calorimeters, 
operating inside a 3.8\unit{T} superconducting solenoid, with data from gas-ionization muon detectors embedded 
in the solenoid flux-return yoke, to build electrons, muons, tau (\PGt) leptons, photons, hadronic jets,
missing transverse  momentum (\ptmiss), 
and  other physics objects~\cite{CMS:2022prd,CMS:2016lmd,CMS:2019ctu}.
Collimated streams of particles arising from the fragmentation of quarks or gluons are called {\sl jets}. 
These jets are identified, and their energies measured, by specialized reconstruction algorithms~\cite{CMS:2017yfk,CMS:2016lmd}.
The missing transverse momentum vector is measured with respect to the incoming proton beams, and it is computed as the negative
vector sum of transverse momenta of all particles in an event.

Several improvements have been introduced into the CMS experiment since the discovery of the Higgs boson in 2012 
(cf.\ Methods section). 

By July 2012, CMS had collected $\lint = 5.1\fbinv$ at a proton-proton ($\Pp\Pp$) collision centre-of-mass energy $\sqrt{s} = 7\TeV$ (in 2011)
and $\lint =5.3\fbinv$ at 8\TeV (in the first half of 2012), with which the Higgs boson was discovered.
By the end of 2012 (Run~1), CMS had collected data corresponding to $\lint=19.7\fbinv$ at $8\TeV$~\cite{CMS:2014fzn}.

In LHC Run~2 (2015--2018), the accelerator delivered collisions at $\sqrt{s} = 13\TeV$.  
At this larger energy the cross section for Higgs boson production increases by a factor of 2.2--4.0, depending
on the production mode~\cite{LHCHiggsCrossSectionWorkingGroup:2011wcg,Dittmaier:2012vm,LHCHiggsCrossSectionWorkingGroup:2013rie,LHCHiggsCrossSectionWorkingGroup:2016ypw}. Physics analyses presented here are based on 2016--2018 data, corresponding to an $\lint$ of up to 138\fbinv (the additional $\sim$2\fbinv recorded in 2015 are not used in this combination).
This enabled not only a reduction of statistical, but also of systematic uncertainties, 
as well as a more precise calibration of the calorimeters and alignment of the tracking detectors.
During Run~2, approximately 8 million Higgs bosons have been produced. 
Many more final states could be studied, since it was possible to separate 
the events by production mode and decay channel, as well as by kinematic properties; and differential distributions could be measured.
Furthermore, improved analysis methods were deployed.

To enable comparison with the more precise experimental results, theoretical calculations have been carried 
out with commensurate improvements in 
accuracy~\cite{LHCHiggsCrossSectionWorkingGroup:2011wcg,Dittmaier:2012vm,LHCHiggsCrossSectionWorkingGroup:2013rie,LHCHiggsCrossSectionWorkingGroup:2016ypw},
involving higher orders in perturbation theory. 

The statistical procedure was developed in preparation for the
search/discovery of the Higgs boson and hasn't changed much
since then. It is based on building a combined likelihood from the
various input channels (cf.\ Methods Section~\ref{sec:stat}). Parameter estimation and limit setting are
performed using a profile likelihood technique with asymptotic
approximation~\cite{ATLAS:2011tau}, taking into account the full correlation
of the systematic uncertainties between individual channels and the
years of data taking.
The different channels included in the combination correlate nuisance
parameters related to the same underlying effect, such as the
uncertainty in the theoretical prediction or the energy scale uncertainty of
the final-state objects. The inclusive signal strength ($\mu$) combination has a total of
$\mathcal{O}(10^4)$ nuisance parameters.
The references to the individual analyses presented in the next
section each contain more details of the
statistical procedure employed for combining the several categories
used, created according to various criteria, such 
as signal-to-background ratios, mass resolutions, and multiplicities of physics objects.

\begin{figure*}[htb]
\centering
    \includegraphics[width=0.99\textwidth]{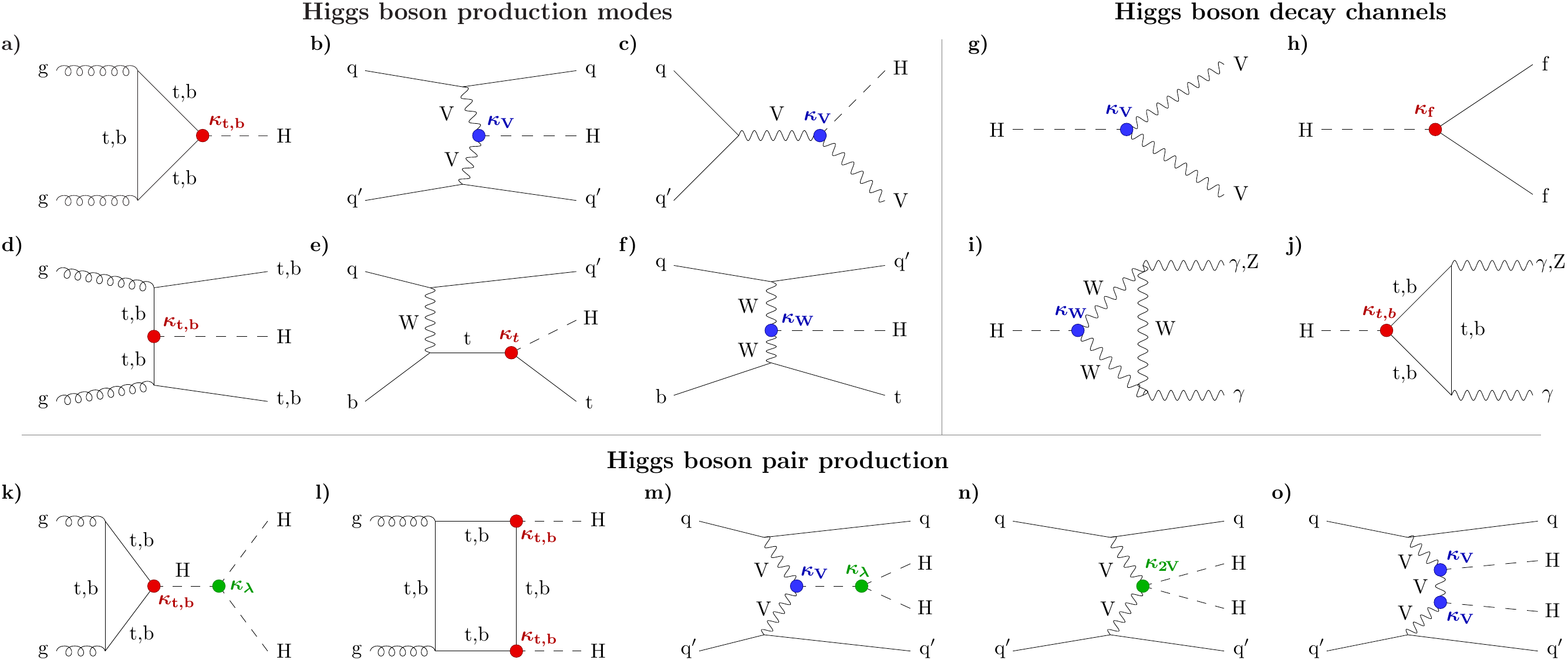}
\caption{
    \figtitle{Feynman diagrams for the leading Higgs boson interactions}
    Higgs boson production in (a) gluon-gluon fusion ($\ggH$), 
(b) vector boson fusion (VBF), (c) associated production with a \PW or \PZ (\PV) boson ($\VH$),
    (d) associated production with a top or bottom quark pair ($\ttH$ or $\bbH$), (e, f) associated production with a single top quark ($\tH$);
    with Higgs boson decays into (g) heavy vector boson pairs, (h) fermion-antifermion pairs, and
  (i, j) photon pairs or $\PZ\PGg$; Higgs boson pair production: (k, l) via  gluon-gluon fusion, and
  (m, n, o) via vector boson fusion. The different Higgs boson interactions are labelled with the coupling modifiers $\kappa$, and highlighted in different colours for Higgs-fermion interactions (red), Higgs-gauge-boson interactions (blue),
  and multiple Higgs boson interactions (green). The distinction between a particle and its antiparticle is dropped.}
\label{feyn-diag}
\end{figure*}

\section{Portrait of the Higgs boson}
\label{higgsboson}
The portrait of the Higgs boson is defined by its production modes, via cross sections, and its decay channels, 
via branching fractions.
For the value of mass measured by CMS $m_\PH = 125.38 \pm 0.14\GeV$ \cite{CMS:2020xrn},
 these are given in Extended Data Table~\ref{Extended-TabXS}~\cite{LHCHiggsCrossSectionWorkingGroup:2016ypw}.

\subsection{Production}
The rate of production of Higgs bosons is given by the product of the instantaneous 
luminosity, measured in units of $\percms$, and the cross section, measured in units of cm$^{2}$.
For $m_\PH =125.38\GeV$, the total cross section for the production of the SM Higgs boson at $\sqrt{s}=13\TeV$ is 
$55.4\pm 2.6\unit{pb}$~\cite{LHCHiggsCrossSectionWorkingGroup:2016ypw}.  (A cross section of $1\unit{pb}$ (picobarn) corresponds to an area 
of $10^{-36}\cmsq$.)
This results in the production of one Higgs boson every 
second at an instantaneous luminosity of $2\times 10^{34}\percms$.  
The dominant production mode in the SM is $\ggH$, where a pair of gluons, 
one from each of the incident protons, fuses, predominantly via a virtual top quark quantum loop. 
This is depicted in Fig.~\ref{feyn-diag}a and represents 87\% of the total cross section.  
The next most important production mode is vector boson fusion (VBF) 
depicted in Fig.~\ref{feyn-diag}b, where a quark from each of the protons radiates 
a virtual vector boson ($\PW$ or $\PZ$), 
which then fuse together to make a Higgs boson. 
Other processes, with smaller cross sections, are: production in association with 
a vector boson or \textit{Higgsstrahlung} ($\VH$) 
depicted in Fig.~\ref{feyn-diag}c, and production in association with top ($\PQt\PH$, $\ttH$) or 
 bottom ($\PQb\PQb\PH$) quarks, 
depicted in Figs.~\ref{feyn-diag}d--f. 
The $\PQb\PQb\PH$ mode has not been studied in the context of the SM Higgs boson because of limited sensitivity. 

Events are categorized according to the signatures particular to each production mechanism. 
For example, they are categorized as VBF produced if there are two high transverse momentum ($\pt$) jets, or 
as $\VH$ produced if there are additional charged leptons ($\ell$) and/or \ptmiss, or  $\ttH$, $\tH$ produced if there are jets identified as coming 
from \PQb quarks, or otherwise $\ggH$ produced.
(The top quark predominantly decays into a \PW boson and a \PQb quark jet.)

\subsection{Decays} 
In the SM, particle masses arise from spontaneous breaking of the gauge symmetry, through gauge couplings to the Higgs field in the case of vector bosons, and Yukawa couplings in the case of fermions. 
The SM Higgs boson couples to vector bosons, with an amplitude proportional to the gauge boson mass squared $m^2_\PV$, and to fermions with an amplitude proportional to the fermion mass $m_\mathrm{f}$. Hence, \eg the coupling is stronger for the 3rd generation of quarks and leptons than for those in the 2nd generation.
The observation of many Higgs boson decays to SM particles and the measurement of their branching fractions are a crucial test of the validity of the
theory. Any sizeable deviation from the predictions 
could indicate the presence of BSM physics. 

The Higgs boson, once produced, rapidly decays into a pair of fermions or a pair of bosons.
In the SM its lifetime is $\tau_{\PH} \approx 1.6 \times 10^{-22}\unit{s}$, and its inverse,
the natural width, is $\Gamma = \hslash / \tau_\PH = 4.14 \pm 0.02\MeV$~\cite{LHCHiggsCrossSectionWorkingGroup:2016ypw}, where $\hslash$ is the reduced Planck's constant. 
The natural width is the sum of all the partial widths, and the ratios of the partial widths to the total width are called branching 
fractions and represent the probabilities for that decay channel to occur.
The Higgs boson does not couple directly to massless particles (\eg the gluon or the photon), 
but can do so via quantum loops (\eg Figs.~\ref{feyn-diag}a, \ref{feyn-diag}i, and~\ref{feyn-diag}j).

By design, the event selections do not overlap among analyses targeting different final states. 
Where the final states are similar, the overlap has been checked and found to be negligible. 

Detailed information on the analyses included in the new combination
along with improvements, and the online and offline  criteria used to select events for the analyses 
can be found in the Methods Section, in Extended Data Tables~\ref{Extended-Tab1} and~\ref{Extended-TabCut}, and in the associated references. 
Online reconstruction is performed in real time as data are being collected. Offline reconstruction is performed later on stored data.
The background-subtracted distributions of the invariant mass of final-state particles in the individual decay channels are shown in Extended Data Figs.~\ref{Extended-Fig3} and~\ref{Extended-Fig4}.
The channels that are used in this combination are listed below.

\noindent  Bosonic decay channels: $\PH \to \PGg\PGg$ (Figs.~\ref{feyn-diag}i,~\ref{feyn-diag}j)~\cite{CMS:2021kom};
 $\PH \to \ZZ \to 4\ell $ (Fig.~\ref{feyn-diag}g)~\cite{CMS:2021ugl};
 $\PH \to \WW \to \ell\PGn\ell\PGn $ (Fig.~\ref{feyn-diag}g)~\cite{CMS:HWW2022};
 $\PH\to \PZ\PGg$ (Figs.~\ref{feyn-diag}i,~\ref{feyn-diag}j)~\cite{CMS:2022ahq}.

 \noindent   Fermionic decay channels: $\PH \to \ditau$, 3rd generation fermion, (Fig.~\ref{feyn-diag}h)~\cite{CMS:2022kdi};
$\PH \to \bb$, 3rd generation fermion, (Fig.~\ref{feyn-diag}h)~\cite{CMS:2017odg,CMS:2018kst,CMS:2018sah,CMS:2018hnq,CMS:2020zge};
$\PH \to \PGm\PGm$, 2nd generation fermion, (Fig.~\ref{feyn-diag}h)~\cite{CMS:2020xwi}. 

\noindent  \ttH/\tH with multileptons (Fig.~\ref{feyn-diag}d and Figs.~\ref{feyn-diag}e,~\ref{feyn-diag}f)~\cite{CMS:2020mpn}.

\noindent Higgs boson decays beyond the SM \cite{CMS:2014fzn}.

\section{Higgs boson pair production}
\label{sec:HH}
The measurement of the pair production of Higgs bosons can probe its
self-interaction $\lambda$.
The pair production modes are shown in Figs.~\ref{feyn-diag}k--o.

In the gluon-gluon fusion mode, there are two leading contributions: in the first (Fig.~\ref{feyn-diag}l), 
two Higgs bosons emerge from a top/bottom quark loop;
in the second (Fig.~\ref{feyn-diag}k), a single virtual Higgs boson, $\PH^*$, emerges from the top/bottom quark 
loop and then decays to two Higgs bosons ($\Pg\Pg \to \PH^*\to \PH\PH$).
Explicit establishment of the latter contribution, a direct manifestation of the Higgs boson's self-interaction, 
would elucidate the strikingly unusual potential of the BEH field.

In the VBF mode, there are three subprocesses that can lead to production of a pair of Higgs bosons:
\begin{itemize}
\item via a virtual Higgs boson (Fig.~\ref{feyn-diag}m); 
\item via a four-point interaction: $\PV\PV \to \PH\PH$ (Fig.~\ref{feyn-diag}n); and
\item via the exchange of a vector boson (Fig.~\ref{feyn-diag}o).
\end{itemize}

The scattering amplitudes of the processes giving rise to Higgs boson pair production via gluon-gluon fusion
(Figs.~\ref{feyn-diag}k,~\ref{feyn-diag}l) are similar in magnitude, but have opposite signs and interfere destructively. 
This makes the overall Higgs boson pair production rate small, rendering its experimental observation challenging.
The SM Higgs boson pair 
production cross section is calculated for $m_H=125\GeV$ to be $32.76^{+1.95}_{-6.83}\unit{fb}$~\cite{Grazzini:2018bsd,Baglio:2020wgt,Dreyer:2018qbw},
three orders of magnitude smaller than the single Higgs boson cross section.

The search for Higgs boson pair production is performed by combining
Higgs bosons candidates reconstructed from different final states~\cite{CMS:2022cpr,CMS:HH4b-comb,CMS:HHbbtautau,CMS:HHleptons,CMS:HHggbb, CMS:HHbbZZ}.
All final states analyzed are defined to be mutually exclusive so that they could be combined as statistically independent observations.

\section{Measurement of the properties of the Higgs boson}
At the time of the Higgs boson discovery~\cite{CMS:2012qbp,CMS:2013btf}, the combination of CMS data gave an observed (obs.) 
statistical significance of 5.0~standard deviations (s.d.) with an expected (exp.) significance of 5.8~s.d. 
Individually, the most sensitive channels, $\PH \to \PGg\PGg$ and $\PH \to \ZZ \to 4\ell$, yielded 4.1~s.d. obs. (2.8~s.d. exp.), and 3.2~s.d.  
obs. (3.8~s.d. exp.) respectively. 

Using all the Run~1 data, it was possible to observe separately the bosonic decay channels with significances of 6.5~s.d. for $\PH \to \ZZ \to 4\ell$, 5.6~s.d. for $\PH \to \PGg\PGg$, 4.7~s.d. for $\PH \to \WW$, and 
3.8~s.d. for the fermionic decay channel $\PH \to \ditau$~\cite{CMS:2014fzn}. 
Earlier, first results of the Higgs boson decay into fermions were presented in Ref.~\cite{CMS:2014suk},
reaching a significance of 3.8~s.d. by combining the $\PH \to \ditau$ and $\PH \to \bb$ decay modes. 
The mass was measured to a precision of $\approx$0.2\%~\cite{CMS:2014fzn}. 
Using the angular distributions of the leptons in the bosonic decay channels, the spin ($J$) and parity
($P$, a parity transformation that effectively turns a phenomenon into its mirror image)
were also found to be compatible with the SM prediction ($J^P=0^+$)
with a large number of alternative spin-parity hypotheses ruled out
at $>$99.9\% confidence level (\CL)~\cite{CMS:2014nkk,ATLAS:2015zhl}. 
The total cross section, combining all of the different decay channels, 
was measured to be in agreement with the SM, with an uncertainty of 14\%~\cite{CMS:2014fzn}.
Each of the VBF, $\VH$, and $\ttH$ production modes was measured at a level of 3~s.d.~\cite{CMS:2014fzn}. 

With Run~2 data, CMS has observed the Higgs boson decaying into a pair of tau leptons 
with a significance of 5.9~s.d.~\cite{CMS:2017khh}, a pair of bottom quarks with a significance of 5.6~s.d.~\cite{CMS:2018kst}, 
and the $\ttH$ production mode at 5.2~s.d.~\cite{CMS:2018uxb}. 
The Higgs boson has also been seen in its decays into muons with a significance of 3~s.d.~\cite{CMS:2020xwi}.
The mass of the Higgs boson has been measured to be $125.38 \pm 0.14$\GeV using the decay channels $\PH \to \PGg\PGg$ and 
$\PH \to \ZZ \to 4\ell$~\cite{CMS:2020xrn}. The natural width of the Higgs boson has been extracted and is found 
to be $\Gamma_{\PH} = 3.2^{+2.4}_{-1.7}\MeV$
by using off-mass-shell and on-mass-shell Higgs boson production~\cite{CMS:2022ley}. On-mass-shell refers to a particle with its physical mass, while
off-mass-shell refers to a virtual particle. 
 
\begin{figure*} [htb]
\centering
\includegraphics[width=0.49\textwidth]{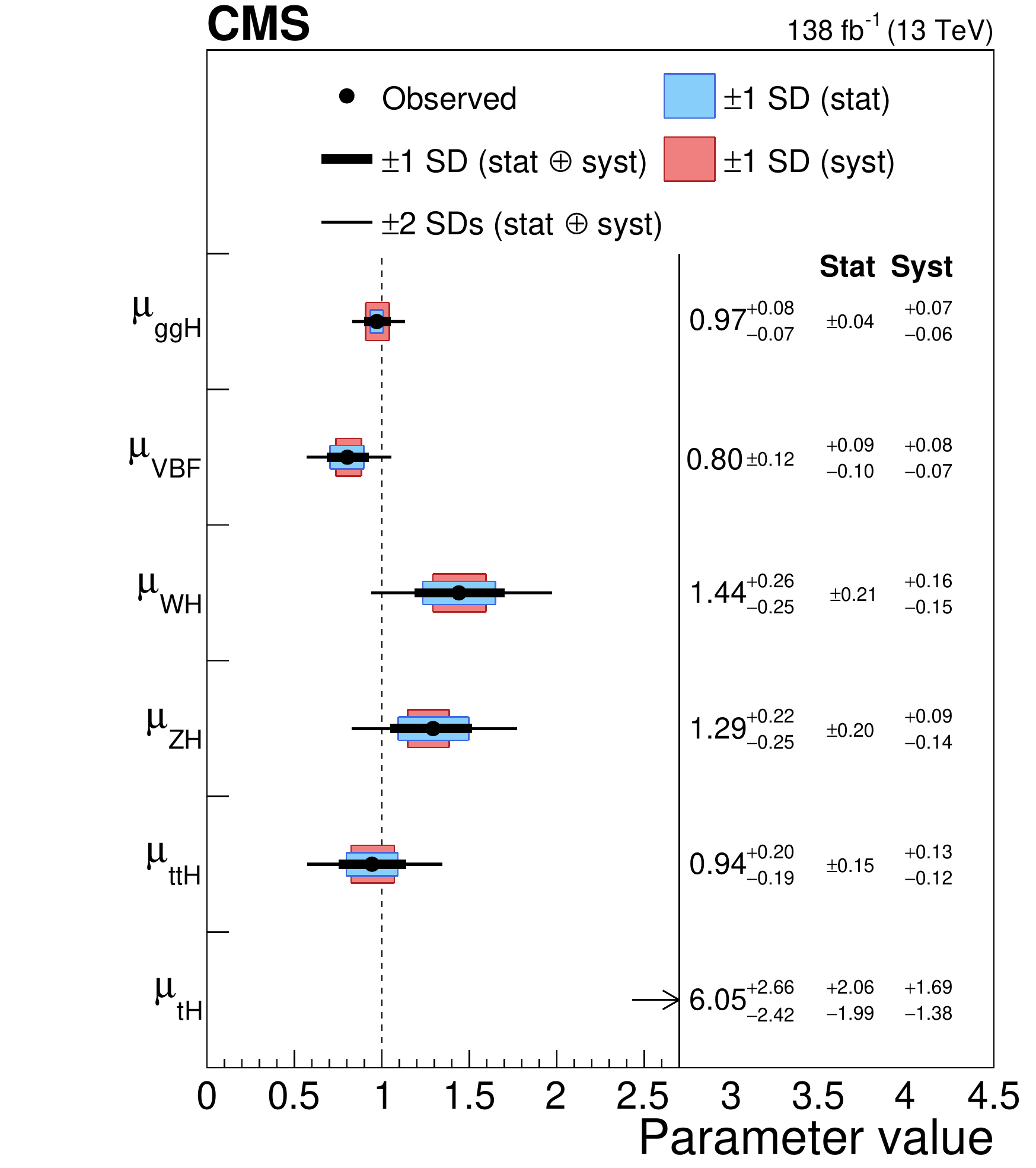}
\includegraphics[width=0.49\textwidth]{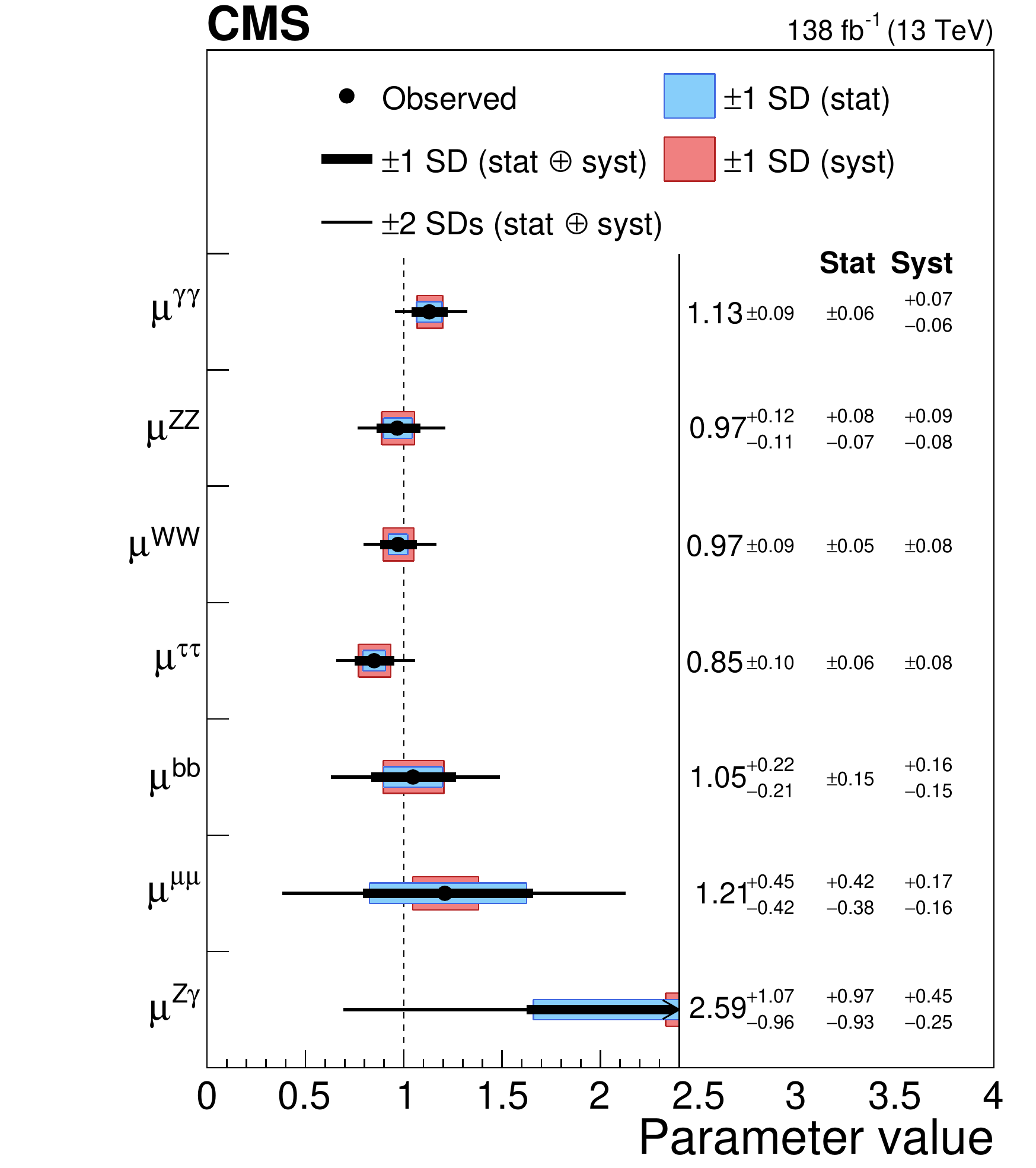}
\caption{
\figtitle{The agreement with the SM predictions for production modes and decay channels.}
    Signal strength parameters extracted for (left) various production modes $\mu_i$, assuming  $\mathcal{B}^f= (\mathcal{B}^f)_\text{SM}$, and (right) decay channels $\mu^f$, 
assuming $\sigma_i = (\sigma_i)_\text{SM}$. The thick (thin) black lines indicate the 1 (2)~s.d. confidence intervals, with the systematic and statistical 
components of the 1~s.d. interval indicated by the red and blue bands, respectively. The vertical dashed line at unity represents the values of $\mu_i$ and $\mu^f$ in the SM. 
The covariance matrices of the fitted signal strength parameters are
shown in Extended Data Fig.~\ref{Extended-FigMuCovs}. The $p$-value with respect to the SM prediction are
3.1\% and 30.1\% for the left and right plot, respectively. The $p$-value
corresponds to the probability that a result deviates as much, or
more, from the SM prediction as the observed one.}
\label{Comb_mu}
\end{figure*}

\subsection{ The \texorpdfstring{$\mu$}{mu}-framework for signal strengths.}
The agreement between the observed signal yields and the SM expectations can be quantified by fitting the data with a model 
that introduces signal strength parameters. These are generically labelled $\mu$, and scale the observed yields with respect 
to those predicted by the SM, without altering the shape of the distributions. The specific meaning of $\mu$ varies depending on the analysis. 
For given initial (i) and final (f) states,
$i \to \PH \to f$, the signal strengths for individual production channels, $\mu_i$, and decay modes, $\mu^f$, are defined as 
$\mu_i = \sigma_i /(\sigma_i)_\text{SM}$ and $\mu^f = \mathcal{B}^f / (\mathcal{B}^f)_\text{SM}$, where $\sigma$ 
is the production cross section and 
$\mathcal{B}$ is the branching fraction. Perfect agreement with SM expectations would yield all $\mu$ equal to one.

A first test of compatibility is performed by fitting all data from production modes and decay channels with a common signal strength parameter,
 $\mu$. At the time of discovery, the common $\mu$ was found to be $0.87 \pm 0.23$. 
 The new combination of all Run~2 data yields $\mu = 1.002 \pm 0.057$, in excellent agreement with the SM expectation.
The uncertainties in the new measurement correspond to an improvement by a factor of 4.5 in precision compared 
to what was achieved at the time of discovery.
At present, the theoretical uncertainties in the signal prediction, the experimental statistical, 
and the systematic uncertainties separately contribute at a similar level, and they are $0.036$, $0.029$, and $0.033$, respectively.

Relaxing the assumption of a common signal strength parameter, and introducing different $\mu_i$ and $\mu^f$, our measurements are 
shown in Fig.~\ref{Comb_mu}.  
The production modes $\ggH$, VBF, $\WH$, $\ZH$, and $\ttH$ are all observed with a significance of 5~s.d. or larger.

\subsection{ The \texorpdfstring{$\kappa$}{k}-framework for coupling modifiers. }
\label{kappaframework}
Beyond-the-SM physics is expected to affect the production modes and decay channels in a correlated way if they are governed by similar interactions. 
Any modification in the interaction between the Higgs boson and, \eg the $\PW$ bosons and top quarks not only would affect the 
$\PH \to \PW\PW$ (Fig.~\ref{feyn-diag}g) or $\PH \to \PGg\PGg$ (Fig.~\ref{feyn-diag}i and~\ref{feyn-diag}j) decay rates, but also the production cross section for the $\ggH$ (Fig.~\ref{feyn-diag}a),  $\WH$ (Fig.~\ref{feyn-diag}c), and VBF (Fig.~\ref{feyn-diag}b) modes. 
To probe such deviations from the predictions of the SM, the $\kappa$-framework \cite{LHCHiggsCrossSectionWorkingGroup:2013rie} is used. 
The quantities, such as $\sigma_i$, $\Gamma^f$, and $\Gamma_\PH$, computed from the corresponding SM predictions,
are scaled by $\kappa_i^2$,  
as indicated by the vertex labels in Fig.~\ref{feyn-diag}. 
As an example, for the decay $\PH \to \PGg\PGg$ proceeding via the loop processes of Fig.~\ref{feyn-diag}i or~\ref{feyn-diag}j, the branching fraction 
is proportional to $\kappa_\PGg^2$ or $(1.26 \kappa_\PW - 0.26 \kappa_\PQt)^2$.
In the SM all $\kappa$ values are equal to one.

\begin{figure*} [htb]
\centering
\includegraphics[width=0.49\textwidth]{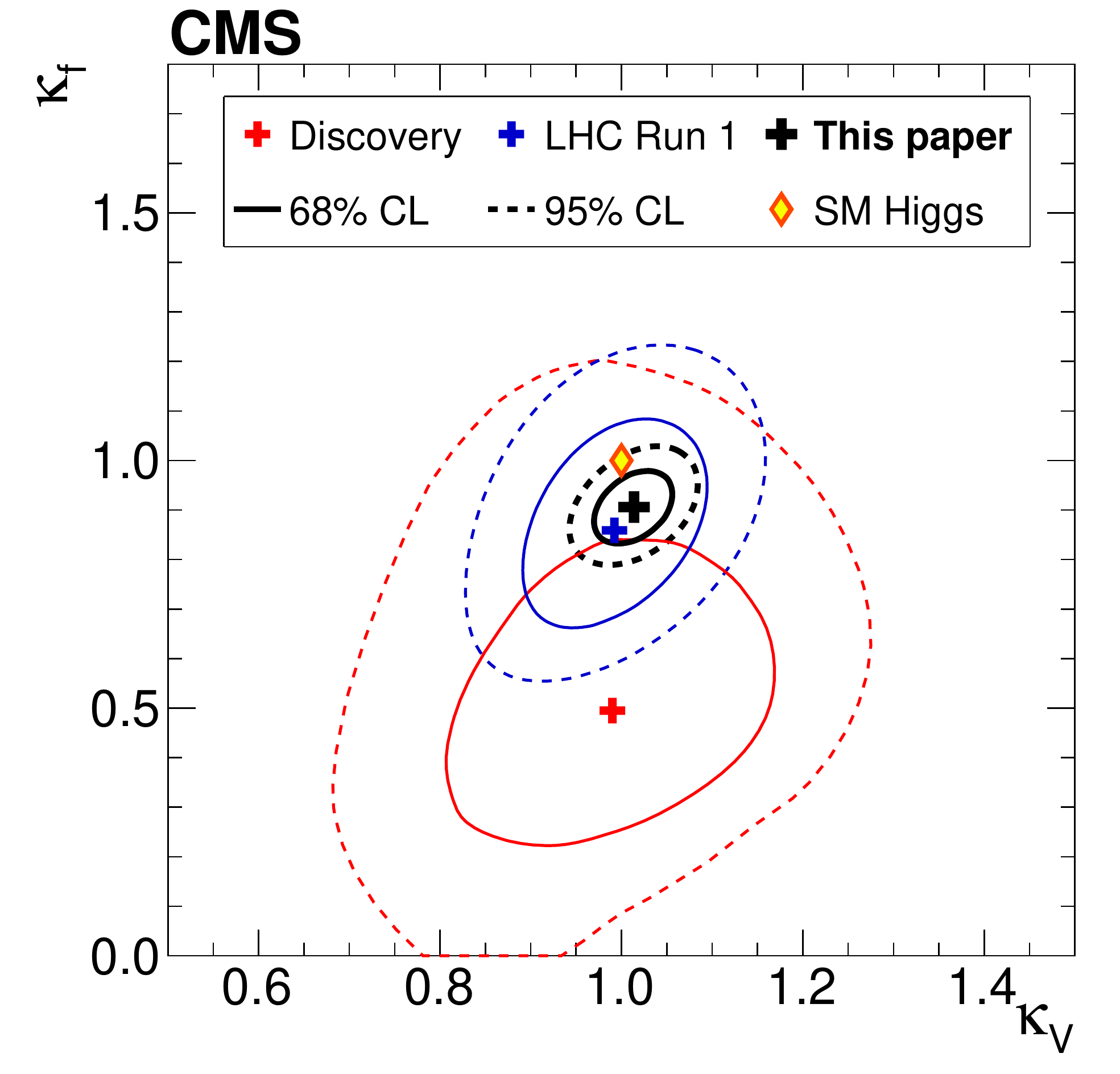}
\includegraphics[width=0.49\textwidth]{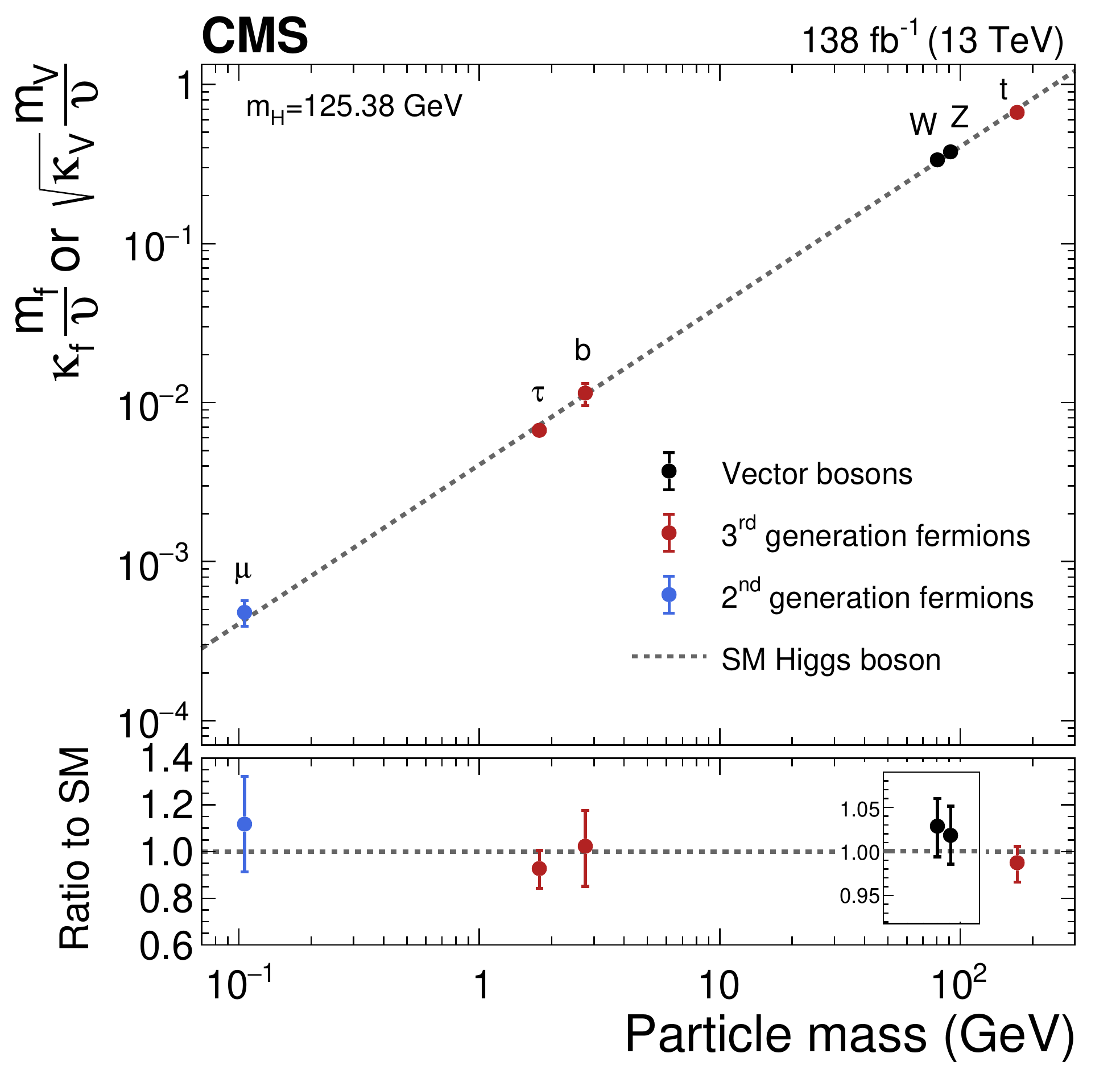}
\caption{
\figtitle{A portrait of the Higgs boson couplings to fermions and vector bosons.}
    (left) Constraints on the Higgs boson coupling modifiers to fermions ($\kappaf$) and heavy gauge bosons ($\kappav$), 
in different data sets: discovery (red), the full LHC Run~1 (blue), and the data presented here (black). 
The SM prediction corresponds to $\kappav = \kappaf = 1$ (diamond marker). (right) The measured coupling modifiers of the Higgs boson to 
fermions and heavy gauge bosons, as functions of fermion or gauge boson mass, where $\upsilon$ is the vacuum expectation value of the BEH field (cf. Methods section~\ref{sec:self}).  
For gauge bosons, the square root of the coupling modifier is plotted, to keep a 
linear proportionality to the mass, as predicted in the SM. The
$p$-value with respect to the SM prediction for the right plot is 37.5\%.}
\label{kappa}
\end{figure*}

A first such fit to Higgs boson couplings introduces two parameters, $\kappav$ and $\kappaf$, scaling the Higgs boson 
couplings to massive gauge bosons and to fermions, respectively. With the limited data set available at the time of 
discovery, such a fit provided first indications for the existence of both kinds of couplings. The sensitivity with present 
data is much improved, and both coupling modifiers are measured to be in agreement, within an uncertainty of 10\%, with 
the predictions from the SM, as shown in Fig.~\ref{kappa}~(left).
 
A second fit is performed to extract the coupling modifiers $\kappa$ for the heavy gauge bosons 
($\kappa_\PW$, $\kappa_\PZ$) and the fermions probed in the present analyses
 ($\kappa_{\PQt}$, $\kappa_{\PQb}$, $\kappa_{\PGt}$, $\kappa_{\PGm}$). Predictions for processes that in the SM occur via loops of 
intermediate virtual particles, \eg Higgs boson production via $\ggH$, or Higgs boson decay to a pair of gluons, 
photons, or $\PZ\PGg$, are computed in terms of the $\kappa_i$ above. The result is shown in Fig.~\ref{kappa}~(right), 
as a function of the mass of the probed particles. 
The remarkable agreement with the predictions of the BEH mechanism over three orders of magnitude of mass
 is a powerful test of the validity of the underlying physics.
Statistical and systematic uncertainties contribute at the same level to all measurements, except 
for $\kappa_{\PGm}$, which still is dominated by the statistical uncertainty. 

\begin{figure*} [htb]
\centering
\includegraphics[width=0.495\textwidth]{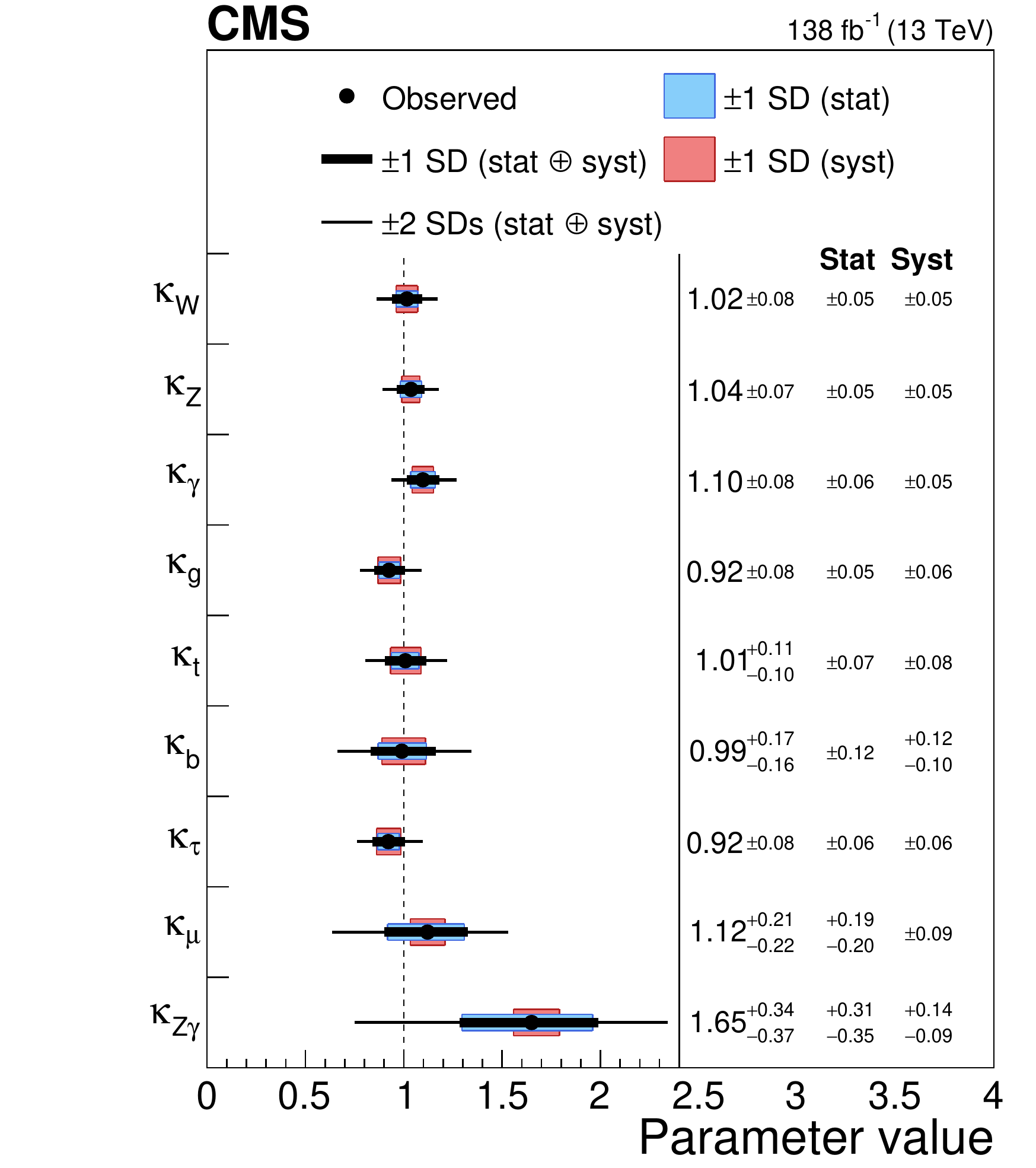}
\includegraphics[width=0.485\textwidth]{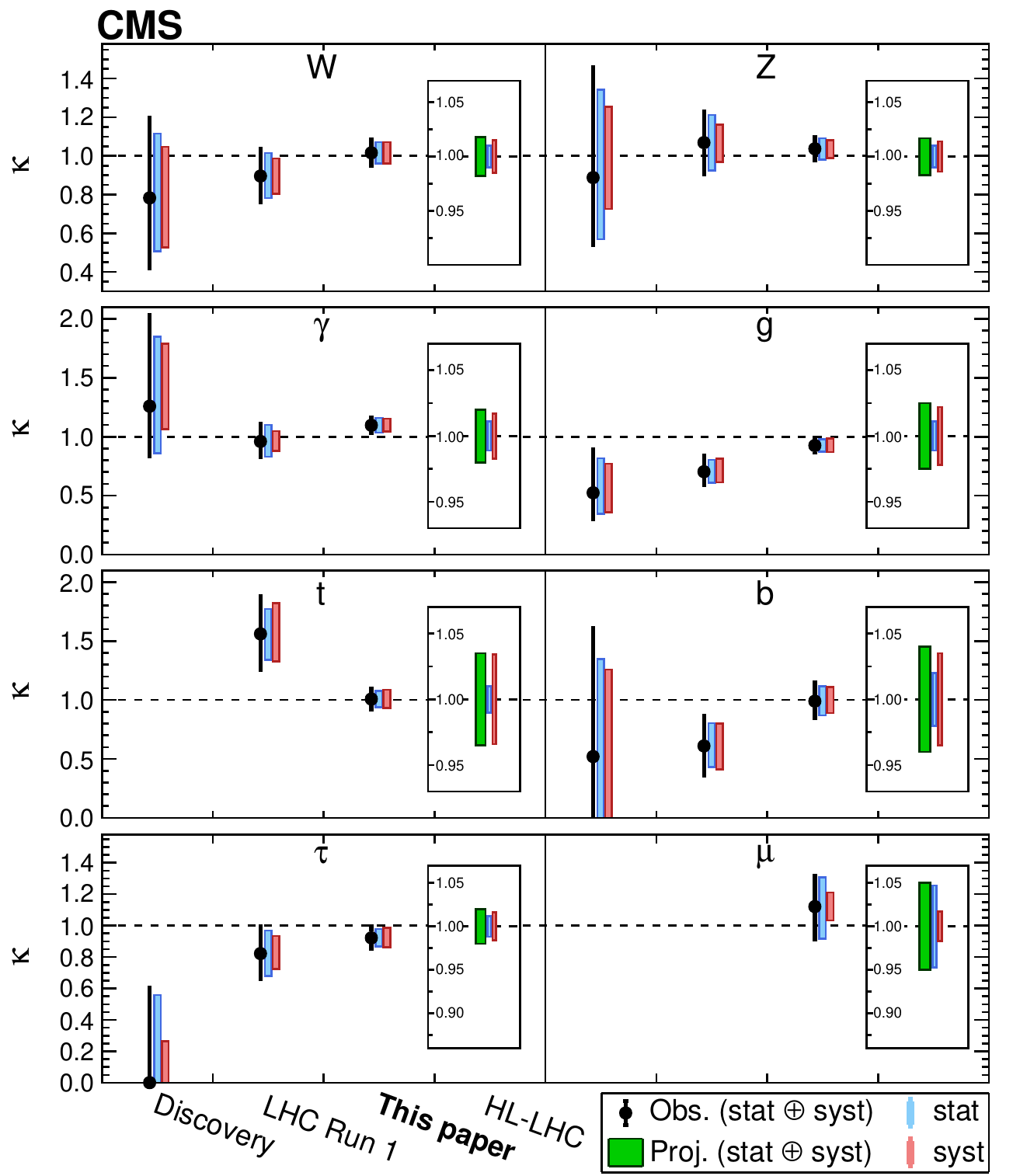} 
\caption{ 
\figtitle{Coupling modifiers measurements and their evolution in time.}
    (left) Coupling modifiers resulting from the fit. The
  $p$-value with respect to the SM prediction is 28\%.
 (right) Observed and projected values resulting from the fit in the $\kappa$-framework
in different data sets: at the time of the Higgs boson discovery, using the full
data from LHC Run~1, in the data set used in this paper, and
the expected 1~s.d. uncertainty at the HL-LHC for $\lint =3000\fbinv$.
The $\PH \to \PGm\PGm$ and $\kappa_{\PQt}$ measurements were not available for earlier data sets due to the lack 
of sensitivity.}
\label{Comb-kappa}
\end{figure*}

In extensions of the SM with new particles, the loop-induced processes may receive additional contributions. 
A more general fit for deviations in the Higgs boson couplings can then be defined by introducing additional modifiers for 
the effective coupling of the Higgs boson to gluons ($\kappa_{\Pg}$), photons ($\kappa_{\PGg}$), and $\PZ\PGg$ ($\kappa_{\PZ\PGg}$).  
Results for this fit are shown in Fig.~\ref{Comb-kappa}~(left). 
Coupling modifiers are probed at a level of uncertainty of 10\%, except for $\kappa_\PQb$ and
 $\kappa_{\PGm}$ ($\approx$20\%), and $\kappa_{\PZ\PGg}$ ($\approx$40\%), and all measured values are compatible with the SM expectations, 
to within 1.5~s.d. These measurements correspond to an increase in precision by a factor of $\approx$5 compared to what 
was possible with the discovery data set. 
Figure~\ref{Comb-kappa}~(right) and Extended Data Fig.~\ref{Extended-FigZ}~(left) illustrate the evolution of several $\kappa$ measurements 
and their uncertainties using the data set:
\begin{itemize}
\item at the time of discovery (July 2012)~\cite{CMS:2012qbp,CMS:2013btf}, 
\item for the full Run~1 (end of 2012)~\cite{CMS:2014fzn},
\item for results presented in this paper, and 
\item expected to be accumulated by the end of the HL-LHC running~\cite{Cepeda:2019klc}, corresponding to $\lint= 3000\fbinv$. The statistical uncertainties have been scaled by $1/\sqrt{\lint}$, the experimental systematic ones by $1/\sqrt{\lint}$ where possible, or fixed at values suggested in Ref.~\cite{Cepeda:2019klc}, whereas the theoretical uncertainties have been halved.
\end{itemize}
A sizeable improvement is expected after HL-LHC operation.
The $\PH \to \PGm\PGm$ measurements were not available for the first two data sets due to the lack of sensitivity.
The evolution of several signal strength measurements $\mu$ are shown in Extended Data Fig.~\ref{Extended-Fig5}.

If new particles exist with masses smaller than $m_\PH$, other decay channels may be open. Examples of such decays could be
into new neutral long-lived particles or into dark matter particles, neither leaving a trace in the CMS detector.
We refer to these as \textit{invisible} Higgs boson decays, which could be inferred from the presence of large \ptmiss
in the direction of the Higgs boson momentum. The events are selected based on other particles accompanying the Higgs boson.
Dedicated searches for such decays~\cite{CMS:2022qva,CMS:2021far,CMS:2020ulv} yielded $\mathcal{B}_\text{Inv.}<0.16$ at 95\% \CL, where $\mathcal{B}_\text{Inv.}$ is the branching fraction to invisible decays.

\section{Results from the search for Higgs boson pair production}
\begin{figure*} [tb]
\centering
\includegraphics[width=0.56\textwidth]{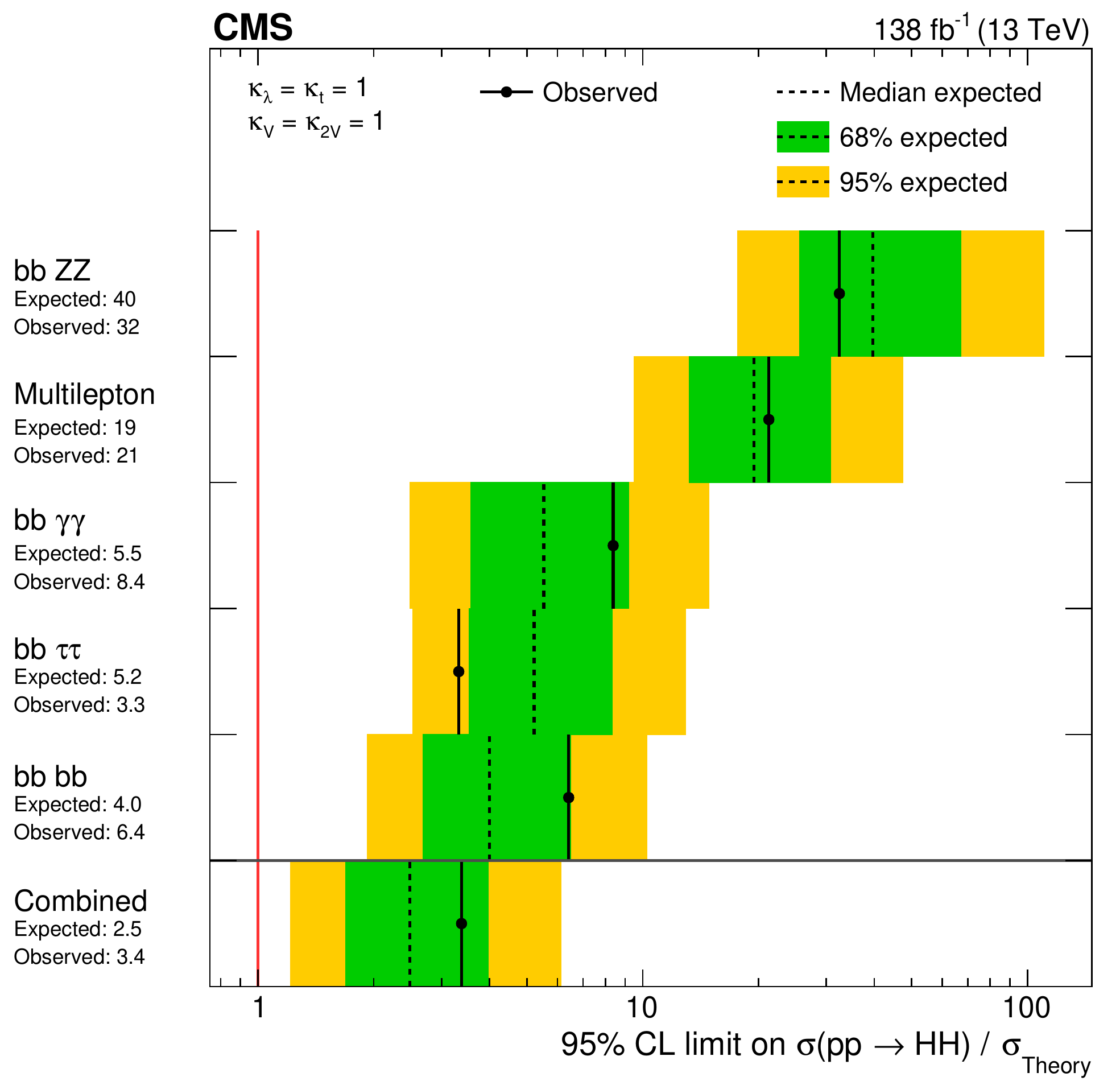} 
\includegraphics[width=0.43\textwidth]{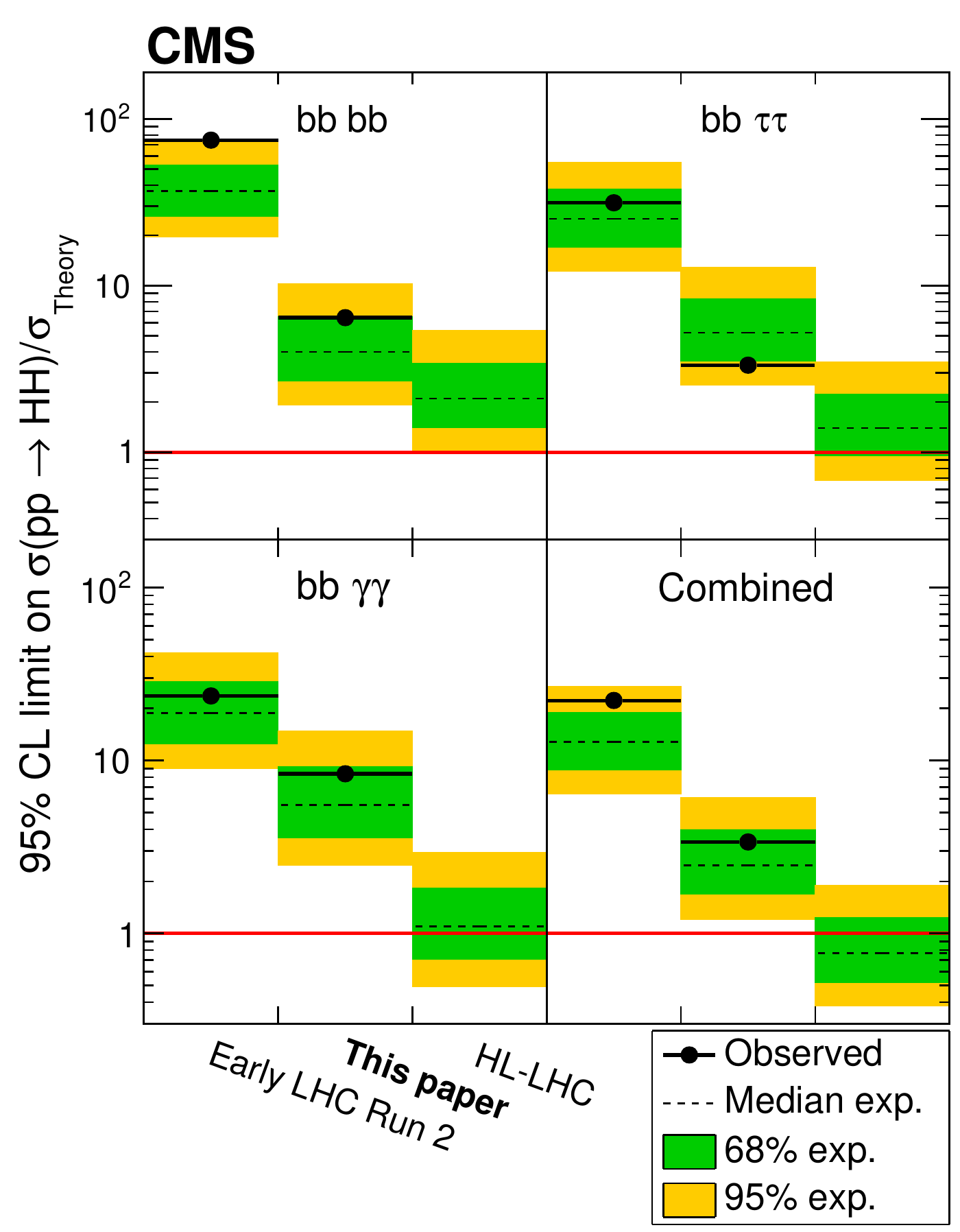} 
    \caption{
\figtitle{Limits on the production of Higgs boson pairs and their time evolution.}
        (left) The expected and observed limits on the ratio of experimentally estimated production cross section 
      and the expectation from the SM ($\sigma_\text{Theory}$) in searches using different final states and their combination.
 The search modes are ordered, from upper to lower, by their expected sensitivities 
from the least to the most sensitive. The overall combination of all searches is shown by the lowest entry.
    (right) Expected and observed limits on $\PH\PH$ production in different data sets: early LHC Run~2 data ($35.9\fbinv$), present results using full LHC Run~2 data ($138\fbinv$), and projections for the HL-LHC ($3000\fbinv$).}
\label{doubleh-limits}
\end{figure*}

The cross section for Higgs boson pair production in the SM is extremely small, 
thus escaping detection at the LHC so far. The results of the search are therefore expressed as 
an upper limit on the production cross section.
Figure~\ref{doubleh-limits}~(left) shows the expected and observed limits on Higgs boson pair production, expressed 
as ratios with respect to the SM expectation, in searches using the different final states and their combination. 
With the current data set, and combining data from all currently studied modes and channels, the Higgs boson pair production 
cross section is found to be less than $3.4$ times the SM expectation at 95\% \CL. 
Figure~\ref{doubleh-limits}~(right) shows the evolution of the limits  
from the three most sensitive modes and the overall combination for: 
the first comprehensive set of measurements using early LHC Run~2 data ($35.9\fbinv$)~\cite{Sirunyan:2018ayu}, 
the present measurements using the full LHC Run~2 data ($138\fbinv$), 
and the projections for the HL-LHC ($3000\fbinv$)~\cite{Cepeda:2019klc}.
The HL-LHC projections are also expressed as limits, 
assuming that there is no Higgs boson pair production. The fact that the combined limit is expected to be below unity shows that the sensitivity is sufficient to establish the existence of the SM $\PH\PH$ production. 
 
\begin{figure*} [htb]
\centering
\includegraphics[width=0.49\textwidth]{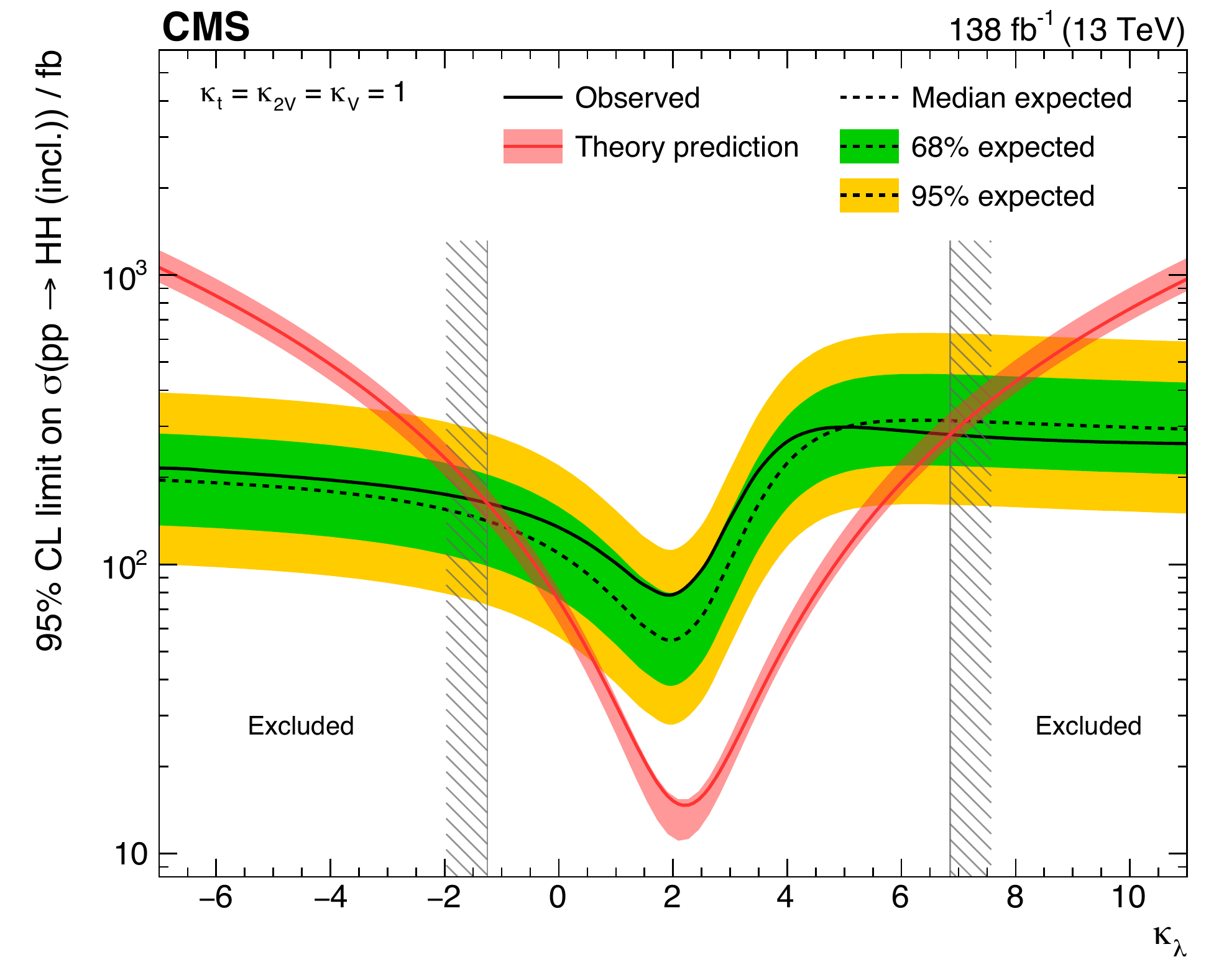} 
\includegraphics[width=0.49\textwidth]{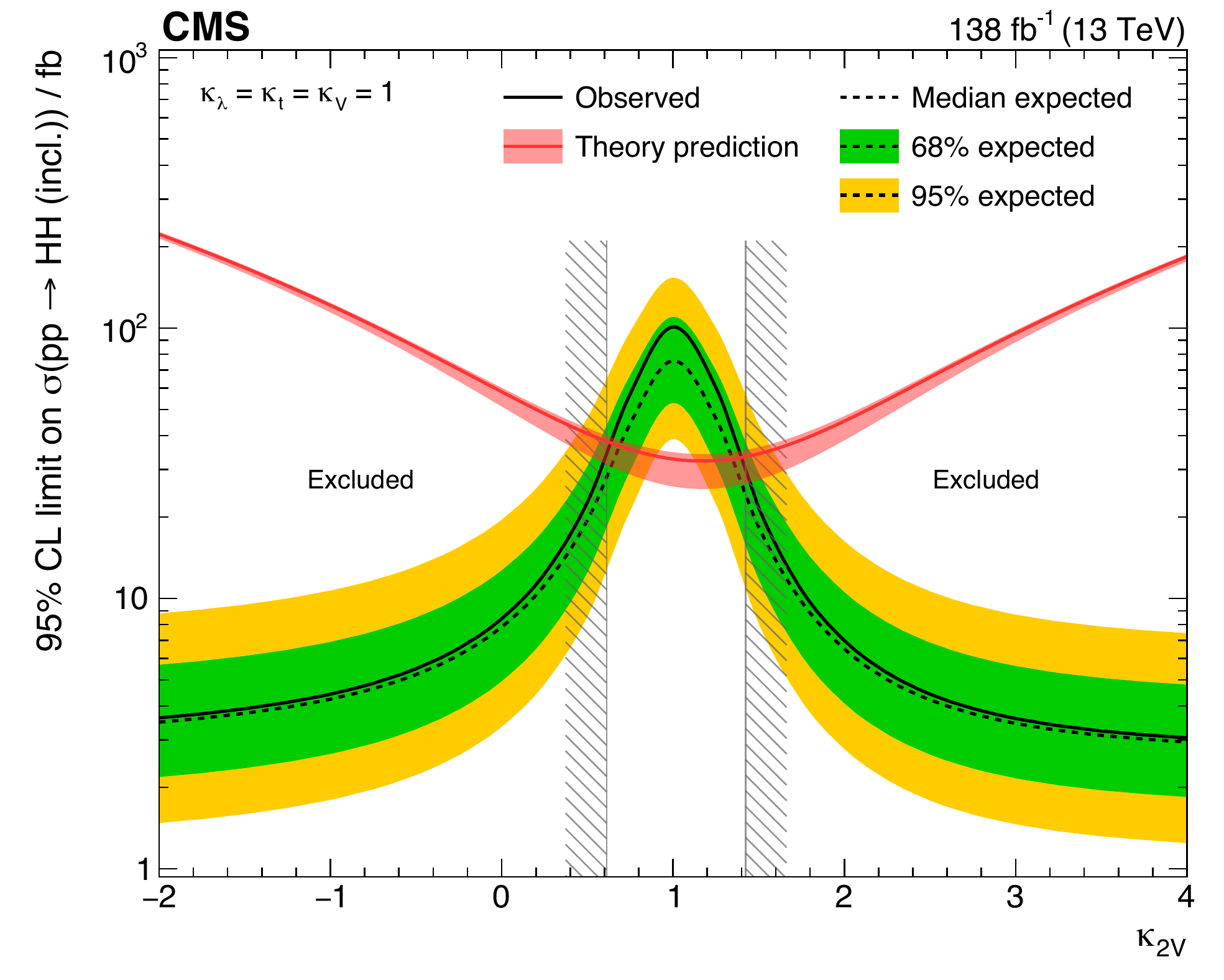} 
\caption{
    \figtitle{Limits on the Higgs boson self-interaction and quartic coupling.}
    Combined expected and observed 95\% \CL upper limits on the $\PH\PH$ production
 cross section for different values of $\kappa_{\lambda}$ (left) and $\kappavv$ (right), assuming the SM values 
 for the modifiers of Higgs boson couplings to top quarks 
 and vector bosons. The green and yellow bands represent, respectively, the 1 and 2 s.d. extensions beyond the expected limit; 
the red solid line (band) shows the theoretical prediction for the $\PH\PH$ production cross section (its 1 s.d. uncertainty). 
The areas to the left and to the right of the hatched regions are excluded at 95\% \CL.}
\label{doubleh-kappa}
\end{figure*}

Figure~\ref{doubleh-kappa} presents the expected and observed experimental limits on the $\PH\PH$ production cross section 
as functions of the Higgs boson self-interaction coupling modifier $\kappal$ and 
the quartic $\VVHH$ coupling modifier $\kappavv$. 
Cross section values above the solid black lines are experimentally excluded at 95\% \CL. 
The red lines show the predicted cross sections as functions of $\kappal$ or $\kappavv$, which exhibit a characteristic dip in the vicinity of 
the SM values ($\kappa = 1$) due to the destructive interference of the contributing production amplitudes, as highlighted in 
Section~\ref{sec:HH}. The experimental limits on the Higgs boson pair production cross section (black lines) also show 
a strong dependence on the assumed values of $\kappa$. This is because the interference between different subprocesses, 
besides changing the expected cross sections, also changes the differential kinematic properties of the two Higgs bosons, 
which in turn affects strongly the efficiency for detecting signal events. With the current data set we can ascertain 
at 95\% \CL that the Higgs boson self-interaction coupling modifier $\kappa_\lambda$ is in the range $-1.24$ to $6.49$, 
while the quartic $\kappavv$ coupling modifier is in the range $0.67$ to $1.38$. 
Figure~\ref{doubleh-kappa}~(right) shows that $\kappavv=0$ is excluded, with a significance of $6.6$~s.d.,  
establishing the existence of the quartic coupling $\VVHH$ depicted in Fig.~\ref{feyn-diag}n.

\section{Current knowledge and future prospects}
The discovery of the Higgs boson in 2012 completed the particle content of the standard model (SM) of elementary
particle physics, a theory that explains visible matter and its interactions in exquisite detail.
The completion of the SM spanned 60 years of theoretical and experimental work.
In the ten years following the discovery, great progress has been made
in painting a clearer portrait of the Higgs boson.

In this paper, the CMS Collaboration reports the most up-to-date combination of results on the properties of the Higgs boson, based on data corresponding to an $\lint$ of up to 138\fbinv, recorded at 13\TeV.
Many of its properties have been determined with accuracies better than 10\%. All measurements made so far are found to be consistent 
with the expectations of the SM. In particular, the overall signal strength parameter has been measured to be 
$\mu=1.002 \pm 0.057$.
It has been shown that the Higgs boson directly couples to bottom quarks, tau leptons, and muons, which had not been observed at the time of the discovery, and 
 also proven that it is indeed a scalar particle.
The CMS experiment is approaching the sensitivity necessary to probe Higgs boson couplings to charm quarks~\cite{CMS:2022psv}. 
The observed (expected) 95\% \CL value for $\kappa_\PQc$ is found to be
$1.1 < \abs{\kappa_\PQc} <5.5$ ($\abs{\kappa_\PQc}<3.40$), the most
stringent result to date.
Moreover, the recent progress in searches for the pair production of Higgs bosons has allowed the setting of tight constraints on the Higgs boson self-interaction strength, and the setting of limits on the Higgs boson pair production cross section not much above twice the expected SM value.  

Much evidence points to the fact that the SM is a low-energy
approximation of a more comprehensive theory.
In connection with the mechanism of spontaneous symmetry breaking,
several puzzles appear:
the so-called naturalness, a technical issue related to the fact that the Higgs boson mass is close to the electroweak scale; 
in relation with cosmology, the metastability of the vacuum state of the SM and the conjectured period of inflation in the early universe; 
the dynamics of electroweak phase transition and its connection to the matter-antimatter asymmetry of our universe. 
These issues motivate attempts at obtaining a deeper understanding of the physics of the Higgs boson.
The impressive progress made over the last decade is foreseen to continue into the next one. 
The current data set is expected to be doubled in size by the middle of this decade, enabling the establishment of rare decays channels such as $\PH \to \PGm\PGm$ and  
$\PH \to  \PZ \PGg$.  Operation with the high-luminosity LHC is expected during the next decade and should yield ten times more data
 then originally foreseen. This should allow the ATLAS and CMS
 experiments to establish the SM Higgs boson pair
 production with a significance of 4~s.d., as well as the Higgs boson coupling to charm quarks, and to search for any exotic decays.
Improvements in experimental techniques and theoretical calculations are also anticipated to continue. 

The CMS experiment is entering the era of precision Higgs physics that will shed light on the physics beyond the SM.

\begin{acknowledgments}
We congratulate our colleagues in the CERN accelerator departments for the excellent performance of the LHC and thank the technical and administrative staffs at CERN and at other CMS institutes for their contributions to the success of the CMS effort. In addition, we gratefully acknowledge the computing centres and personnel of the Worldwide LHC Computing Grid and other centres for delivering so effectively the computing infrastructure essential to our analyses. Finally, we acknowledge the enduring support for the construction and operation of the LHC, the CMS detector, and the supporting computing infrastructure provided by the following funding agencies: BMBWF and FWF (Austria); FNRS and FWO (Belgium); CNPq, CAPES, FAPERJ, FAPERGS, and FAPESP (Brazil); MES and BNSF (Bulgaria); CERN; CAS, MoST, and NSFC (China); MINCIENCIAS (Colombia); MSES and CSF (Croatia); RIF (Cyprus); SENESCYT (Ecuador); MoER, ERC PUT and ERDF (Estonia); Academy of Finland, MEC, and HIP (Finland); CEA and CNRS/IN2P3 (France); BMBF, DFG, and HGF (Germany); GSRI (Greece); NKFIH (Hungary); DAE and DST (India); IPM (Iran); SFI (Ireland); INFN (Italy); MSIP and NRF (Republic of Korea); MES (Latvia); LAS (Lithuania); MOE and UM (Malaysia); BUAP, CINVESTAV, CONACYT, LNS, SEP, and UASLP-FAI (Mexico); MOS (Montenegro); MBIE (New Zealand); PAEC (Pakistan); MES and NSC (Poland); FCT (Portugal); MESTD (Serbia); MCIN/AEI and PCTI (Spain); MOSTR (Sri Lanka); Swiss Funding Agencies (Switzerland); MST (Taipei); MHESI and NSTDA (Thailand); TUBITAK and TENMAK (Turkey); NASU (Ukraine); STFC (United Kingdom); DOE and NSF (USA).  

\hyphenation{Rachada-pisek} Individuals have received support from the Marie-Curie programme and the European Research Council and Horizon 2020 Grant, contract Nos.\ 675440, 724704, 752730, 758316, 765710, 824093, 884104, and COST Action CA16108 (European Union); the Leventis Foundation; the Alfred P.\ Sloan Foundation; the Alexander von Humboldt Foundation; the Belgian Federal Science Policy Office; the Fonds pour la Formation \`a la Recherche dans l'Industrie et dans l'Agriculture (FRIA-Belgium); the Agentschap voor Innovatie door Wetenschap en Technologie (IWT-Belgium); the F.R.S.-FNRS and FWO (Belgium) under the ``Excellence of Science -- EOS" -- be.h project n.\ 30820817; the Beijing Municipal Science \& Technology Commission, No. Z191100007219010; the Ministry of Education, Youth and Sports (MEYS) of the Czech Republic; the Stavros Niarchos Foundation (Greece); the Deutsche Forschungsgemeinschaft (DFG), under Germany's Excellence Strategy -- EXC 2121 ``Quantum Universe" -- 390833306, and under project number 400140256 - GRK2497; the Hungarian Academy of Sciences, the New National Excellence Program - \'UNKP, the NKFIH research grants K 124845, K 124850, K 128713, K 128786, K 129058, K 131991, K 133046, K 138136, K 143460, K 143477, 2020-2.2.1-ED-2021-00181, and TKP2021-NKTA-64 (Hungary); the Council of Science and Industrial Research, India; the Latvian Council of Science; the Ministry of Education and Science, project no. 2022/WK/14, and the National Science Center, contracts Opus 2021/41/B/ST2/01369 and 2021/43/B/ST2/01552 (Poland); the Funda\c{c}\~ao para a Ci\^encia e a Tecnologia, grant CEECIND/01334/2018 (Portugal); the National Priorities Research Program by Qatar National Research Fund; MCIN/AEI/10.13039/501100011033, ERDF ``a way of making Europe", and the Programa Estatal de Fomento de la Investigaci{\'o}n Cient{\'i}fica y T{\'e}cnica de Excelencia Mar\'{\i}a de Maeztu, grant MDM-2017-0765 and Programa Severo Ochoa del Principado de Asturias (Spain); the Chulalongkorn Academic into Its 2nd Century Project Advancement Project, and the National Science, Research and Innovation Fund via the Program Management Unit for Human Resources \& Institutional Development, Research and Innovation, grant B05F650021 (Thailand); the Kavli Foundation; the Nvidia Corporation; the SuperMicro Corporation; the Welch Foundation, contract C-1845; and the Weston Havens Foundation (USA).
\end{acknowledgments}

\bibliography{auto_generated} 
\clearpage
\appendix
\numberwithin{figure}{section}
\numberwithin{table}{section}

\section{Methods}
\subsection{Large Hadron Collider project and the Higgs boson}
The primary goals of the LHC and its two general-purpose experiments, ATLAS and CMS, are to: 
\begin{itemize}
\item elucidate the mechanism of EW symmetry breaking and
        find the associated particle, which in the SM
        of particle physics is the Higgs boson \cite{Englert:1964et,Higgs:1964ia,Higgs:1964pj}, and, 
\item search for BSM physics.
\end{itemize}

The necessity to study the wide range of processes in Fig.~\ref{feyn-diag} largely drove the design of the ATLAS and CMS experiments.
The production cross sections and the decay branching fractions for a
SM Higgs boson with a mass of 125.38\GeV are shown in
Extended Table~\ref{Extended-TabXS}.

The LHC~\cite{Evans:2008zzb} is designed to accelerate protons to an energy
up to $7\TeV$ by powerful electric
fields generated in superconducting radio frequency cavities and guided around
their circular orbits by strong ($8.3\unit{T}$) superconducting dipole
magnets in tubes under very high vacuum.
The counterrotating LHC beams are organized in $\approx$2800 bunches
comprising ${>}10^{11}$ protons per bunch, separated by $25\unit{ns}$,
leading to a bunch crossing rate of $\approx$32\unit{MHz}. The two proton beams are
brought into collision at the center of the four LHC experiments.
In Run~2,
$\Pp\Pp$ interaction rates of 2\unit{GHz} were reached.
Multiple pairs of protons interact in each bunch crossing,
the average number ranging from 21 in 2012 to 32 in 2018. These are superposed on the triggered 
interaction and are labelled \textit{pileup}.

\subsection{The CMS experiment}
\subsubsection{Design criteria and the SM Higgs boson}
In the early 1990s, during the design phase of the Compact Muon Solenoid (CMS) experiment,
considerable emphasis was placed on the identification and measurement
of high-energy electrons, photons, and muons, as these particles were
expected to play an important role in the search for the SM Higgs boson and in the search for BSM physics. 

As the rate of production of energetic muons at high-luminosity
hadron colliders is very large, the online selection of events
using muons is a particularly formidable task. The muon momentum 
has to be measured in real time and a momentum threshold placed to limit
the rate. This requires a high bending power (high magnetic field) and
an adequately precise and robust measurement of the trajectory of muons. This consideration
determined the starting point of the design of CMS, and by implication the choice, size, and the power of the analyzing
magnet.
The next design priority was driven by the search for the Higgs boson
via its decay $\PH \to \PGg\PGg$, requiring an excellent
EM calorimeter (ECAL). The muon system and the ECAL were to be
complemented by a precision inner tracking system,
immersed in a high magnetic field, giving 
good momentum resolution and a hadron calorimeter (HCAL) that provided an almost full calorimetric coverage (\eg for the search for the Higgs 
boson if its mass turned out to be larger than 500\GeV). 

\subsubsection{The CMS detector}
\begin{figure*}[hbt]
  \centering
  \includegraphics[width=0.8\textwidth]{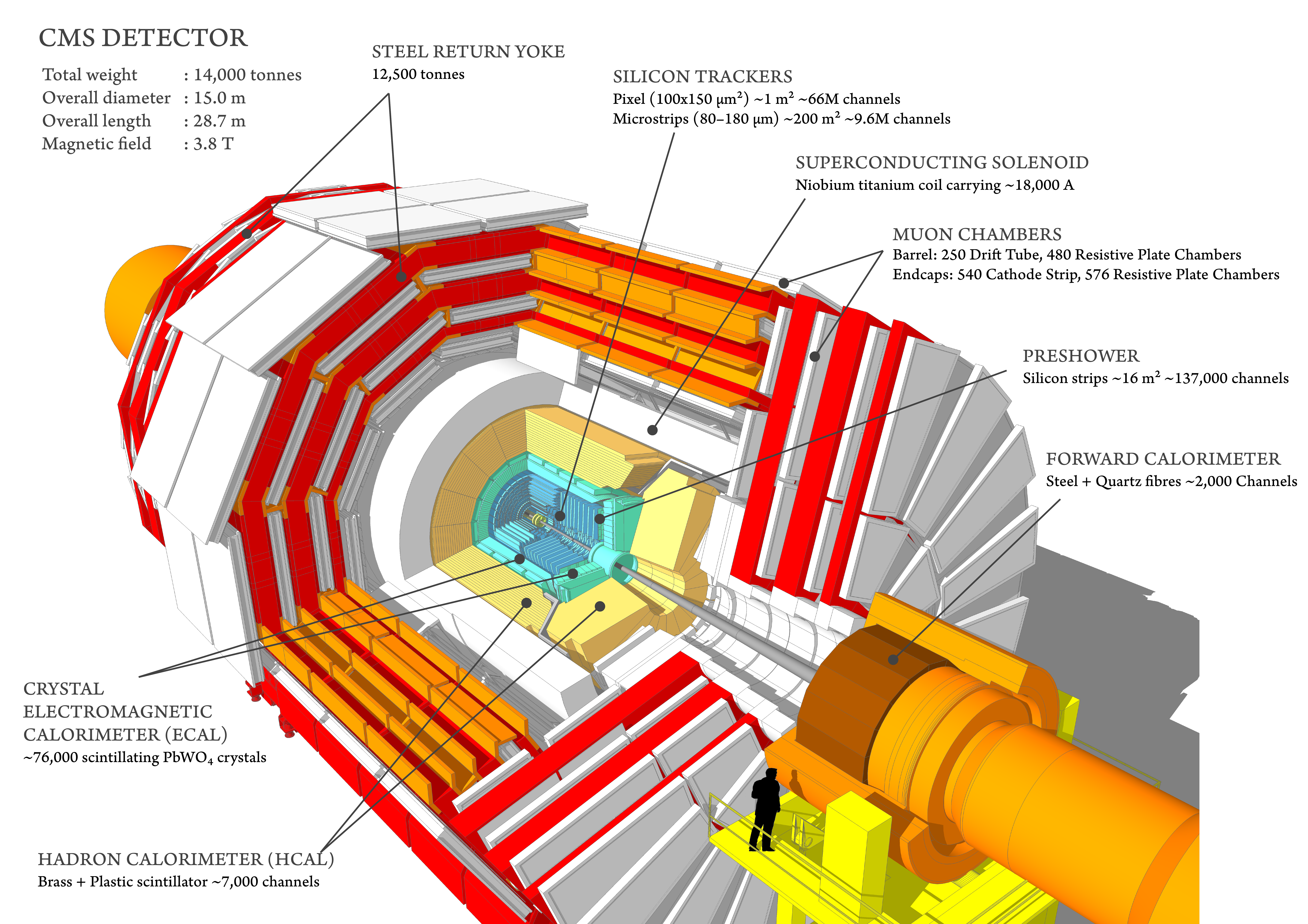}
  \caption{
  \figtitle{The CMS detector at the CERN LHC.}
      Schematic longitudinal cut-away view of the CMS detector, showing the different layers around the LHC beam axis, 
  with the collision point in the centre.}
  \label{Extended-Fig1}
\end{figure*}

The longitudinal cut-away view of the CMS detector
is shown in Extended Data Fig.~\ref{Extended-Fig1}. The CMS detector comprises four principal layers: the inner tracker,
the ECAL, the HCAL, and the muon
system. The various types of detecting elements and their channel counts are also indicated. 
Physics objects (\eg electrons, photons, muons, quark or gluon jets, etc.)
are identified by different combinations of the patterns of energy deposits and/or traces in these four layers.

The defining choice and the central element of the CMS detector is the long (13\meter), large inner diameter ($\approx$6\meter),
state-of-the-art high-field ($3.8\unit{T}$) superconducting solenoid, generating the magnetic field both for the
inner tracker and the muon system. The large size of the solenoid allows the inner tracker and almost all the calorimetry 
to be installed inside the solenoid.

\paragraph{Inner tracking}
Particles emerge from the interaction region into the inner tracker,
housed in a cylindrical volume with a length of 5.8\meter and a diameter of
2.5\meter. The particles first encounter the pixel detector, configured in three (four)
cylindrical layers of silicon sensors in the barrel region, and two (three) disks in the endcap
region before (after) 2017. The pixel detector is surrounded by 10 concentric
layers of silicon sensors
in the barrel region, with 10\cm long or 20\cm long silicon microstrips,
and 12 vertical planes in each endcap region. Points are measured with an accuracy 
of ${\approx}15\mum$ in the bending plane. The geometric coverage extends down to
angles of $9^\circ$ from the beamline.

\paragraph{Electromagnetic and hadron calorimeters}
The ECAL employs dense lead tungstate
scintillating crystals. Each crystal has a length of ${\approx}23\cm$ that is sufficient to
contain the full energy of high-energy electron and photon showers.
The amount of generated or collected light is proportional to the energy of the incident particle.
The fine transverse size of the crystals means that the energy of an EM shower is
distributed over a cluster of crystals ranging from 9 ($3{\times}3$) to 25
($5{\times}5$) crystals. The geometric coverage of the ECAL goes down to about $6^{\circ}$ from the beamline. 

The HCAL, comprising $\approx$7000 channels, is a 
sandwich of $\approx$5\cm thick brass absorber plates and $\approx$4\mm
thick scintillator plates. The charged particles in the shower, generated in the absorber
plates,  traverse the scintillator plates and produce light that is collected and guided by fibres to the photodetectors.
The geometric coverage of the HCAL goes down to about $6^{\circ}$ from the
beamline. This coverage is augmented by the very forward calorimeter, comprising
an iron absorber with quartz fibres embedded in a matrix arrangement.
The relativistic charged particles in the showers traverse the
fibres and generate Cherenkov light, a part of which is guided by the
fibres to the photodetectors.
This calorimeter extends the calorimetric coverage down to an angle of ${\approx}0.75^\circ$ from the beamline.
The thickness of hadron calorimetry is sufficient to absorb almost all of the energy of high energy hadrons.

\paragraph{Muon system}
Muons (and neutrinos) are the only particles that normally reach the
muon system. All other particles deposit almost all of their energies in the
calorimeters, and hence are said to have been absorbed. In addition to the
measurements inside the inner tracker, the momentum of muons is
measured a second time in gas-ionization chambers.
These chambers are organized in four \textit{stations} that measure several points,
to a precision of $\approx$150\mum, and generate track segments
whose direction is measured online with an angular precision
of ${\approx}5\unit{mrad}$. An independent set of gas-ionization chambers, provide a signal timing resolution 
of ${\approx}3\unit{ns}$ aiding the triggering process.
The instrumented geometric coverage of the muon system goes down to an angle of $10^\circ$ from the beamline.

\paragraph{Event selection}
As the resources needed to record data for later use from all
$\approx$32 million beam crossings per second would be prohibitively costly, specific filters
(known as triggers) are used to select the most interesting ones.
An online two-tiered trigger system~\cite{CMS:2020cmk,CMS:2016ngn} is deployed, with the first tier (Level-1) being hardware-based 
and the second one (high-level or HLT) being software-based.
The Level-1 uses custom hardware that processes coarse
information from the calorimeters or the muon chambers to select around
100 thousand crossings of interest per second, corresponding to a reduction of a factor of $\approx$400.
Crossings of interest are selected if the energy deposits in the calorimeters
or the momentum of muons, are
above predefined thresholds. Upon the issuance of a Level-1 trigger,
and after a fixed latency of just under 4\mus, all data from the
\textit{triggered} crossing are off-loaded from the pipeline memories
in the $\approx$100 million on-detector electronics channels. These data,
after suitable treatment in electronics housed
in the underground \textit{services} cavern, are sent up 100\meter to the surface as fragments
on $\approx$1000 optical fibres and
fed into a commercial telecommunication \textit{switch}.
The switch takes the individual fragments, puts them together,
\ie \textit{builds} the event, and feeds the event into the next available CPU
core, in a computer farm of some 50\,000~CPU cores. There, in real time,
full-event physics-grade software algorithms, optimized for fast processing, 
reconstruct physics objects and select for permanent storage some 1000 events or crossings per second,
based on topological and kinematic information (see Extended Table~\ref{Extended-TabCut}).

\paragraph{Event reconstruction}
The CMS experiment generates a large amount of collision and simulated data. 
To handle, store, and analyze all these data required the development of the worldwide LHC distributed computing grid (wLCG), 
providing universal access to data for all CMS Collaboration members.

The data from the stored events are transferred to the Tier-0 centre housed on CERN's main site, 
where a first processing stage is performed. The result of this stage is then distributed to seven other major 
centres worldwide, labelled Tier-1 centres, for offline analysis. The Tier-1s are designed to carry out tasks 
of further reconstruction of the collision data with improved calibration and alignment of the various CMS subdetectors,
while the generation and reconstruction of Monte Carlo event samples 
is carried out both at the Tier-1 centres and smaller university-based locations, labelled Tier-2 centres.

The particle-flow (PF) algorithm~\cite{CMS:2017yfk} reconstructs and identifies each individual particle in an event,
with an optimized combination of information from the various elements of the CMS detector.
The energy of photons is obtained from the measurements in the ECAL.
The energy of electrons is determined from a combination of the electron momentum at the primary
interaction vertex as determined by the tracker, and the energy in the corresponding cluster of crystals,
including the energy sum of all bremsstrahlung photons spatially compatible with originating from the electron track.
The momentum of muons is derived from the curvature of the corresponding track. The energy of charged hadrons
is determined from a combination of their momentum measured in the tracker, and the matching ECAL and
HCAL energy deposits.
The energy of neutral hadrons is obtained from the corresponding corrected ECAL and HCAL energies. 

Hadronic jets, arising from quarks or gluons, are created from all the particles reconstructed by the PF algorithm 
within a cone of half-angle of ${\approx}25^\circ$, centred on the axis determined by the vectorial sum of the momenta of all particles in the jet. 

\paragraph{Improvements of the CMS detector}
Several improvements have been introduced into the CMS experiment since the discovery of the Higgs boson in 2012. These included:
\begin{itemize}
  \item The replacement, in late 2016, of the silicon pixel detector,
    with a new one comprising four concentric layers in the barrel
    region, at radii of 29, 68, 109, and 160\mm, and six endcap
    disks placed at $\pm34$, $\pm41$, and $\pm51\mm$ from the interaction point, along the beamline.
    The new configuration leads to an improvement in the reconstruction of the secondary vertices and in the quality of tagging of $\PQb$ quarks. 
    The sensitivity of $\PH \to \bb$ analysis is found to be improved by a factor of 2.
\item The replacement of photodetectors in HCAL
    (hybrid photodiodes replaced by silicon photomultipliers)
        and implementation of more precise timing, allowing a reduction of accidental or instrumental backgrounds, \eg stray or out-of-time particles. 
\item The installation in 2013 and 2014 of chambers in the fourth
  endcap muon station that were left out for Run~1. 
\item The upgrade of the Level-1 trigger hardware prior to LHC Run~2
  to improve the selection of physics events of interest. The trigger rate from background
  processes is reduced and the trigger efficiency improved for a wide
  variety of physics signals. In the muon system, new trigger
  processor boards deploy powerful commercial field-programmable gate arrays (FPGAs). A time-multiplexed
  architecture was introduced that enabled data from all the
  calorimetry in each crossing to be pushed into a single FPGA of the type used in the muon trigger system.
   The FPGAs allow sophisticated and innovative algorithms to be implemented and evolved as conditions change.
\item In the data acquisition system a new switch was installed and the CPU power of the computer
  farm increased. The whole fabric of the distributed
  computing systems was upgraded to allow more events to be stored
  (at least 1000 events/s instead of the initially foreseen 100 events/s).
\end{itemize}

\subsection {Offline event analysis} 
The principal physics objects are required to have transverse momenta or energies
above a set threshold. The thresholds are lowered for the second, or
any further, objects.
Typical values of these thresholds are listed in Extended
Table~\ref{Extended-TabCut}.

Lepton and photons resulting from the decays of Higgs bosons are expected to be unaccompanied by other particles; 
they are said to be \textit{isolated}. Isolation criteria are imposed by requiring no additional energetic particles 
within a cone of ${\approx}20^\circ$ opening angle around the object's direction. 
Particles, other than from decays of \PQb and \PQc quarks or $\tau$ leptons, are expected to emerge directly from the primary 
interaction vertex, defined as 
the vertex corresponding to the $\Pp\Pp$ collision identified by the online selection.

Increased use of regression and classification algorithms implemented using machine learning (ML) methods, 
such as deep neural networks (DNNs) and boosted decision trees (BDTs), led to a simultaneous increase in purity 
and in efficiencies of identification and reconstruction of physics objects (electrons, muons, photons, \PQb quarks, tau leptons, jets, 
and \ptmiss), and improvements in the calibration of related kinematic observables.

All analyses make extensive use of Monte Carlo simulation of the
signal and background processes.
The CMS detector is precisely described in software code that is used
to generate Monte Carlo event samples.
Multiple interactions are included, that match the distribution of the
number of pileup interactions observed in data.
All the simulated event samples are then processed through the same
chain of software programs and procedures as are collision data.
Simulated samples are used to evaluate or
determine geometric acceptances, energy/momentum/mass
resolutions, as well as for online and offline particle identification and
reconstruction efficiencies, and for training for the many BDT algorithms and DNNs. 

\subsection{Higgs boson decay channels: notes}
Distributions of the invariant mass of final-state particles in the individual decay channels are shown in 
Extended Data Figs.~\ref{Extended-Fig3} and~\ref{Extended-Fig4}.

\subsubsection{Bosonic decay channels}
\paragraph{\texorpdfstring{$\PH \to \PGg\PGg$}{H to gamma+gamma}}
The signal is extracted by measuring the narrow 
signal peak over a smoothly falling background distribution~\cite{CMS:2021kom}. Despite its small branching fraction (0.23\%), 
this mode is a sensitive one due to the excellent precision in the measurement of the energies of photons. 
The diphoton invariant mass resolution is $\sigma_{m_{\PGg\PGg}}/m_{\PH} \approx 1\%$.   
All the principal production modes can be studied (\ggH, VBF, \VH, \ttH, and \tH).
The background largely consists of an irreducible one from quantum chromodynamics (QCD) production of two photons.
There is also a reducible background where one or more of the reconstructed photon candidates originate from misidentification of
jet fragments, that is dominated by QCD Compton scattering from quarks. 

\paragraph{\texorpdfstring{$\PH \to \ZZ \to 4\ell$}{H to ZZ to 4l}}
The study of this decay channel uses the distinctive decay of the
$\PZ$ bosons to charged leptons ($\ell$) leading to a final state with
4$\Pe$, or 4$\PGm$, or 2$\Pe$2$\PGm$ \cite{CMS:2021ugl}. 
The signal appears as a narrow peak on top of a smooth and small background. The momentum (energy) measurement of muons (electrons) 
is precise enough to give an invariant mass resolution with $\sigma_{m_{4\ell}} /m_{\PH} \approx 1\%$. 
The background comprises
an irreducible part arising from the nonresonant production of two
$\PZ$ bosons or $\PZ\PGg^*$, and a reducible part from the production of $\PZ$+jets
and top pair events, where the jets originate from heavy quarks,
and thus could contain charged leptons, or are misidentified
as charged leptons. The event yield for this process is tiny due to
the small branching fractions of $\PH \to \ZZ$ (2.71\%) and subsequent $\PZ \to \ell\ell$ (3.37\% per lepton type) decays. 
To enhance the signal over background and to categorize events, discriminants exploiting the production and decay 
kinematics expected for the signal and background events based on a matrix element likelihood approach
are used together with the invariant mass of the particle.

Extended Data Fig.~\ref{Extended-Event}~(upper) shows a display of a candidate $\PH \to \PZ \PZ \to \Pe\Pe \PGm\PGm$ event 
produced in $\Pp\Pp$ collisions at a centre-of-mass energy $\sqrt{s} = 13\TeV$ and recorded in the CMS detector.

\paragraph{\texorpdfstring{$\PH \to \WW \to \ell\PGn\ell\PGn$}{H to WW to 2l+2nu}}
Two high-\pt $\ell$ and large \ptmiss characterize this final state~\cite{CMS:HWW2022} and benefit  
 from the $\PH \to  \WW$ decay having one of the largest branching fractions ($\sim$22\%).
Due to the presence of two neutrinos, the computation of the $\WW$ invariant mass is not possible. 
However, an associated variable, the transverse mass, \mT,
can be computed from the $\ptvec$ of the charged leptons and 
the $\ptvecmiss$. 
The square of transverse mass for a collection of particles $[P_i]$  
is defined as $\mT^2([P_i]) = (\sum \abs{ \vec{p}_{\mathrm{T},i}})^2 - \abs{\sum \vec{p}_{\mathrm{T},i}}^2$.
The dominant background arises from irreducible nonresonant $\WW$ production and is estimated
from data. The channel has a good sensitivity
to the $\ggH$ and VBF production processes. In the analysis, $3\ell$
and $4\ell$ categories are also included, sensitive to
production of the Higgs boson in association
with a leptonically decaying vector boson. The analysis does not
target the $\ttH$ and $\tH$ production modes,
which are covered by a dedicated analysis discussed in Section~\ref{sec:tth}.

\paragraph{\texorpdfstring{$\PH\to \PZ\PGg$}{H to Z+gamma}}
The signal is sought as a peak over a smoothly falling 
 background distribution~\cite{CMS:2022ahq}. This analysis targets decays of the \PZ\ boson into 2$\Pe$  or 2$\PGm$. 
 To increase the sensitivity to the signal, the events are divided into
different categories based on the production mode.
Multivariate analysis (MVA) techniques are used to further categorize regions with high and low signal-to-background ratios. 
The dominant background arises from Drell--Yan dilepton production in
association with an initial-state photon.

\subsubsection{Fermionic decay channels}
\paragraph{\texorpdfstring{$\PH \to \ditau$}{H to tau+tau}}
Four different ditau final states are studied~\cite{CMS:2022kdi}: 
$\Pe\PGm$, $\Pe\tauh$, $\PGm\tauh$, and $\tauh\tauh$, where $\tauh$ refers to a hadronically decaying 
tau lepton. 
The analysis of this decay channel targets the $\ggH$, VBF, and $\VH$ production modes.
The identification of $\tauh$ candidates uses DNN discriminants to
reject quark and gluon jets misidentified as $\tauh$. In order to separate the $\PH \to \ditau$ signal events from the
sizeable contribution of irreducible $\PZ \to \ditau$ events,  the likelihood 
estimate of the reconstructed mass of the $\ditau$  system is used. This analysis does not
target $\ttH$ production,
which is covered by the dedicated analysis discussed in Section~\ref{sec:tth}.

\paragraph{\texorpdfstring{$\PH \to \bb$}{H to bb}}
The $\PH \to \bb$ decay channel has by far the largest branching fraction 
of all the decay channels considered, with around 60\% of Higgs bosons decaying in this way. 
The background from QCD production of pairs of \PQb jets is very large, hence 
final states with special characteristics have been chosen to enhance the signal-to-background
ratio~\cite{CMS:2017odg,CMS:2018kst,CMS:2018sah,CMS:2018hnq,CMS:2020zge}.

To select jets most likely to originate from $\PQb$ quarks, a DNN
algorithm is used~\cite{CMS-DP-2018-058,Bols:2020bkb}.
It provides a continuous discriminant score, which combines
information typical of $\PQb$ quark jets, such as the presence of tracks
displaced from the primary vertex, identified secondary vertices, and
the presence of low-\pt leptons in the jet.
The threshold on the discriminant score is set such that the
misidentification rate for light ($\PQu$, $\PQd$, $\PQs$) quarks
or gluons is low. For example, setting this misidentification rate at
0.1\% gives a 50\% efficiency for $\PQb$ quark
jet identification when applied to jets in top quark-antiquark events. 

The $\VH$ production mode uses the presence of one or more leptons
from the decay of the vector boson, or large \ptmiss. In the signal-sensitive region, DNNs are
used to separate
the signal from the background dominated by QCD multijet production.

The $\ttH$ and $\tH$ production modes are included in the combination and 
MVA techniques are used to separate the signal from the large multijet backgrounds.
This analysis uses the 2016 data set.

Lastly, an inclusive analysis  is included that targets Higgs bosons produced with large \pt~\cite{CMS:2020zge}. 
In this kinematic region the signal to background ratio is larger. 
The two $\PQb$ jets from decays of highly Lorentz-boosted Higgs bosons are close in space and appear in the detector as a 
single broad jet with distinctive internal structure.

\begin{figure*}[htb]
\centering
\includegraphics[width=0.8\textwidth]{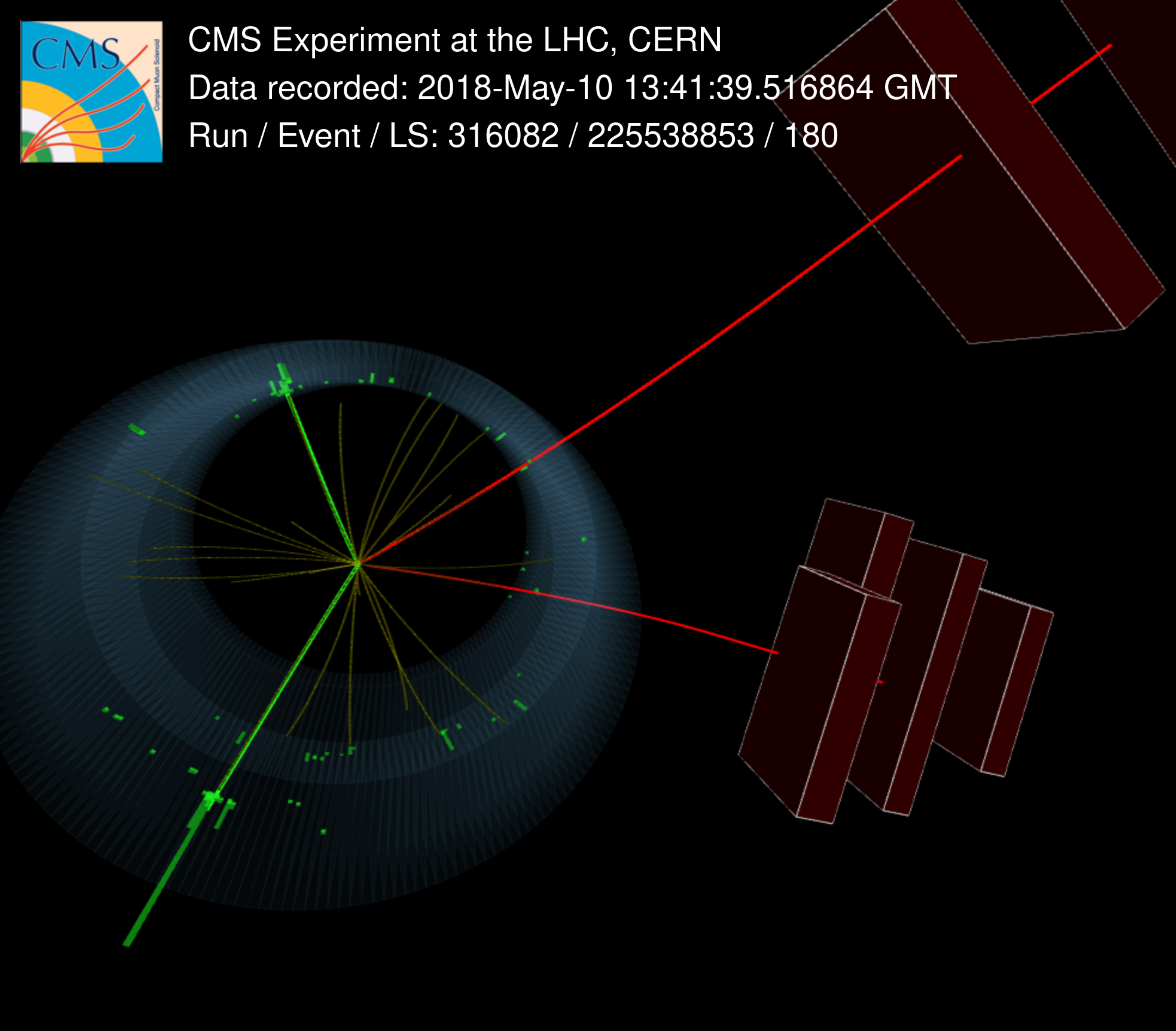} 
\includegraphics[width=0.8\textwidth]{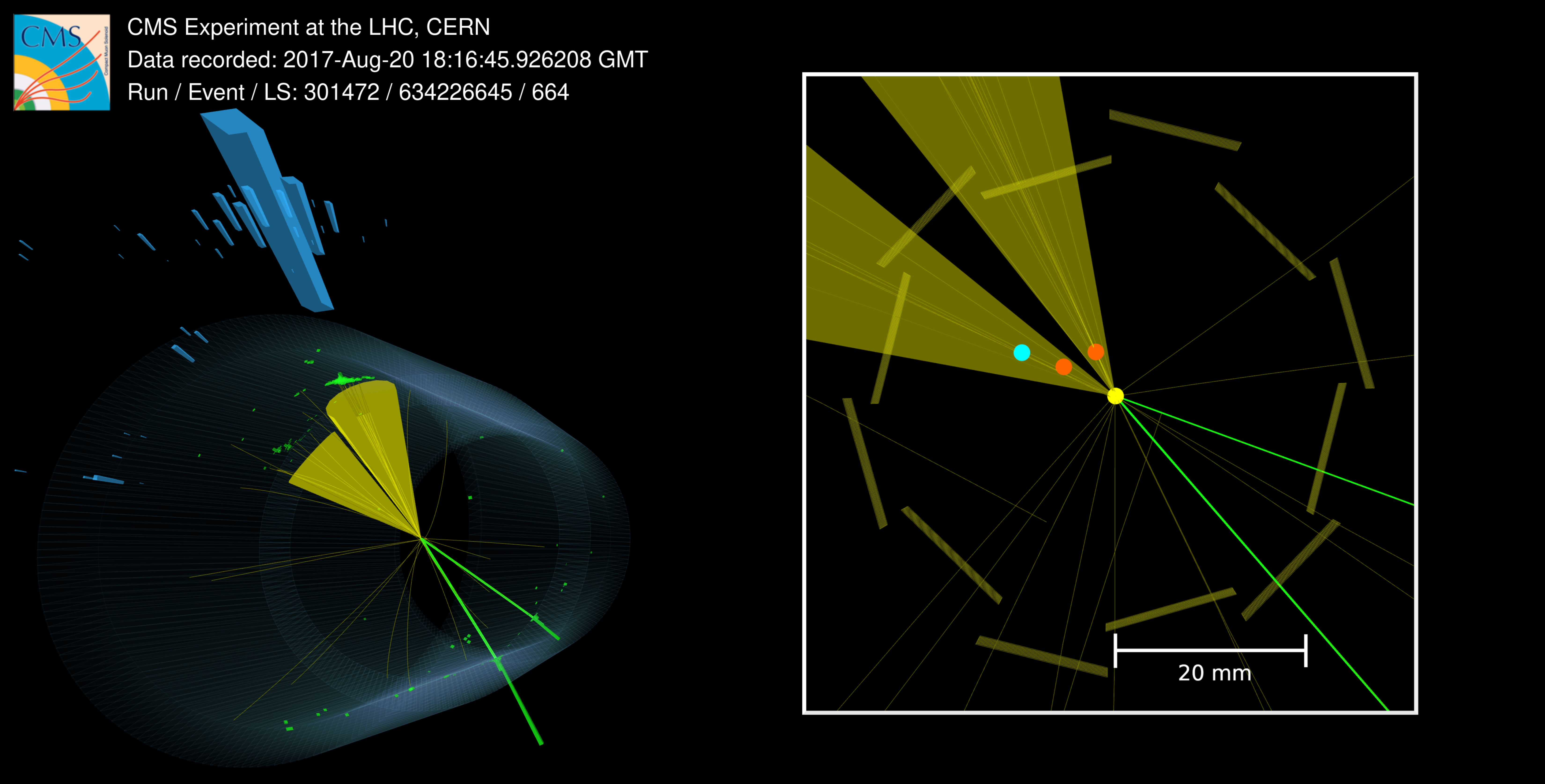}
\caption{
\figtitle{Higgs boson candidate events.}
    (upper)  An event display of a candidate $\PH \to \ZZ \to \Pe\Pe \PGm\PGm$. 
(lower) An event display of an $\PH \to \bb$ candidate produced in association with a \PZ boson decaying into an electron-positron 
pair, in $\Pp\Pp$ collisions at $\sqrt{s}=13\TeV$ recorded by CMS. The charged-particle tracks, as reconstructed in the inner tracker, 
are shown in yellow; the electrons are shown in green, the energy deposited by the electrons in the ECAL is shown as large green towers, 
the size of which is proportional of the amount of energy deposited; the blue towers are indicative of the energy deposits in the HCAL, 
while the red boxes are the muon chambers crossed by the muons (red tracks); the yellow cones represent the reconstructed jets. 
(lower, inset) The zoom into the collision region shows the displaced secondary vertices (in red) of the two \PQb quarks 
decaying away from the primary vertex (in yellow). One of the bottom hadrons decays into a charm hadron that moves away from the secondary vertex 
 before decaying ($\PQb \to \PQc \to \PX$; vertex in cyan).}
\label{Extended-Event}
\end{figure*}

\begin{figure*}[htb]
\centering
\includegraphics[width=0.49\textwidth]{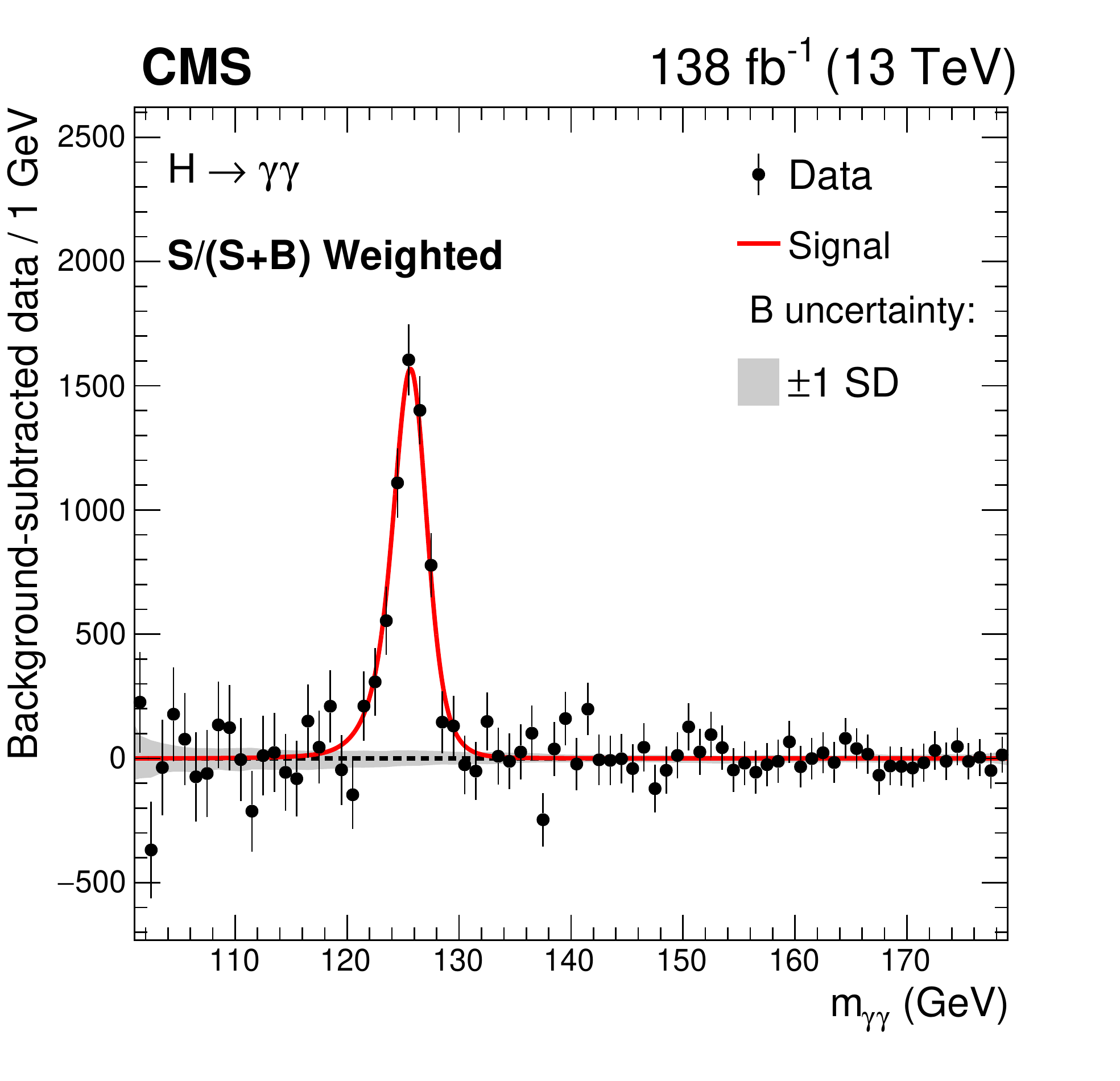} 
\includegraphics[width=0.49\textwidth]{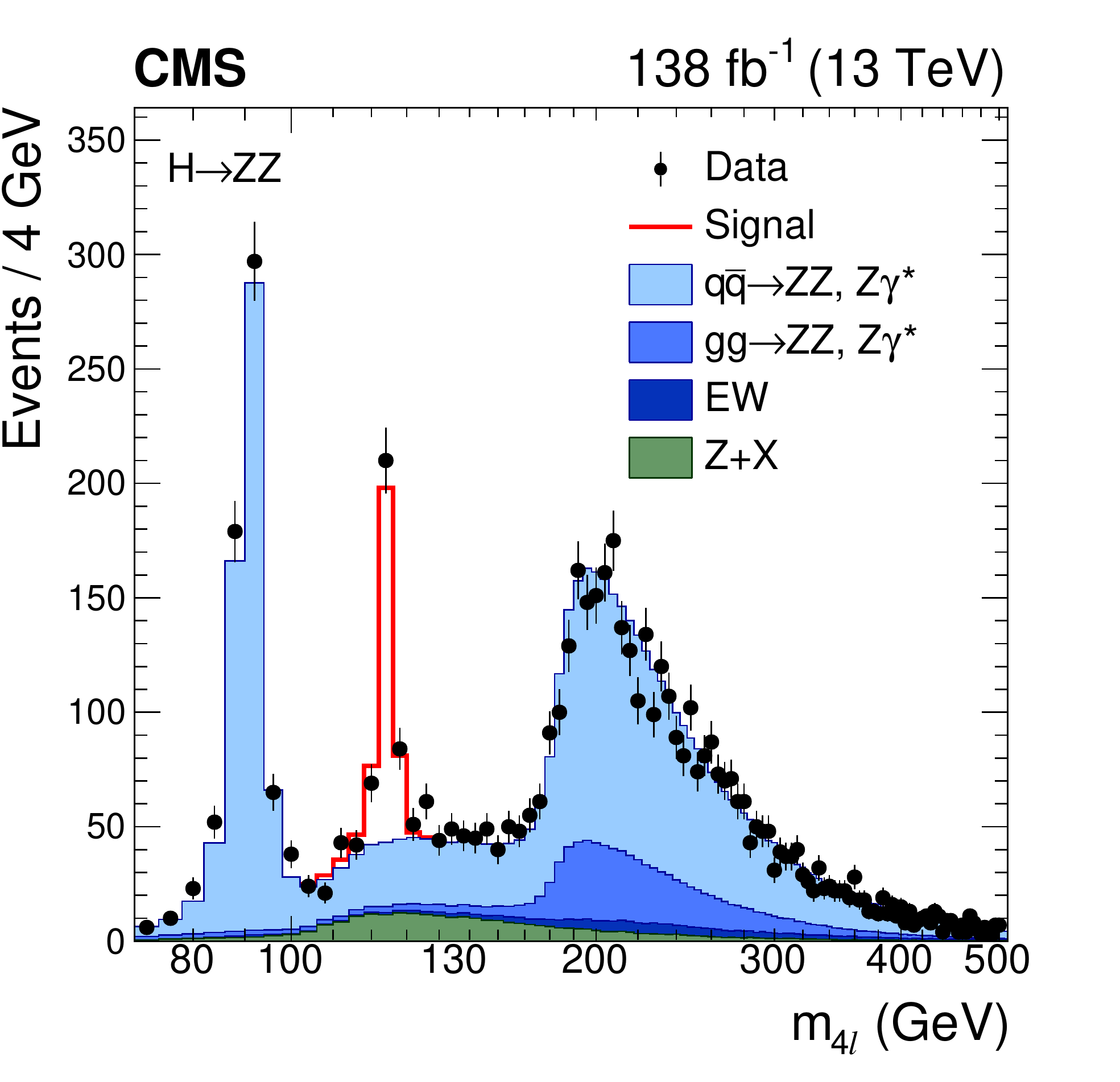}
\includegraphics[width=0.49\textwidth]{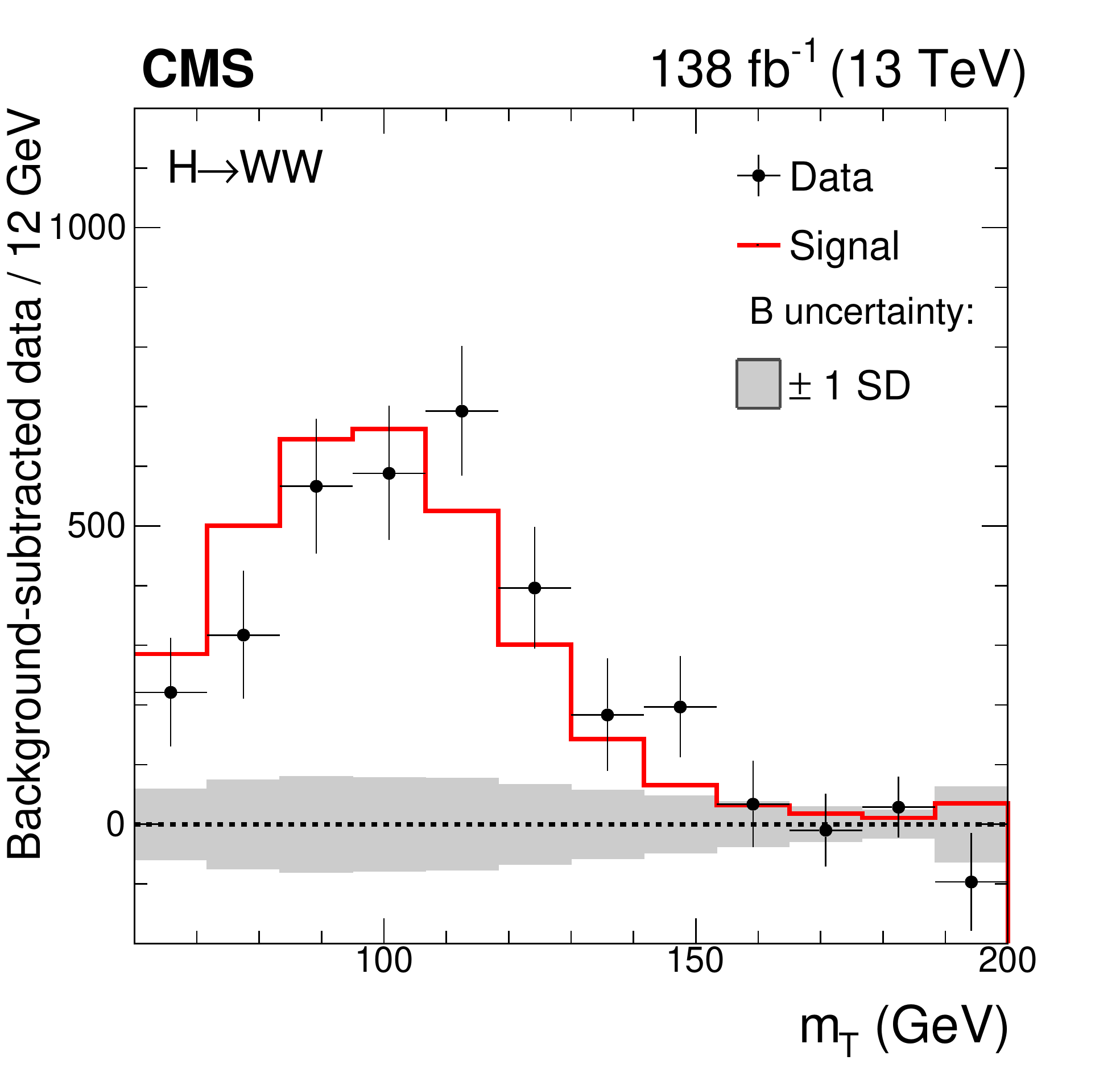}
\includegraphics[width=0.49\textwidth]{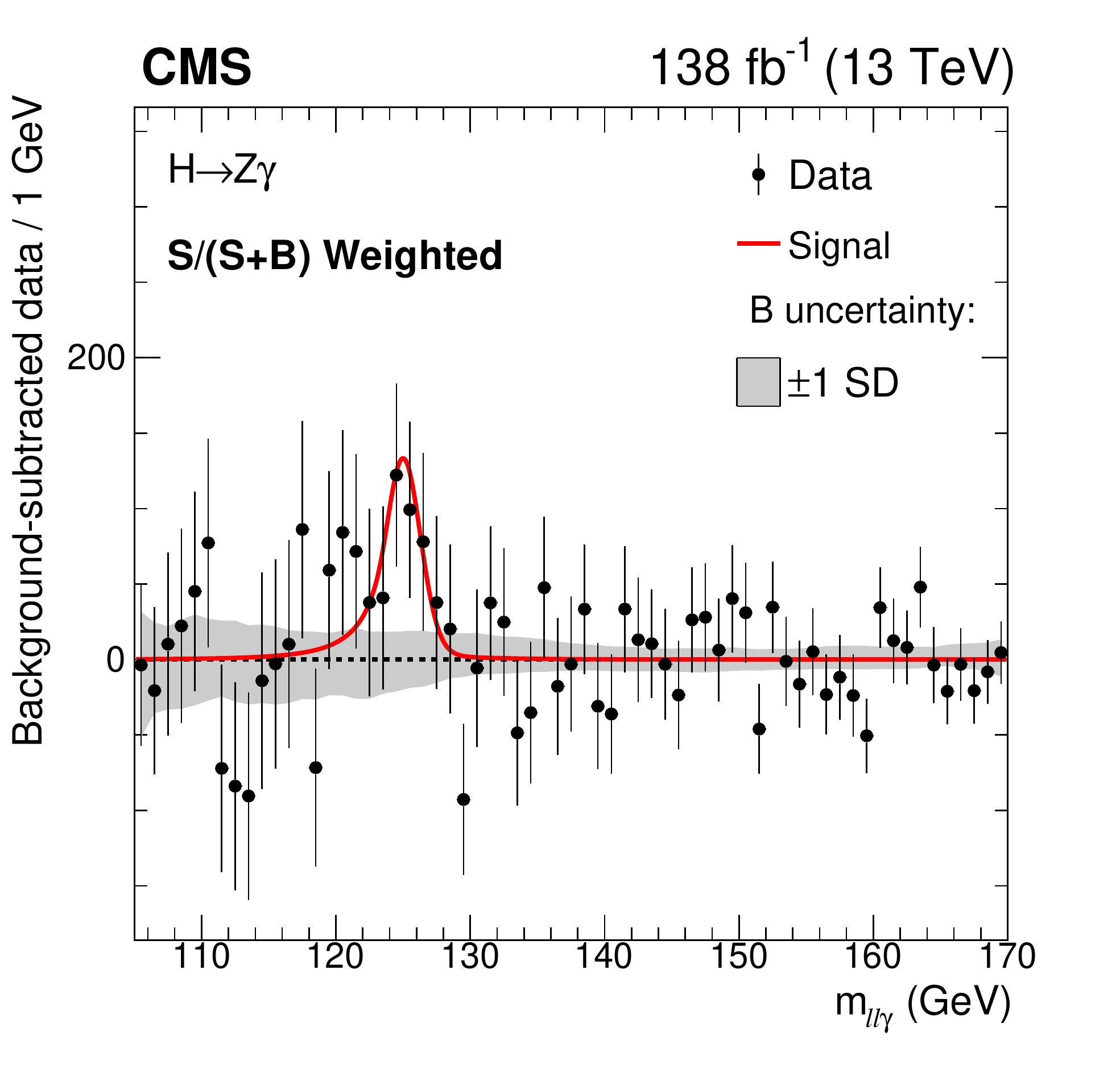}
\caption{ 
\figtitle{Higgs boson mass peak in diboson decay channels.}
    (upper~left) The background-subtracted diphoton invariant mass distribution targeting the study of the decay channel $\PH \to \PGg\PGg$.
(upper~right) The invariant mass distribution of four charged leptons targeting the study of the decay channel $\PH \to \ZZ \to 4\ell$. 
(lower~left) The background-subtracted transverse mass \mT distribution targeting
  the study of the decay channel $\PH \to \WW$. 
(lower~right) The background-subtracted $\ell\ell\PGg$ invariant mass distribution targeting the study of the decay channel $\PH \to \PZ\PGg$.
The SM prediction for the signal (red line) is scaled by the
value of $\mu$, as estimated in the dedicated analysis for that channel, and computed for $m_\PH=125.38 \GeV$. The grey band around zero shows the 1 s.d. uncertainty in the background subtraction. }
\label{Extended-Fig3}
\end{figure*}

\begin{figure*}[htb]
\centering
\includegraphics[width=0.3\textwidth]{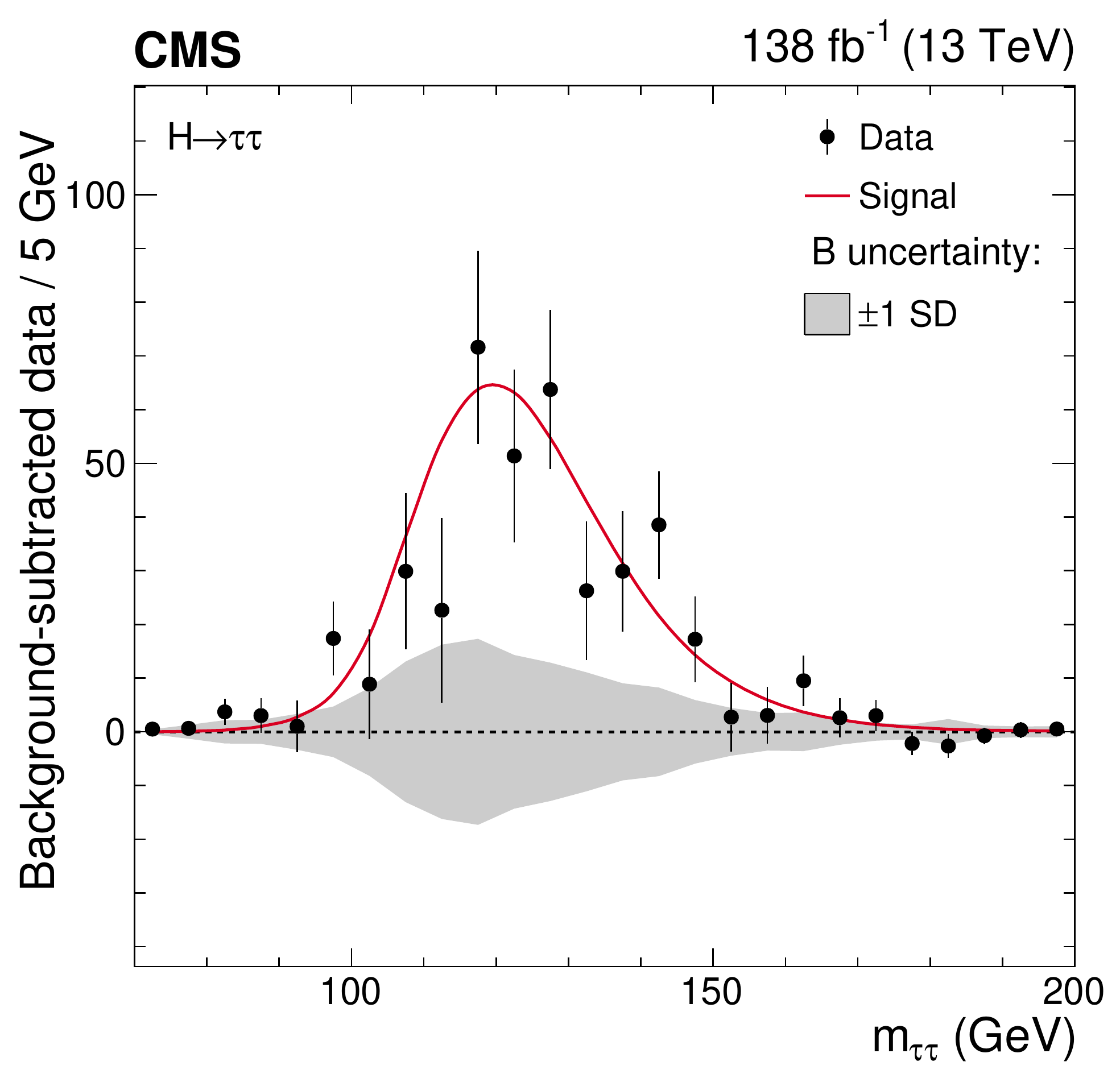}
\includegraphics[width=0.3\textwidth]{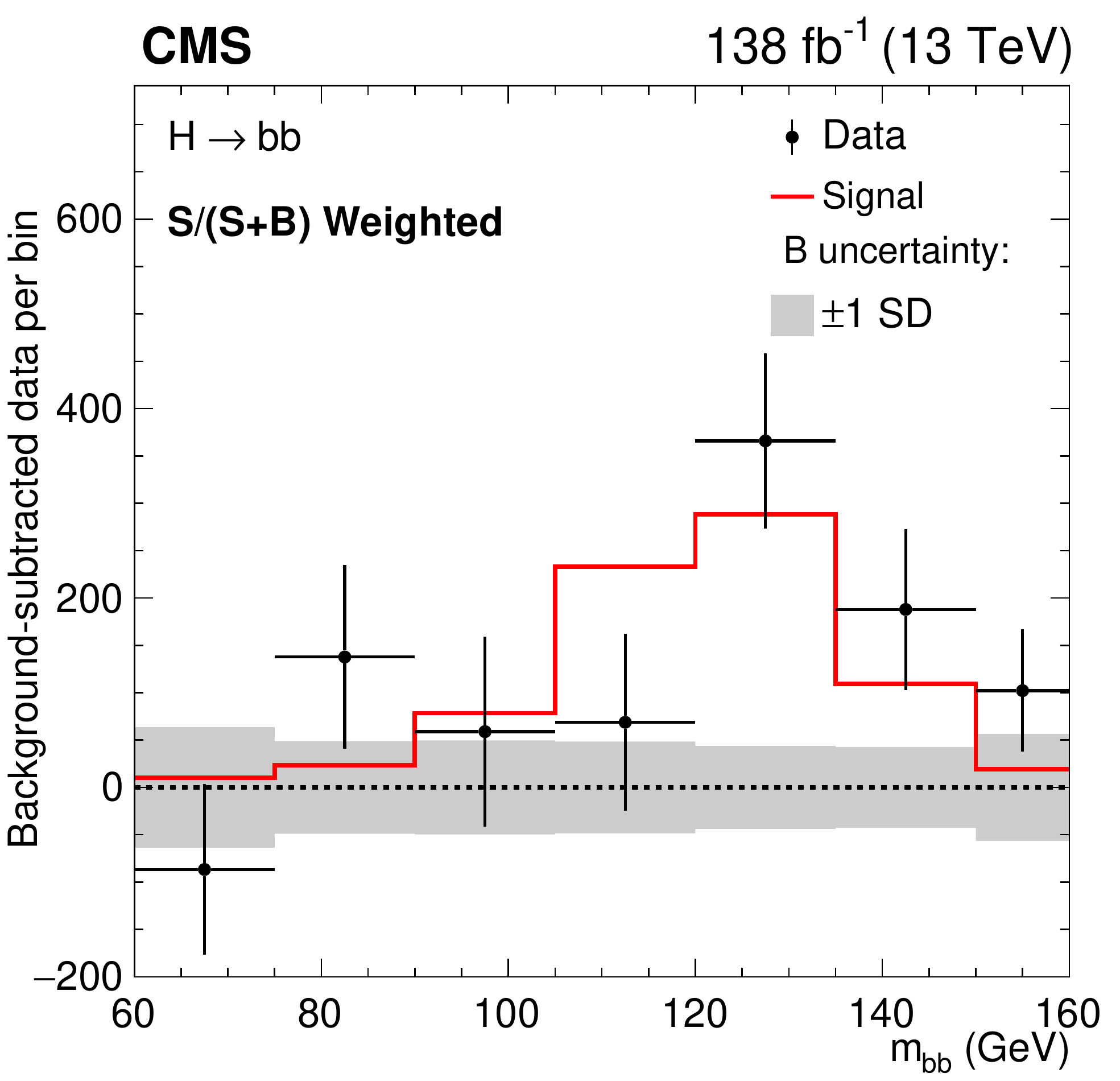}
\includegraphics[width=0.3\textwidth]{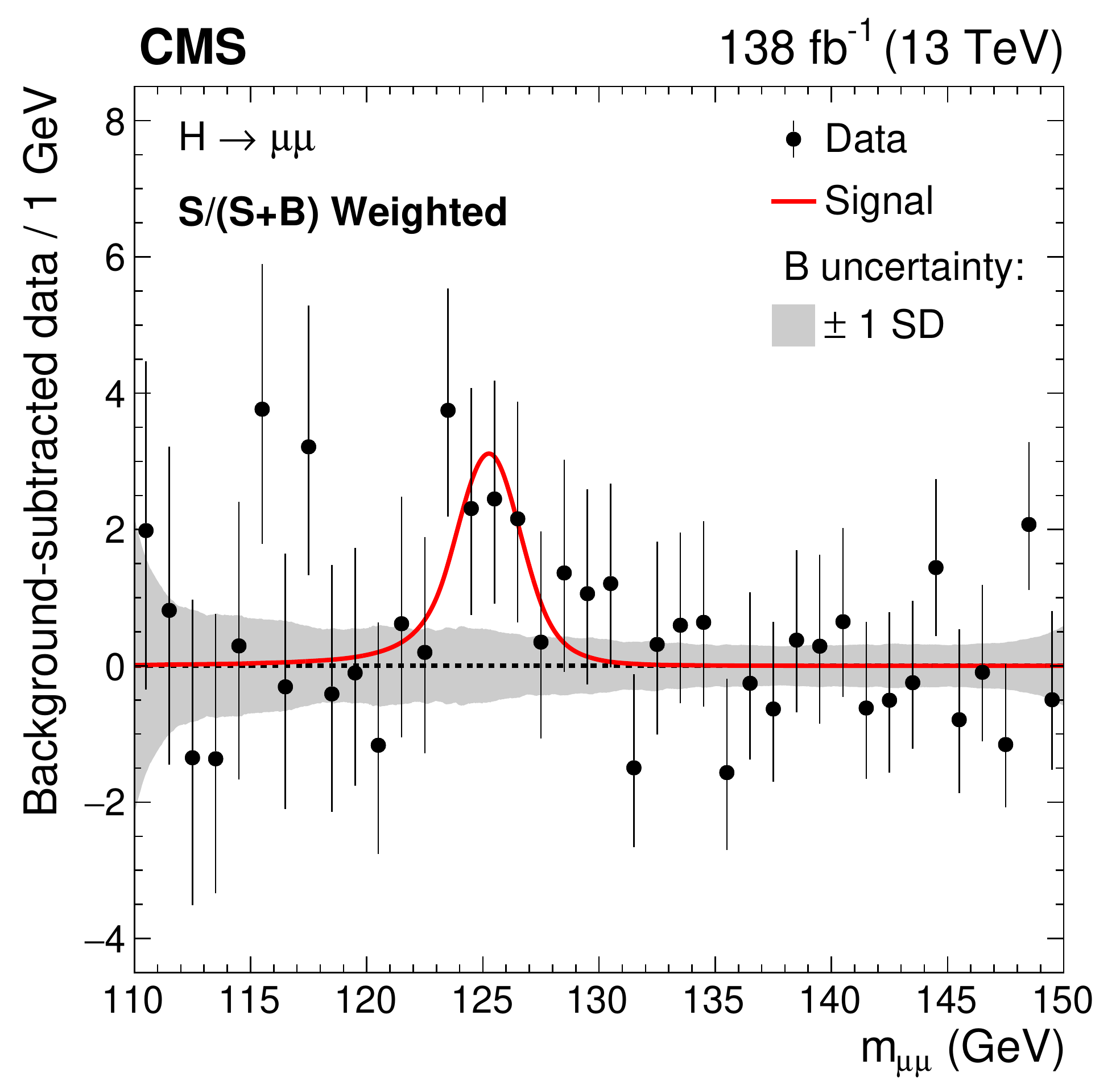}
\caption{ 
\figtitle{Higgs boson mass peak in difermion decay channels.}
    The background-subtracted diparticle invariant  mass distribution targeting the study of the decay channel (left)  $\PH \to \ditau$,
(center) $\PH \to \bb$, (right)  $\PH \to \PGm\PGm$.  The SM prediction for the signal (red line) is scaled by the
value of $\mu$, as estimated in the dedicated analysis for that channel, and computed for $m_\PH=125.38 \GeV$. The grey  band around zero shows
 the 1 s.d. uncertainty in the background subtraction.}
\label{Extended-Fig4}
\end{figure*}

Extended Data Fig.~\ref{Extended-Event} (lower) shows a display of a candidate $\PH \to \bb$ event 
produced in $\Pp\Pp$ collisions at a centre-of-mass energy $\sqrt{s} = 13\TeV$ and recorded in the CMS detector. 

\paragraph{\texorpdfstring{$\PH \to \PGm\PGm$}{H to mu+mu}}
The signal is searched for as a peak in the dimuon mass distribution, over a smoothly falling background~\cite{CMS:2020xwi}.
The dimuon invariant mass resolution is $\sigma_{m_{\PGm\PGm}}/m_{\PH} \approx 1\%$. 
The analysis of this decay channel targets the $\ggH$, VBF, $\VH$, and
$\ttH$ production modes, and is most sensitive in the first two
modes.
The largest background in this decay channel comes from Drell--Yan
dimuon production in which an off-shell $\PZ^*$ boson decays to a pair of muons.
Events are split into production modes based on their kinematical
properties. To improve the sensitivity of the analysis,
MVA techniques are used in each of these different categories. 

\paragraph{\texorpdfstring{$\PH \to \PQc\PQc$}{H to cc}}
{\tolerance=800
  The analysis of this final state in the \VH\ production mode (Fig.~\ref{feyn-diag}c) has recently been
  presented~\cite{CMS:2022psv} but has not been included in the
 present combination. The analysis yields $\sigma(\VH) \mathcal{B}(\PH \to \PQc \PQc) < 0.94\unit{pb}$ at 95\% \CL. The observed 95\% \CL interval (expected upper limit) 
 for $\kappa_{\PQc}$  is found to be $1.1 < \abs{\kappa_{\PQc}} < 5.5$ ($\abs{\kappa_{\PQc}}< 3.4$), the most stringent to date. A search for $\PZ \to \PQc \PQc$ in \VZ events is used to validate the analysis strategy and yields a first observation of this decay channel, at a hadron collider, with a significance of 5.7 s.d.
\par}

\subsubsection{ttH/tH with multileptons}
\label{sec:tth}
The $\ttH$ (Fig.~\ref{feyn-diag}d) and $\PQt\PH$ (Figs.~\ref{feyn-diag}e,~\ref{feyn-diag}f) production channels, which probe the coupling of the Higgs boson 
to the top quarks, are studied in the case where the Higgs boson and the top quarks subsequently decay into final states with several 
leptons~\cite{CMS:2020mpn}, supplementing dedicated studies of the $\PH \to \PGg\PGg$, $\PH \to \ZZ \to 4\ell$, and $\PH \to \bb$ decay modes.

This analysis uses a categorization based on the number of leptons and/or $\tauh$ candidates to target both
the different Higgs boson final states and $\PQt\PQt$ decay channels.
Categories with at least two leptons, or one lepton and two $\tauh$
candidates, target cases where at least one top quark decays via a
leptonically decaying $\PW$ boson.
Categories with one lepton and one $\tauh$, or with no leptons and two
$\tauh$ candidates are used to target events
in which both top quarks decay via hadronically decaying $\PW$ bosons.
This analysis is sensitive to the $\PH \to \WW$, $\PH \to \ditau$, and
$\PH \to \ZZ$ decay channels. Several MVA techniques are employed to better
separate the $\ttH$ and $\tH$ production modes.

\subsubsection{Higgs boson decays beyond the SM}
In addition to the invisible Higgs boson decays discussed in Section~\ref{kappaframework}, other BSM decays are possible, into \textit{undetected} particles. 
That is, these particles may or may not leave a trace in the CMS detector, 
but we do not have dedicated searches looking for these signatures. Nevertheless, the presence of undetected decays can be inferred indirectly from a reduction in
 the branching fraction for SM decays (or by an increase in the total Higgs boson width). In this interpretation, the total width becomes
 $\Gamma_\PH = \sum \Gamma_f (\kappa) / (1 - \mathcal{B}_\text{Inv.} - \mathcal{B}_\text{Undet.})$, where $\mathcal{B}_\text{Undet.}$ is the branching fraction to undected particles.

To probe invisible or undetected decays of the Higgs boson, another fit can be performed, including $\mathcal{B}_\text{Inv.}$ and 
$\text{B}_\text{Undet.}$ as additional floating parameters, while imposing as an upper bound on 
$\kappa_\PW$ and $\kappa_\PZ$ their SM values, also valid in most proposed extensions of the SM~\cite{Gunion:1990kf,Englert:2014uua}. 
As can be seen from Extended Data Fig. \ref{Extended-FigZ}~(right), $\mathcal{B}_\text{Inv.}$ and $\mathcal{B}_\text{Undet.}$ 
are found to be consistent with zero. The 95\% \CL upper limit on  $\mathcal{B}_\text{Undet.}$ is found to be ${<} 0.16$, with only small 
changes to the other $\kappa_i$ fitted values, as shown in Extended Data Fig.~\ref{Extended-FigZ}~(right). 
The measurement of the width \cite{CMS:2022ley} of the Higgs boson will be used in the future to constrain
these quantities without imposing bounds on $\kappa_\PW$ and $\kappa_\PZ$. 

\subsection{Statistical analysis}
\label{sec:stat}
The statistical framework used to build the combination of all the channels is based on an established combined likelihood
method (\cite{ATLAS:2011tau} and references therein),
and briefly detailed in this section.

Given the enormous number of $\Pp\Pp$ collisions produced at the LHC and the relatively small probability that one of those
collisions will produce a signal-like event, the observations in data are described by Poisson probability functions, 
$\mathcal{P}(k\vert\lambda)= \mathrm{e}^{-\lambda} \lambda^k/ k!$, where $k$ is the observed number of events, and the parameter
$\lambda$ is the expected number of events in a particular bin or region of one or more of the discriminating distributions used
for each channel entering the combination. The combined likelihood is obtained from the product of the individual
Poisson probability functions, accounting for the observed data and expected number of events across all channels. 

The  parameters $\lambda$ are functions of the model parameters of interest: ${\mu}$,
which represent the Higgs boson couplings or signal strengths,  and
nuisance parameters $\theta$,
which model
the effect of systematic uncertainties on the predicted signal and background contributions. Additional terms are included in the
combined likelihood to represent constraints on the nuisance parameters due to external measurements, such as energy and
momentum scale calibrations or an integrated luminosity determination. 
The measurements reported in this paper are determined using the profile likelihood ratio
$q(\mu) = -2 \ln \mathcal{L}(\mu,\hat{\theta}_\mu)/\mathcal{L}(\hat{\mu},\hat{\theta})$
where $\hat{\mu}$ and $\hat{\theta}$ are the values of the parameters of interest and nuisance parameters that maximize
the likelihood $\mathcal{L}(\mu,\theta)$, and $\hat{\theta}_{\mu}$ are the values of the nuisance parameters that maximize the
likelihood for a fixed value of $\mu$.  
The compatibility between a given set of measurements and their corresponding SM predictions is reported as a $p$-value, derived
from the difference between $q_\text{SM}$ and $q(\hat{\mu})$. Expected
intervals are derived from
the \textsl{Asimov} data set, in which the nuisance parameters are set to their maximum likelihood estimator values.

The modified likelihood ratio test statistic
$\tilde{q}(\mu) = -2 \ln[\mathcal{L}(\mu,\hat{\theta}_\mu)/\mathcal{L}(\hat{\mu},\hat{\theta})]$
with a constraint  $0\le \hat{\mu} \le \mu$
is used to set 95\% CL upper
limits on signal strengths and production cross sections
using the ``CL$_s$ criterion''~\cite{ATLAS:2011tau}.

All the reported confidence intervals, confidence regions, and
$p$-values are obtained assuming various asymptotic approximations for the distributions of the
(modified) likelihood ratio test statistic~\cite{Cowan:2010js}. The validity of the
asymptotic assumptions has been routinely checked in the context of individual analyses whenever the event yields are small
or particular validity conditions are not met.

\subsection{Signal strengths \texorpdfstring{$\mu$}{mu}: production channels and decay modes}
For a Higgs boson produced in mode $i$ and decaying into a
final state $f$, the signal event yields are proportional to
$\sigma_i \mathcal{B}^f$, where  $\sigma_i$ is the production cross section
and $\mathcal{B}^f$ the decay branching fraction.
The branching fraction is in turn given by $\mathcal{B}^f = \Gamma^f /\Gamma_\PH$, where $\Gamma^f$ is  the partial decay width
in the final state $f$ and $\Gamma_\PH$ the total natural width of the Higgs boson.

Fits are performed under different assumptions:
\begin{itemize}
\item per overall single signal strength, yielding $\mu=1.002 \pm 0.057$,
\item per production channel signal strengths ($\mu_i = \sigma_i /\sigma_i^{\text{SM}}$ with
  $\mathcal{B}^f = \mathcal{B}^f_{\text{SM}}$), Fig.~\ref{Comb_mu}~(left),
\item per decay mode signal strengths ($\mu^f =\mathcal{B}^f/\mathcal{B}^f_{\text{SM}}$, with $\sigma_i =\sigma_i^{\text{SM}}$), Fig.~\ref{Comb_mu}~(right), and 
\item with a free parameter per individual combination of production modes and decay channels, as illustrated in Extended Data Fig.~\ref{Extended-FigMuPD}.
\end{itemize}

\begin{figure*}[htb]
\centering
    \includegraphics[width=0.45\textwidth]{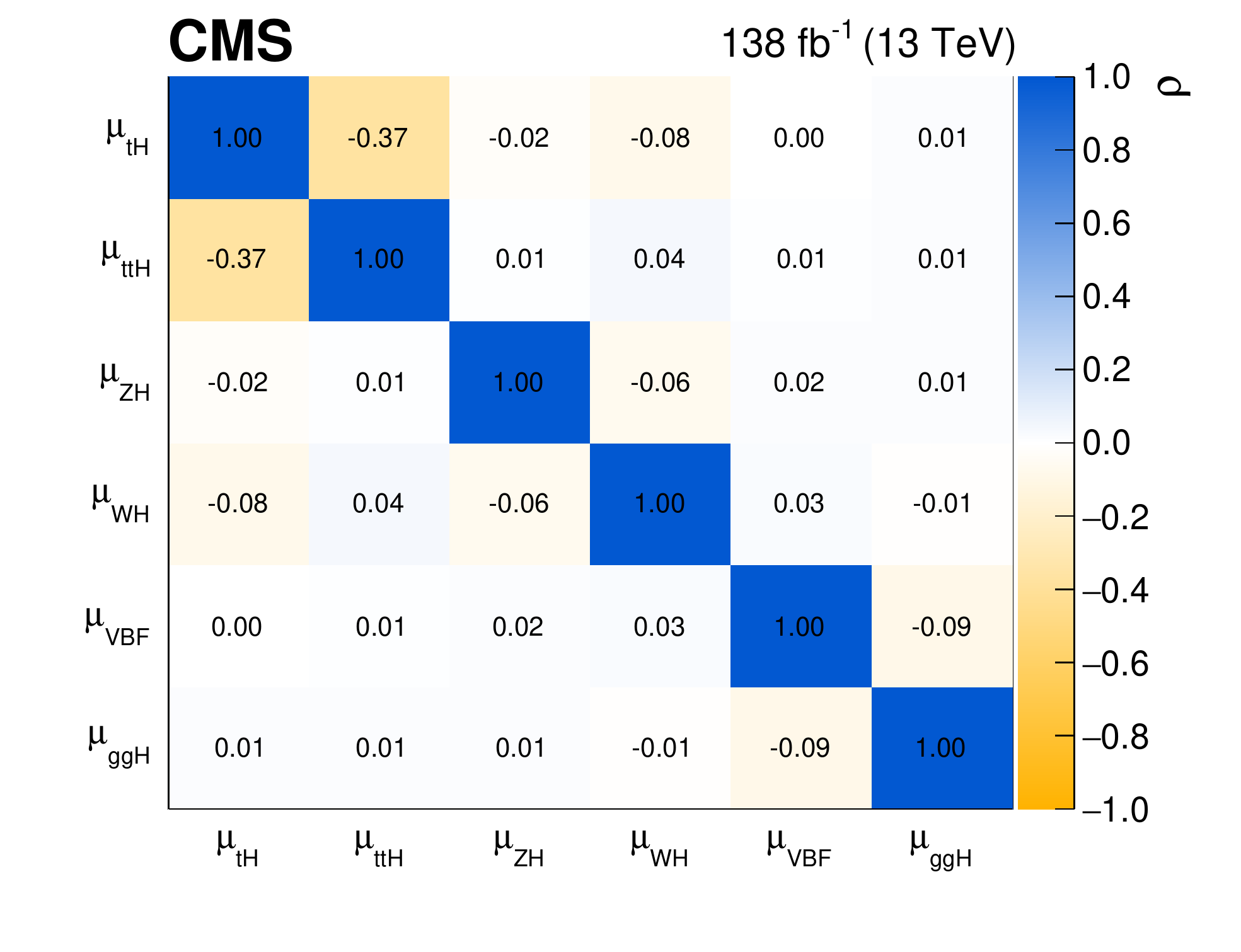}
    \includegraphics[width=0.45\textwidth]{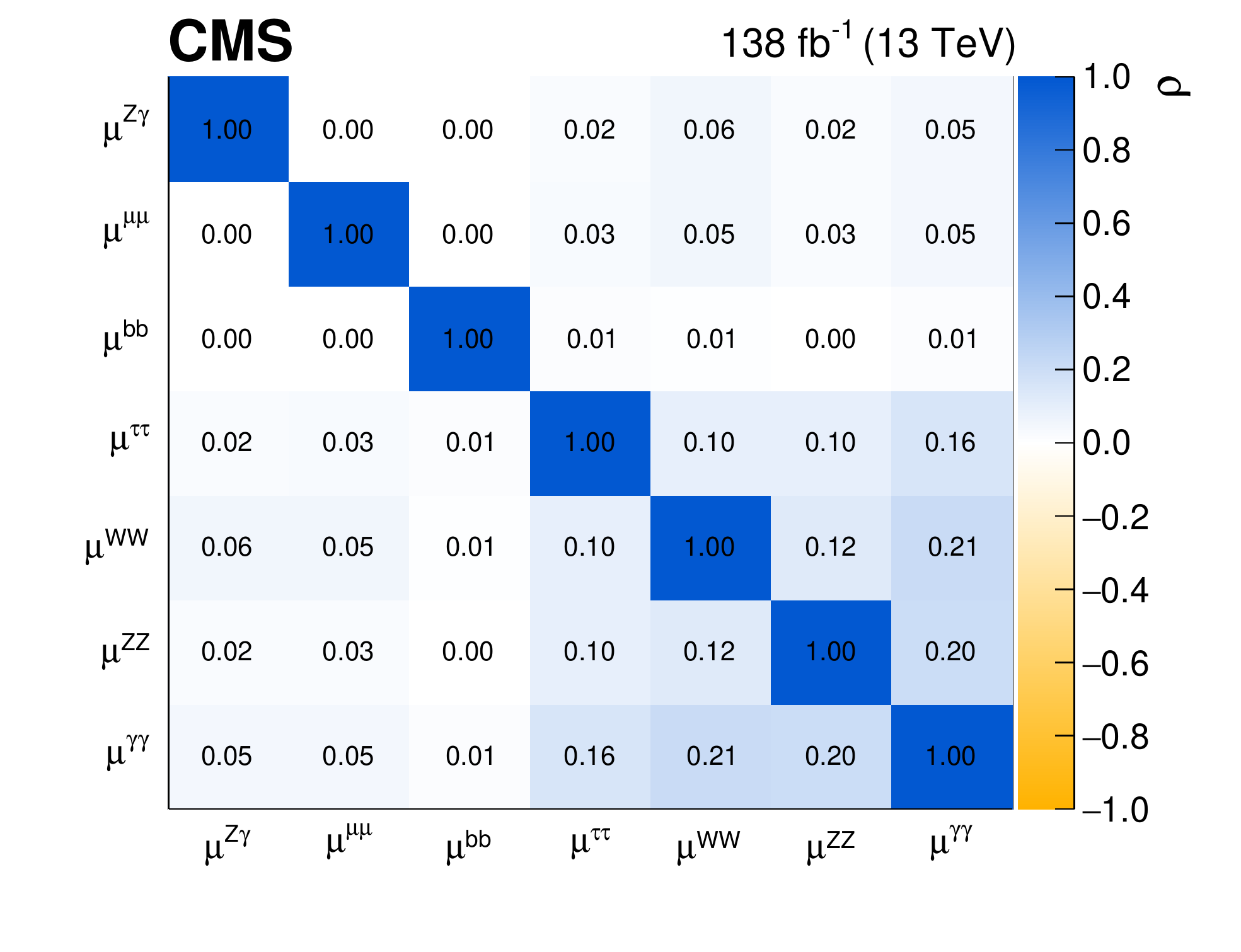}
    \caption{
    \figtitle{Correlations between the measurements of different couplings.}
        Covariance matrices for the fits of the signal strength parameters per production mode $\mu_i$ (left) and 
    per decay mode $\mu^f$ (right). The values of the correlation coefficients, $\rho$, are indicated both in text 
    and in the color scale.}
\label{Extended-FigMuCovs}
\end{figure*}

\begin{figure*}[htb]
\centering
\includegraphics[width=0.99\textwidth]{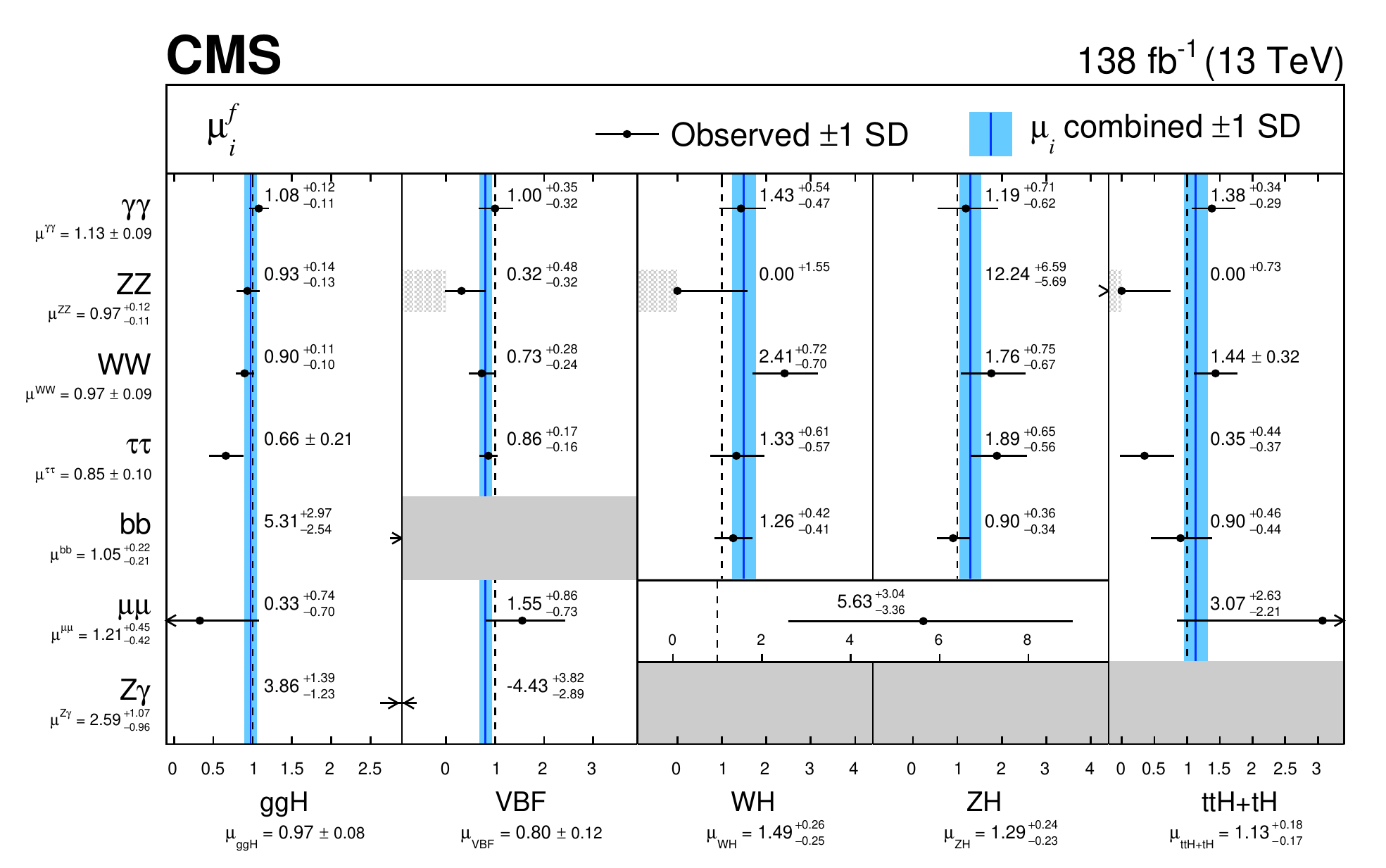}
\caption{
\figtitle{The agreement with the SM predictions in Higgs boson production and decay.}
    Signal strength parameters per individual production mode and decay channel $\mu_i^f$, and combined per production 
mode $\mu_i$ and decay channel $\mu^f$. In this fit, \ttH and \tH are
considered together and the $\mu_i$ results are slightly different
from those of Fig.~\ref{Comb_mu}~(left). The dashed vertical lines at
1 represent the SM value. Light grey shading indicates that $\mu$ is
contained to be positive. Dark grey shading indicates the absence of
measurement. The $p$-value with respect to the SM prediction is 5.8\%.}
\label{Extended-FigMuPD}
\end{figure*}

The covariance matrices for the fitted signal strengths per production mode $\mu_i$ and 
per decay channel $\mu^f$ are shown in Extended Data Fig.~\ref{Extended-FigMuCovs}.

\subsection{Notes on self-interaction strength}
\label{sec:self}
The potential energy of the BEH field ($\phi$) is given by 
$V(\phi) = \frac{1}{2} m_\PH^2  \phi^2+ \sqrt{\lambda/2} m_\PH  \phi^3 + \frac{1}{4} \lambda \phi^4 $. 
The first term accounts for the mass of the Higgs boson $m_\PH$.
The second term represents the Higgs boson self interaction, of strength $\lambda$. 
In the SM, $\lambda = m_\PH^2 / (2 \upsilon^2) $ (where the vacuum expectation value of the BEH field, corresponding to its minimum, is $\upsilon = 246\GeV$) and it can be measured 
via the study of Higgs boson pair production.
The third term represents the interaction of four Higgs bosons at a point, a process that is even rarer than its pair production.
Knowledge of the exact shape of the potential $V$ is crucial for understanding
the phase transition that occurred in the early universe and its consequences~\cite{Isidori:2001bm}. 

The search for Higgs boson pair production is performed by combining Higgs boson pairs, each with differing decay modes. 
The decay modes that have been used are $\bb$, $\ditau$, and $\WW$~\cite{CMS:2022cpr,CMS:HH4b-comb,CMS:HHbbtautau,CMS:HHleptons}, benefitting from the large branching fractions, 
and $\PGg\PGg$~\cite{CMS:HHggbb} and $\ZZ \to 4\ell$~\cite{CMS:HHbbZZ}, benefitting from the presence of narrow mass peaks, thus improving the 
signal-to-background ratio. All final states analyzed are defined to be mutually exclusive so that they could be 
properly combined as statistically independent observations. 

Measurements of Higgs boson pair production are used to constrain the Higgs boson self-interaction strength $\lambda$.
Several combinations of individual Higgs boson decay modes are
employed in this search.
The highest rate for Higgs boson pair production and decays occurs
when both Higgs bosons decay to
$\PQb$ quark pairs, $\PH\PH \to \bb \bb$, corresponding to $\approx$35\% of all the possible $\PH\PH$ decays in the SM. 

The search in the $4\PQb$ decay mode~\cite{CMS:2022cpr,CMS:HH4b-comb} is performed separately under the
assumptions that
$m_{\PH^*} \gg 2 m_\PH$ or not. In the case $m_{\PH^*} \gg 2 m_\PH$, each Higgs
boson is energetic (and hence said to be boosted), such that its decay products,
\eg $\PQb$ quark jets, merge and appear as one \textit{broad} jet, but
with a distinctive internal structure.
In the latter case, all four $\PQb$ quark jets rarely
overlap, and hence said to be resolved.  

Another group of analyses targets the $\PH\PH$ final states where one
$\PH$ decays to $\PQb$ quarks and the other to $\ditau$~\cite{CMS:HHbbtautau},
$\PGg\PGg$~\cite{CMS:HHggbb}, or $\ZZ \to 4\ell$~\cite{CMS:HHbbZZ}. Analyses targeting a set of
multileptons final states
with \ptmiss are $\PH\PH \to (\WW) (\WW)$, 
$(\WW)(\ditau)$, or $(\ditau)(\ditau)$~\cite{CMS:HHleptons}, where hadronic tau lepton decays are also included.

A fit to Higgs boson pair production data can be used to simultaneously constrain $\kappal$ and $\kappavv$, 
as shown in Extended Data Fig.~\ref{Extended-FigKL}~(left).

Measurements of single Higgs boson production and decay can also be
used to constrain $\kappal$ as quantum corrections
to the SM Higgs boson production modes and decay channels 
depend on $\kappal$~\cite{Degrassi:2016wml,Maltoni:2017ims}.
These corrections have been derived~\cite{CMS-Note-kappa} for the different production and decay modes entering the combination, 
as shown in Extended Data Table~\ref{Extended-Tab1}.

The values of $\kappal$ extracted from single and pair Higgs boson production are shown in Extended Data Fig.~\ref{Extended-FigKL}~(right).

\subsection{Upgrade of the CMS experiment for HL-LHC running}
To exploit the full potential of the LHC, the accelerator and its
experiments will be upgraded.
The HL-LHC will operate at an instantaneous luminosity of $5 \times
10^{34}\percms$. The intention is to collect ten times more data than
the $300\fbinv$ foreseen in the initial LHC phase.
This means that the integrated radiation levels will be
correspondingly larger. 

The physics to be studied drives the technical
choices for the upgrade. The physics goals are:
\begin{itemize}
\item precise measurements of the properties of the Higgs boson and its self-coupling, to elucidate further the physics of EW symmetry breaking;
\item search for BSM physics; and
\item selected precision SM measurements. 
\end{itemize}

The translation of these physics goals into experimental design goals requires:
\begin{itemize}
  \item 
  The construction of a new higher-granularity, more radiation-hard silicon tracker.
The design of the new front-end electronics will allow information from the inner tracker to
participate in the Level-1 trigger. The size of the individual
detecting elements will be decreased leading to about ten times
larger number of electronics channels. All components inside the tracker (silicon sensors, front-end
electronics, 10\unit{Gb/s} data links, etc.) will have to withstand
integrated doses of up to 500\unit{Mrad} and fluences of $10^{16}$ ($1\MeV$ equivalent neutrons)/\cmsq.
The geometric coverage of the inner tracker will be increased, extending it down an angle of $2^\circ$ from the beamline. 
\item The replacement of other components affected by radiation.
Principally these are the endcap calorimeters and the ECAL front-end electronics.
The endcap calorimeters will be replaced with a
new high-granularity ``imaging''
calorimeter with precision timing. It will be based on $600\meter^2$ of
silicon sensors with detecting cells of sizes of
0.5 to 1.0\cmsq. Regions in this calorimeter will reach
integrated doses of up to 500\unit{Mrad}
and fluences of $10^{16}$ ($1\MeV$ equivalent neutrons)/$\mathrm{cm}^2$. The new front-end electronics for the
ECAL barrel will allow data
from each crystal to be sent to the calorimeter Level-1 trigger
processor, instead of the sum
of 25~crystals today, and which will allow better measurement
of the timing of the impact of electrons or photons. 

\item Higher-bandwidth Level-1 and high-level triggers. 
        Information from the inner trackers will be used at Level-1.
        The Level-1 trigger latency will be increased from 4
        to over $12\mus$, requiring corresponding changes in the front-end electronics, allowing more processing time leading to a
        purer selection of events.
        The output rate from the Level-1 processors will be increased from 100 to 750\unit{kHz} and correspondingly the number of
        events stored for later analysis will be increased from 1 to 10\unit{kHz}. 
        
\item The introduction of precision timing detectors.
A new set of detectors will be installed in the barrel and endcap
regions, covering a region down to an angle of 9$^\circ$ from the
beamline.
The precision timing of photons (in the barrel region)  and charged tracks will greatly improve the localization of the correct
interaction vertex. At HL-LHC, on the average, some 140 pairs of
protons are expected to interact
in each crossing, spread over a time characterized by $\sigma \approx
200\unit{ps}$. Furthermore, suppression of energy can be
carried out that is not consistent in time with the interaction of interest.
\end{itemize}

The upgraded CMS experiment at HL-LHC will be more powerful than the current one. 
Uncertainties in many measurements of the properties of the Higgs boson are expected 
to approach the percent level, benefitting from the anticipated larger event samples, 
reduced experimental systematic uncertainties, and more accurate theoretical calculations. 

\subsection{Theoretical references}
The theoretical works used in our analyses can be found in the  LHC Higgs
Cross Section Working Group reports~\cite{LHCHiggsCrossSectionWorkingGroup:2011wcg,Dittmaier:2012vm,LHCHiggsCrossSectionWorkingGroup:2013rie,LHCHiggsCrossSectionWorkingGroup:2016ypw} and in the following Refs.~\cite{Gritsan:2016hjl,Gritsan:2020pib,Demartin:2016axk,Frederix:2015fyz,Ball:2017nwa,Gainer:2014bta,Mattelaer:2016gcx,Alioli:2010xa,Melnikov:2006kv,Czakon:2011xx,Kidonakis:2013zqa,Li:2012wna,Grazzini:2017ckn,Lindert:2017olm,Kallweit:2015fta,Kallweit:2015dum,Becker:2669113,Dreyer:2016oyx,Bonciani:2015eua,Gehrmann:2015dua,Heinrich:2017kxx,Heinrich:2019bkc,Jones:2017giv,Heinrich:2020ckp,Buchalla:2018yce,Grazzini:2018bsd,Dreyer:2018qbw}.

\subsection{Data availability}
Tabulated results are provided in the \href{https://dx.doi.org/10.17182/hepdata.127765}{HEPData record} for this analysis. Release and preservation of data used by the
CMS collaboration as the basis for publications is guided by the \href{https://cms-docdb.cern.ch/cgi-bin/PublicDocDB/RetrieveFile?docid=6032&filename=CMSDataPolicyV1.2.pdf&version=2}{CMS data preservation, re-use and open acess policy}.

\subsection{Code availability}
The CMS core software is publicly available on GitHub (\url{https://github.com/cms-sw/cmssw}).

\subsection{Ethics declarations}
\noindent\textbf{Author contributions}
All authors have contributed to the publication, being variously involved in the design and the construction of the detectors, in writing software, calibrating subsystems, operating the detectors and acquiring data, and finally analysing the processed data. The CMS Collaboration members discussed and approved the scientific results. The manuscript was prepared by a subgroup of authors appointed by the collaboration and subject to an internal collaboration-wide review process. All authors reviewed and approved the final version of the manuscript.

\noindent\textbf{Competing interests}
The authors declare no competing interests.

\section{Extended data}

\begin{table*} [hb]
\centering
\topcaption{\figtitle{The SM Higgs production cross sections and branching fractions.} Theoretical cross sections for each production mode and branching fractions for the decay channels, 
at $\sqrt{s}= 13$\TeV and for $m_{\PH} = 125.38$\GeV  \cite{LHCHiggsCrossSectionWorkingGroup:2016ypw}. }
\cmsTableInt{
\begin {tabular}{ c  d{-1} @{$\,\pm\,$}l   |c  d{-1}@{$\,\pm\,$}l }
Production mode & \multicolumn{2}{c|}{Cross section (pb)} &  Decay channel &  \multicolumn{2}{c}{Branching fraction (\%)}     \\ 
\hline 
$\ggH$    &    48.31   &  2.44    &   $\bb$       & 57.63  & 0.70\\
VBF       &   3.771    & 0.807    &   $\WW$       & 22.00  & 0.33 \\
$\PW\PH$  &  1.359     & 0.028    &   $\Pg\Pg$    & 8.15   & 0.42\\
$\PZ\PH$  & 0.877      & 0.036    &   $\ditau$    & 6.21   & 0.09 \\
$\ttH $   & 0.503      & 0.035    &   $\PQc\PQc$  & 2.86   & 0.09 \\
$\bbH$    & 0.482      & 0.097    &   $\ZZ$       & 2.71   & 0.04\\
$\PQt\PH$ &   0.092    & 0.008    &   $\PGg\PGg$  & 0.227  & 0.005\\
          &\multicolumn{2}{c|}{ }&   $\PZ\PGg$   & 0.157  & 0.009\\
          &\multicolumn{2}{c|}{ }&   $\PQs\PQs$  & 0.025  & 0.001\\
          &\multicolumn{2}{c|}{ }&   $\PGm\PGm$  & 0.0216 & 0.0004 \\

\multicolumn{3}{c|}{\includegraphics[width=0.39\textwidth]{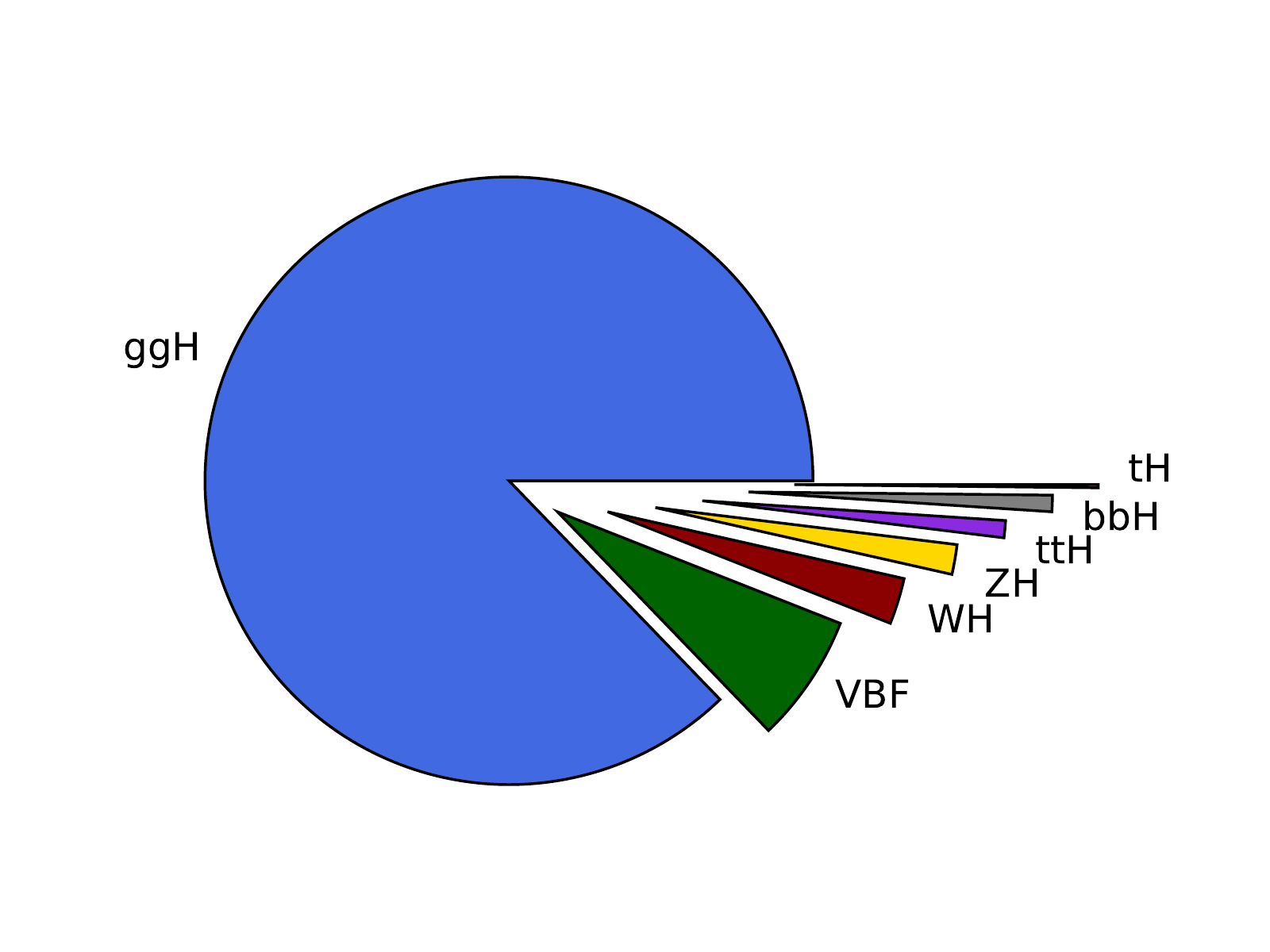}} &  
\multicolumn{3}{c}{\includegraphics[width=0.39\textwidth]{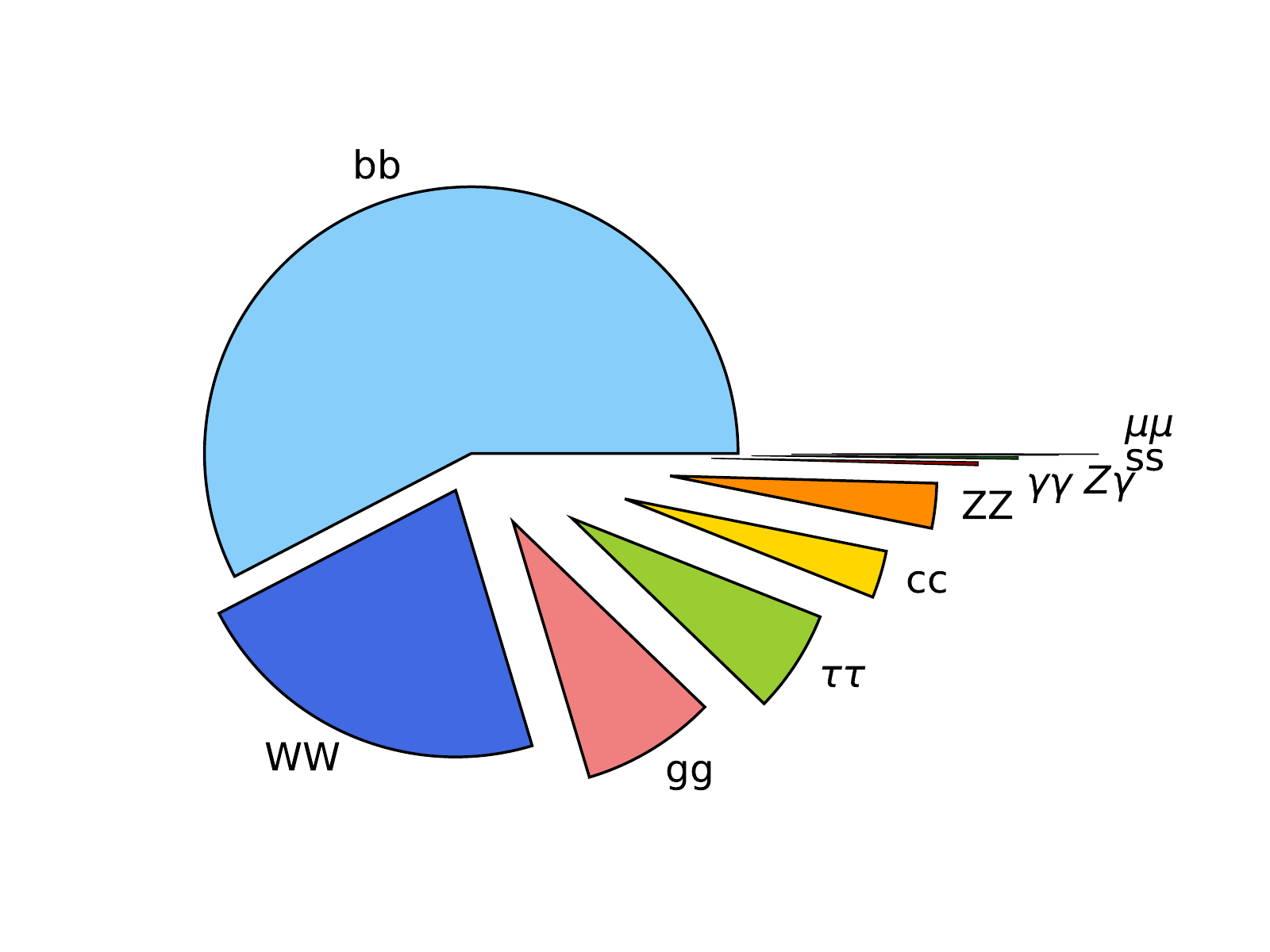}} \\
\hline 
\end{tabular}
}
\label{Extended-TabXS}
\end{table*}

\begin{table*}[tp!]
  \centering
	\topcaption{
        \figtitle{Summary of the Higgs boson analyses included in this paper.}
	The analysis and decay channels are indicated in the first two columns, with the third column containing
        the production mechanism and kinematic regions targeted	by each analysis.
        All analyses, apart from $\ttH$ in the $\PH \to \bb$ final state (2016 data only)
        and $\VH$ in the $\PH \to \bb$ final state (2016--2017 data), use the full data set collected in Run~2.  
        The various symbols are as follows: $\ell$ is $\Pe$ or $\PGm$, jet (j), di-jet mass ($m_\text{jj}$), number of jets ($N_\text{j}$), 
        same-sign (SS) of electric charge, 
        hadronic decay of the tau lepton ($\tauh$).}
	\label{Extended-Tab1}
	\cmsTable{
	\begin{tabular}{ccl}
	
	Analysis & Decay tags  & Production tags   \\
	\hline 
	\multicolumn{3}{l}{Single Higgs boson production} \\
	\hline\\ 
	\multirow{5}{*}{$\PH \to \PGg\PGg$~\cite{CMS:2021kom}} &  & $\ggH$, $\pt(\PH)$ $\times$ $N_\text{j}$ bins  \\
	 
	 &   & VBF/VH hadronic, $\pt(\PH \text{jj})$ bins      \\
	  &   & WH leptonic, $\pt(\PV)$ bins      \\
	 &   &  ZH leptonic \\
	 
	 & \multirow{-5}{*}{$\PGg\PGg$}  & $\ttH$ $\pt(\PH)$ bins, $\PQt\PH$       \\ [\cmsTabSkip]
	
	\multirow{5}{*}{$\PH \to \ZZ \to 4 \ell$~\cite{CMS:2021ugl}} & & $\ggH$, $\pt(\PH)$ $\times$ $N_\text{j}$ bins       \\
	 
	 & & VBF, $m_\text{jj}$ bins               \\
	 
	 & & V$\PH$ hadronic        \\
	 
	 & & V$\PH$ leptonic, $\pt(\PV)$ bins      \\
	 
	& \multirow{-5}{*}{4$\PGm$, $2\Pe2\PGm$, 4$\Pe$}  & $\ttH$  \\ [\cmsTabSkip]
        
	\multirow{5}{*}{$\PH \to \PW \PW \to \ell\PGn\ell\PGn$~\cite{CMS:HWW2022} } &
        $\Pe\PGm / \Pe\Pe / \PGm\PGm$ & $\ggH$ $\leq$ 2-jets        \\
	 
    &   & VBF \\
    & \multirow{-2}{*}{$\PGm\PGm$+jj/$\Pe\Pe$+jj/$\Pe\PGm$+jj}  & V$\PH$ hadronic \\
	 
	& $3\ell$  & $\PW\PH$ leptonic     \\
	 
	& $4\ell$  & $\PZ\PH$ leptonic  \\ [\cmsTabSkip]

  \multirow{2}{*}{$\PH \to \PZ \PGg$~\cite{CMS:2022ahq}}  &   & $\ggH$ \\ 
       &   \multirow{-2}{*}{$\PZ\PGg$} & VBF \\ [\cmsTabSkip]
	 
	\multirow{4}{*}{$\PH \to \ditau$~\cite{CMS:2022kdi}} &  & $\ggH$, $\pt(\PH)$ $\times$ $N_\text{j}$ bins        \\
	 
	&  & V$\PH$ hadronic      \\
	&  & VBF       \\
	 
	& \multirow{-4}{*}{$\Pe\PGm, \Pe\tauh, \PGm\tauh, \tauh\tauh$}  & V$\PH$, high-$\pt(\PV)$  \\ [\cmsTabSkip]
	
	\multirow{4}{*}{$\PH \to \bb$~\cite{CMS:2017odg,CMS:2018kst,CMS:2018sah,CMS:2018hnq,CMS:2020zge}} & $\PW(\ell\PGn)\PH(\bb)$  & $\PW\PH$ leptonic      \\
	 
	& $\PZ(\PGn\PGn)\PH(\bb)$, $\PZ(\ell\ell)\PH(\bb)$  & $\PZ\PH$ leptonic \\
	  
					 & & $\ttH$, $\tt \to$ 0, 1, 2$\ell+\mathrm{jets}$    \\
	& \multirow{-2}{*}{$\bb$} & $\ggH$, high-$\pt(\PH)$ bins               \\ [\cmsTabSkip]
 
  \multirow{2}{*}{$\PH \to \PGm\PGm$~\cite{CMS:2020xwi}} &	 & $\ggH$   \\ 
	&\multirow{-2}{*}{$\PGm\PGm$}& VBF  \\ [\cmsTabSkip]

	$\ttH$ production & $2\ell$\,SS, $3\ell$, $4\ell$,  &     \\
	 
	 with $\PH\to\mathrm{leptons}$~\cite{CMS:2020mpn} & $1\ell + \tauh$, $2\ell$\,SS$+1\tauh$, $3\ell+1\tauh$  & \multirow{-2}{*}{$\ttH$}   \\ [\cmsTabSkip]
	
	\multirow{4}{*}{$\PH \to$ Inv.~\cite{CMS:2021far,CMS:2020ulv}} &  \multirow{4}{*}{$\ptmiss$}  & $\ggH$ \\ 
		 & &  VBF \\ 
			 &  & $\PV\PH$ hadronic \\ 
				& & $\PZ\PH$ leptonic \\ [\cmsTabSkip]

	\hline
	\multicolumn{3}{l}{Higgs boson pair production} \\
	\hline\\
	$\PH\PH \to  \bb\bb$~\cite{CMS:2022cpr,CMS:HH4b-comb} 	& $\PH(\bb)\PH(\bb)$ 	 & $\ggH\PH$, VBF$\PH\PH$ (resolved, boosted) \\ [\cmsTabSkip]
	$\PH\PH \to \bb\ditau$~\cite{CMS:HHbbtautau}   & $\PH(\bb)\PH(\ditau)$ & $\ggH\PH$, VBF$\PH\PH$ \\  [\cmsTabSkip]
	$\PH\PH \to$ leptons~\cite{CMS:HHleptons} & $\PH(\WW)\PH(\WW)$, $\PH(\WW)\PH(\ditau)$, $\PH(\ditau)\PH(\ditau)$ & $\ggH\PH$, VBF$\PH\PH$  \\ [\cmsTabSkip]
	$\PH\PH \to \bb\PGg\PGg$~\cite{CMS:HHggbb} 	& $\PH(\bb)\PH(\PGg\PGg) $ & $\ggH\PH$, VBF$\PH\PH$ \\  [\cmsTabSkip]
  $\PH\PH \to \bb\ZZ$~\cite{CMS:HHbbZZ}	& $\PH(\bb)\PH(\ZZ)$ 	 & $\ggH\PH$ \\[\cmsTabSkip]
        \hline
	\end{tabular}
	}
\end{table*}

\begin{table*} [htb]
\centering
\topcaption{
    \figtitle{Summary of the event selections.}
    Some of the typical selection criteria used in the trigger (online selection) and in offline analysis for some of the final states and for leading (1) and
  subleading (2) particles. The \ptmiss is a measure of the imbalance in energy in the plane transverse to the colliding proton beams.}
\cmsTable{
\begin {tabular}{l  c  c  c}

Analysis & 	Physics objects  & 	Trigger selections threshold [\GeVns{}]  & Kinematic requirements [\GeVns{}]	\\ \hline \\
$\PH \to \PGg\PGg$          &  $\PGg$ 					     & $\pt^{\PGg} (1/2) > 30/18 $   & $\pt^{\PGg} > 35/25 $ \\ [\cmsTabSkip]
$\PH \to \ZZ \to 4 \ell$    &  $\PGm$, $\Pe$
                                           & $\pt^{\PGm} (1/2) > 23/8 $   & $\pt^{\PGm} > 5 $ \\
& 			                   &  $\pt^{\Pe} (1/2)> 17/8 $     & $\pt^{\Pe} > 7  $ \\ [\cmsTabSkip]
$\PH \to \PW \PW \to \ell \PGn \ell \PGn$ &  $\PGm$, $\Pe$,  \ptmiss     &  $\pt^{\PGm} (1/2) > 23/12 $   & $\pt^{\PGm} (1/2) > 25/13 $ \\
				         &   &  $\pt^{\Pe} (1/2) > 23/12  $     & $\pt^{\Pe} (1/2) > 25/13 $ \\
& 				&   & \ptmiss $> 20 $ \\ [\cmsTabSkip]
$\PH \to \PZ\PGg$             &  $\PGm$, $\Pe$, $\PGg$ & $\pt^{\PGm} (1/2) > 17/8  $ &  $\pt^{\PGm}  (1/2) > 20/10 $\\
                                                 &    & $\pt^{\Pe} (1/2) > 23/12 $ &  $\pt^{\Pe} (1/2) > 25/15 $\\
                                                &  &   &  $\pt^{\PGg} > 15 $\\ [\cmsTabSkip]
$\PH \to \ditau$	       &  $\PGm$, $\Pe$,  $\tauh$          &     $\pt^{\PGm} > 20 $ &  $\pt^{\PGm} > 20 $\\
				         & 			     &$\pt^{\Pe} > 24 $     & $\pt^{\Pe} > 25 $ \\ 
& 			   &$\pt^{\tauh} > 35 $     & $\pt^{\tauh} > 40 $ \\  [\cmsTabSkip]
$\PH \to \bb$       & $\PGm$, $\Pe$, \ptmiss,  jet    				&       $\pt^{\PGm} > 22 $   & $\pt^{\PGm} > 25  $ \\	
& 			                   &  $\pt^{\Pe} > 32 $     & $\pt^{\Pe} > 30 $ \\
& 				&   $\ptmiss > 120 $ & $\ptmiss > 170 $ \\
& 				&  $E_{\text{j}} >330 $  & $E_{\text{j}} > 450 $ \\ [\cmsTabSkip]
$\PH \to \PGm\PGm$       &  $\PGm$   &    $\pt^{\PGm} > 24 $   & $\pt^{\PGm} (1/2) > 26/20 $ \\ [\cmsTabSkip]
\hline
\end{tabular}
}
\label{Extended-TabCut}
\end{table*}

\begin{figure*}[htb]
\centering

\includegraphics[width=0.48\textwidth]{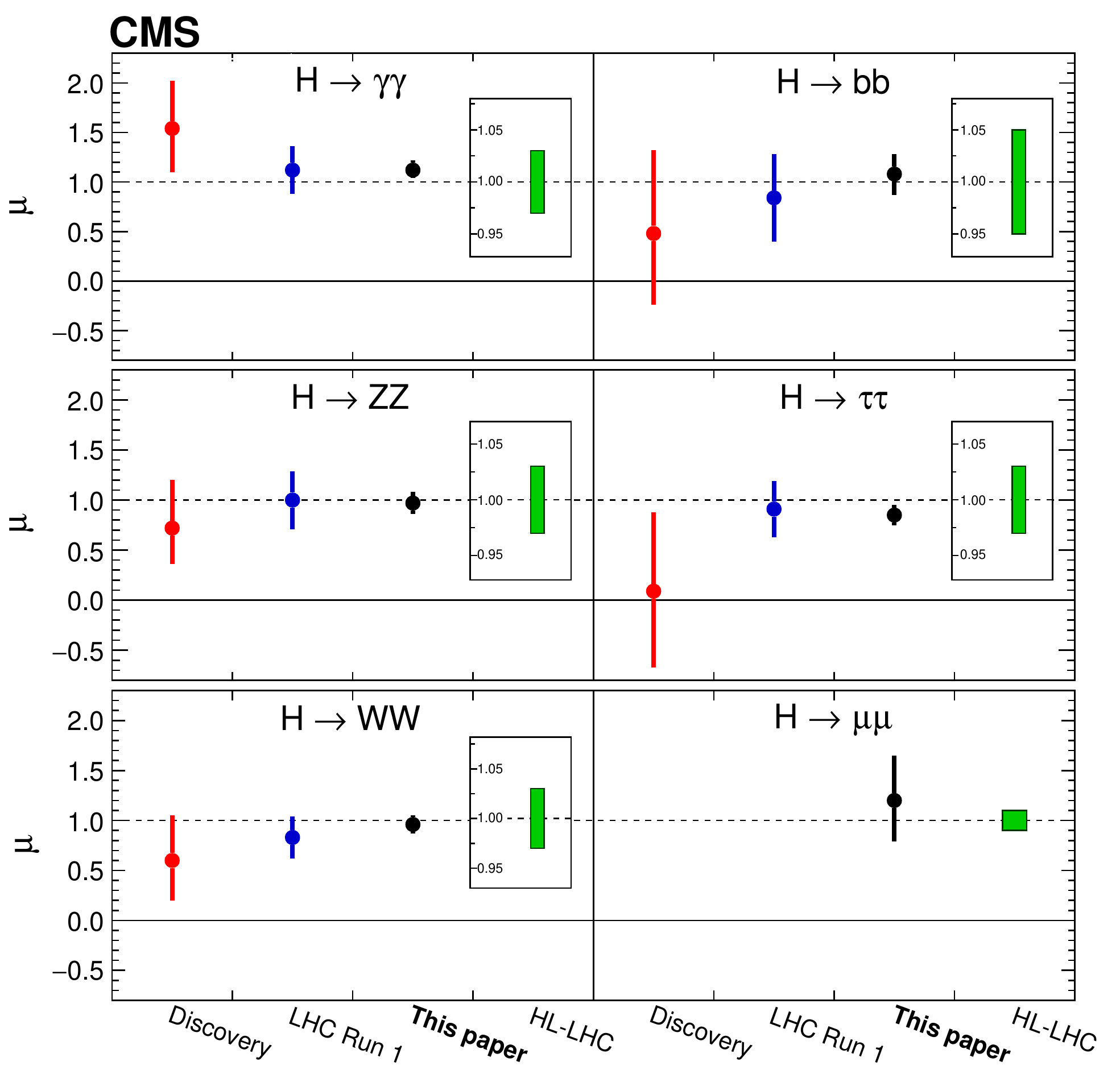}
\caption{
\figtitle{Time evolution of the signal strength measurements and their precision.}
    Comparison of the signal strength parameter $\mu$ fit results in different data sets; in each panel, from left to right: 
at the time of the Higgs boson discovery, using the full data from LHC Run~1, in the data set analyzed for this paper, and the expected 1~s.d. uncertainty for HL-LHC for $\lint=3000\fbinv$. The  $\PH \to \PGm\PGm$ measurements were not available for the earlier data sets due to the lack of sensitivity.}
\label{Extended-Fig5}
\end{figure*}

\begin{figure*}[htb]
\centering
\includegraphics[width=0.485\textwidth]{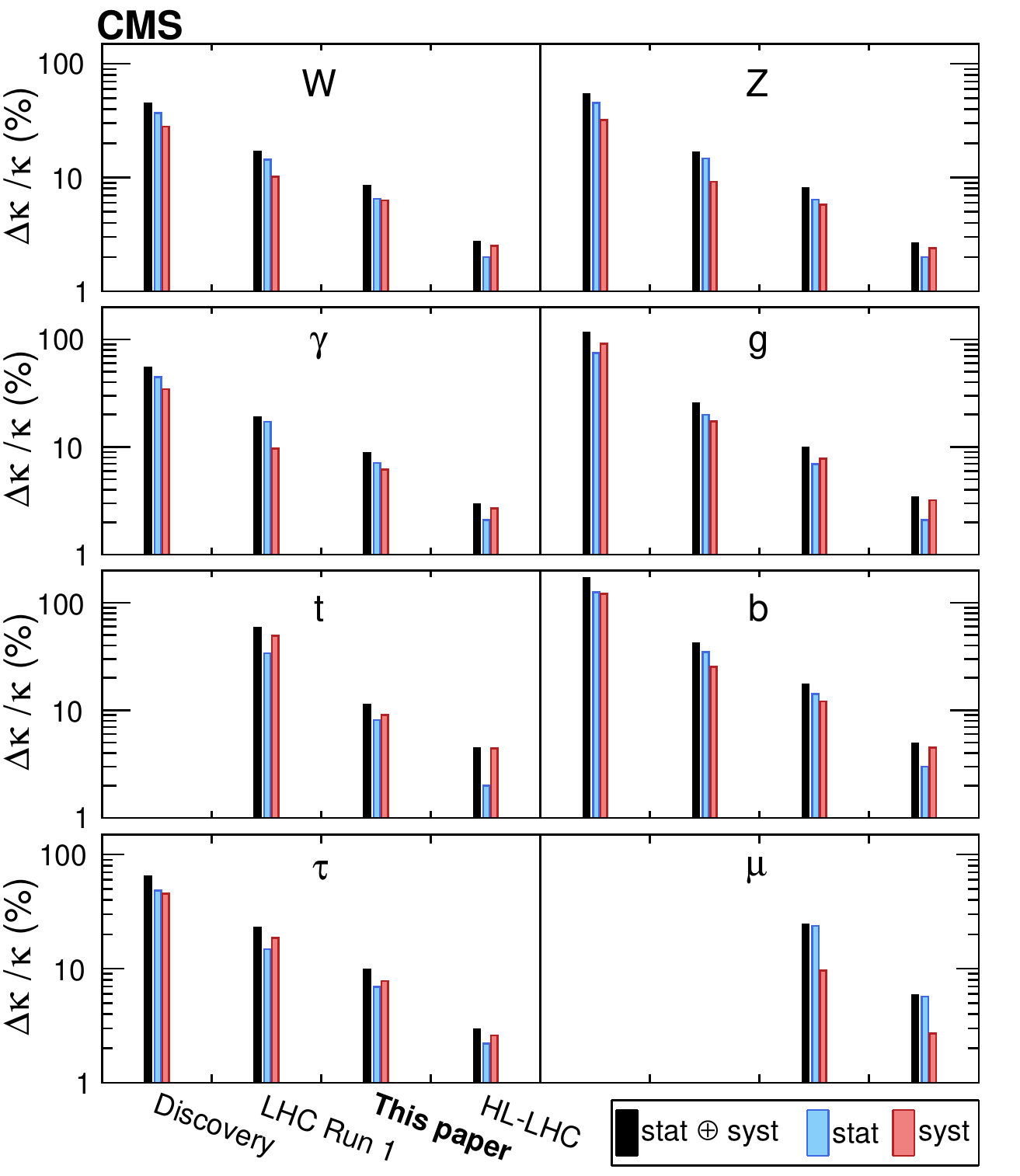}
\includegraphics[width=0.495\textwidth]{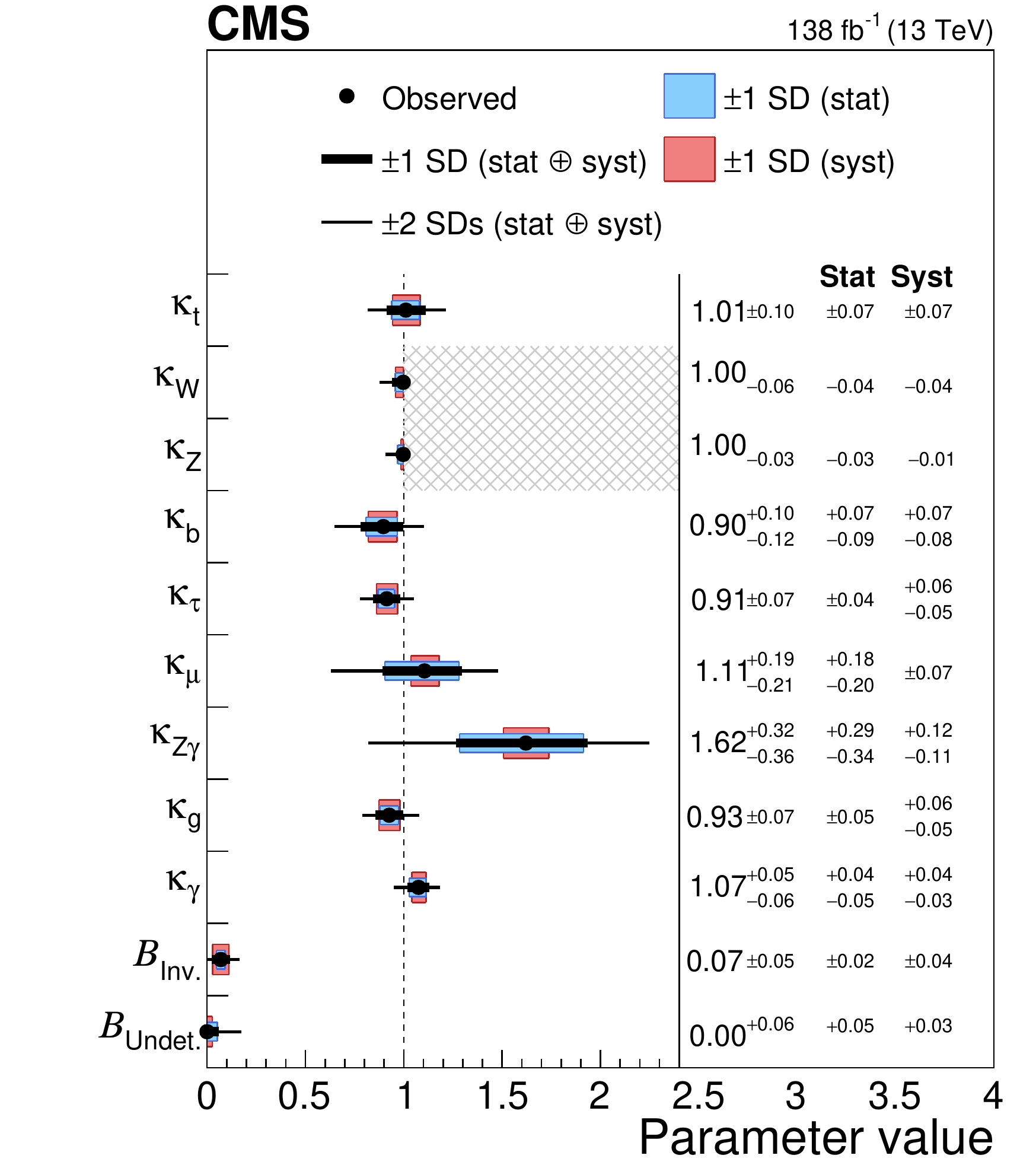}
    \caption{
\figtitle{Time evolution of the coupling measurements and their precision.}
        (left) Comparison of the expected 1~s.d. uncertainties in the $\kappa$-framework fit including coupling modifiers 
        for both tree-level and loop-induced Higgs boson 
interactions, in different data sets: at the time of the Higgs boson discovery, using the full
data from LHC Run~1, in the data set used in this paper, and
the expected 1 s.d. uncertainty for HL-LHC for $\lint =3000\fbinv$. 
        (right) Results of a fit to the coupling modifiers $\kappa$ allowing both invisible and the undetected decay modes, with the SM value used as an upper bound on both $\kappa_\PW$ and $\kappa_\PZ$.
        The thick (thin) black lines indicate the 1 (2) s.d. confidence
        intervals, with the systematic and statistical components of
        the 1~s.d. interval indicated by the red and blue bands,
        respectively. The $p$-value with respect to the SM prediction is 33\%.
    }
\label{Extended-FigZ}
\end{figure*}

\begin{figure*}[htb]
  \centering
\includegraphics[width=0.43\textwidth]{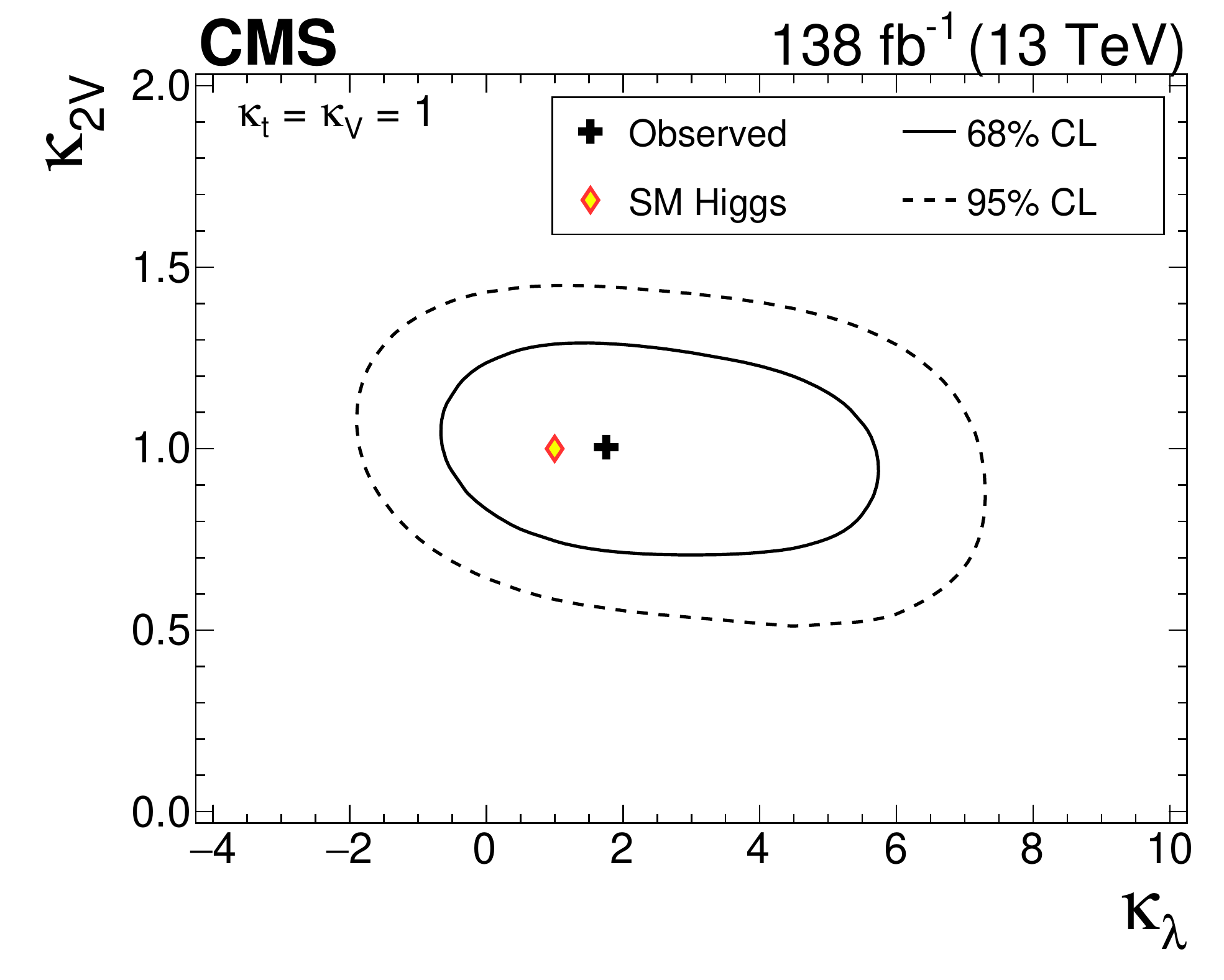}
\includegraphics[width=0.56\textwidth]{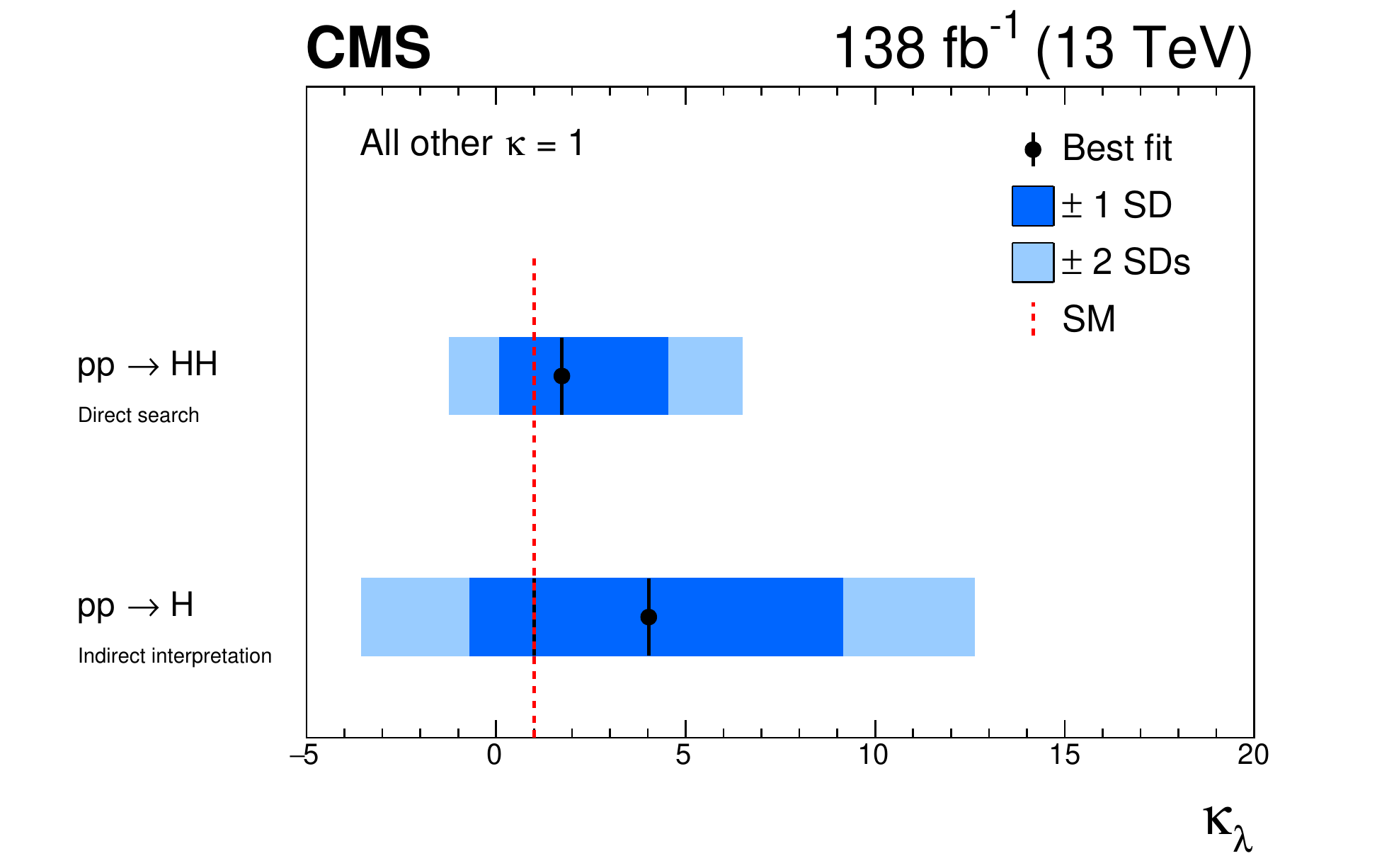}
    \caption{
  \figtitle{Constraints on Higgs boson self-interaction and quartic coupling.}
    (left) Constraints on $\kappal$ and $\kappavv$ from the production of Higgs boson pairs.
    (right) Constraint on the Higgs boson self-coupling modifier $\kappal$
from single and pair production of Higgs boson(s).
    }
\label{Extended-FigKL}
\end{figure*}
\cleardoublepage \section{The CMS Collaboration \label{app:collab}}\begin{sloppypar}\hyphenpenalty=5000\widowpenalty=500\clubpenalty=5000\input{HIG-22-001-authorlist.tex}\end{sloppypar}
\end{document}

%% file: HIG-22-001-authorlist.tex
\cmsinstitute{Yerevan~Physics~Institute, Yerevan, Armenia}
{\tolerance=6000
A.~Tumasyan
\par}
\cmsinstitute{Institut~f\"{u}r~Hochenergiephysik, Vienna, Austria}
{\tolerance=6000
W.~Adam\cmsorcid{0000-0001-9099-4341}, J.W.~Andrejkovic, T.~Bergauer\cmsorcid{0000-0002-5786-0293}, S.~Chatterjee\cmsorcid{0000-0003-2660-0349}, K.~Damanakis\cmsorcid{0000-0001-5389-2872}, M.~Dragicevic\cmsorcid{0000-0003-1967-6783}, A.~Escalante~Del~Valle\cmsorcid{0000-0002-9702-6359}, P.S.~Hussain\cmsorcid{0000-0002-4825-5278}, M.~Jeitler\cmsAuthorMark{1}\cmsorcid{0000-0002-5141-9560}, N.~Krammer\cmsorcid{0000-0002-0548-0985}, L.~Lechner\cmsorcid{0000-0002-3065-1141}, D.~Liko\cmsorcid{0000-0002-3380-473X}, I.~Mikulec\cmsorcid{0000-0003-0385-2746}, P.~Paulitsch, F.M.~Pitters, J.~Schieck\cmsAuthorMark{1}\cmsorcid{0000-0002-1058-8093}, R.~Sch\"{o}fbeck\cmsorcid{0000-0002-2332-8784}, D.~Schwarz\cmsorcid{0000-0002-3821-7331}, S.~Templ\cmsorcid{0000-0003-3137-5692}, W.~Waltenberger\cmsorcid{0000-0002-6215-7228}, C.-E.~Wulz\cmsAuthorMark{1}\cmsorcid{0000-0001-9226-5812}
% ...
\par}
\vskip\cmsinstskip
\dag:~Deceased\\
$^{1}$Also at TU Wien, Vienna, Austria\\
% ...
